\documentclass[preprint,review,12pt]{elsarticle}

\usepackage{fullpage}
\usepackage{amssymb}
\usepackage{amsmath}
\usepackage{multirow}
\usepackage[breaklinks = true,colorlinks = true]{hyperref}
\usepackage{breqn}
\usepackage{tikz}
\usetikzlibrary{arrows,positioning,shapes.geometric}
\usepackage{caption}
\usepackage{subcaption}

\begin{document}

\begin{frontmatter}

\title{A stable SPH with adaptive B-spline kernel}

\author[a]{Saptarshi Kumar Lahiri}

\author[a]{Kanishka Bhattacharya}

\author[a]{Amit~Shaw\corref{cor1}}
\ead{abshaw@civil.iitkgp.ac.in}

\author[a]{{L. S. Ramachandra}}

\cortext[cor1]{Corresponding author}

\address[a]{Civil Engineering Department, Indian Institute of Technology Kharagpur, West Bengal, India}

\begin{abstract}
\textit{Tensile instability}, often observed in smoothed particle hydrodynamics (SPH), is a numerical artefact that manifests itself by unphysical clustering or separation of particles. The instability originates in estimating the derivatives of the smoothing functions which, when interact with material constitution may result in negative stiffness in the discretized system. In the present study, a stable formulation of SPH is developed where the kernel function is continuously adapted at every material point depending on its state of stress. B-spline basis function with a variable intermediate knot is used as the kernel function. The shape of the kernel function is then modified by changing the intermediate knot position such that the condition associated with instability does not arise. While implementing the algorithm the simplicity and computational efficiency of SPH are not compromised. One-dimensional dispersion analysis is performed  to understand the effect adaptive kernel on the stability. Finally, the efficacy of the algorithm is demonstrated through some benchmark elastic dynamics problems. 
\end{abstract}


\end{frontmatter}


\section{Introduction}
\label{sec1}
The smoothed particle hydrodynamics (SPH) \citep{lucy1977numerical,gingold1977smoothed} is a particle-based method, evolved as an alternative approach for numerical simulation of many physical processes which are otherwise difficult to model through mesh-based methods. This includes problems in astrophysics \citep{springel2010smoothed, monaghan2012smoothed}, fluid dynamics \citep{xu2013sph, monaghan2013simple}, impact mechanics \citep{Johnson1996, mehra2006high, shaw2009heuristic}, large deformation \citep{shaw2015beyond}, fracture \citep{benz1995simulations, Rabczuk2003,chakraborty2013pseudo, chakraborty2017computational, islam2017computational} etc. In SPH, the spatial discretization through particles (instead of elements) provides relief from the numerical difficulty associated with element distortion. Moreover, the absence of explicit connectivity between particles makes SPH better equipped to deal with discontinuity and material separation. However, all these advantages are often overshadowed by some computational pitfalls which may result in inaccurate and sometime unstable computation if not dealt with properly. The \textit{tensile instability} is perhaps the major concern among them.

Such instability in SPH was first witnessed by \citet{schuessler1981comments} in gas dynamics problems and subsequently by \citet{phillips1985numerical} in magnetohydrodynamics problems. Therein the authors observed a prominent clustering or separation of particles, which were not in the line with the realistic physical behaviour of the system. Almost after a decade since the instability was first reported, \citet{swegle1995smoothed} performed a stability analysis of SPH. In their analysis, they used the cubic B-spline kernel function which is most common in SPH computations. For such kernel, if smoothing length is taken same as the particle spacing the instability occurs explicitly under tensile state of stress and thus coined as the \textit{tensile instability}. However, depending on the kernel function and smoothing length (with respect to the inter particle spacing) the instability may also occur under compressive stresses. As per the analysis by \citet{swegle1995smoothed}, the instability is due to the manifestation of a negative stiffness resulting from the interaction of kernel function and the material constitution. Therein, it was also shown that the \textit{tensile instability} cannot be removed by increasing artificial viscosity, though the artificial viscosity may control, to some extent, the perturbation growth rate.  

In the literature several corrective measures are suggested to deal with the \textit{tensile instability}. \citet{dyka1995approach, dyka1997stress} proposed a stress point approach, where the field variables such as density, stress and energy are evaluated at some prescribed stress points placed in between the actual SPH particles. Velocities are calculated at the particles and mapped to the stress points using the moving least square (MLS) technique. The concept is also extended to higher dimension in \citep{randles2000normalized}. The approach estimated better accuracy in one dimensional problems. However, tracking, velocity mapping and subsequent updating of stress points make the approach computationally intensive especially in higher dimension. Through an unified stability analysis, \citet{belytschko2000unified} highlighted that the stress points are not effective if Eulerian kernel is used. They also suggested to use a Lagrangian kernel (with or without stress points) for a stable computation. Total Lagrangian formulation is also adopted in \citep{guenther1994conservative, swegle1994analysis}. Though the total Lagrangian formulation does not suffer any \textit{tensile instability}, the difficulty arises while modelling large deformation and discontinuities in computational domain due to material fracture. Moreover the particle distribution near the highly deformed region becomes very distorted which may yield poor convergence rate \citep{khayyer2011enhancement}. \citet{monaghan2000sph, gray2001sph} proposed the concept of artificial stress. Therein an artificial stress term is introduced in the momentum Equation, which produces a short range repulsive force between the neighbouring particles in order to keep them in a stable configuration. The parameters which controls the intensity for the artificial stress are determined through a dispersion analysis. Although this approach is widely accepted, the form and associated parameters of artificial stress depend on the physical process to be studied, governing Equations and also the choice of kernel function. Over the years few other approaches are also proposed to overcome the \textit{tensile instability}, such as conservative smoothing \citep{guenther1994conservative, swegle1994analysis, hicks2004conservative}, corrected kernel estimate \citep{dilts1999moving,chen1999improvement} etc.  

All the corrective measures mentioned above either are computationally intensive or require some artificial penalty force that need to be tuned depending on the problem. Moreover these approaches generally suppress the instability, instead of completely removing them from the system. As the \textit{tensile instability} is ascribed to the shape of the kernel function, use of adaptive kernel would be more suitable and intrinsic to the root cause of the instability. In the present study an adaptive algorithm is developed where the shape of the kernel function is continuously modified such that the condition associated with the instability does not arise. B-spline basis function constructed over a variable knot vector is taken as the kernel in the proposed formulation. The shape of the kernel is adapted by changing the location of the intermediate knots. The theoretical basis of the algorithm is discussed. One dimensional dispersion analysis is also performed to investigate the effect of adaptive kernel on the stability. Through some benchmark elastic dynamics problem it shown that the algorithm can remove the \textit{tensile instability} in a computationally simpler and efficient way. 

\section{SPH and Tensile Instability}
For a better comprehension of the development to follow, a brief account of SPH (conservation Equations and discretization) and the \textit{tensile instability} are provided in this section. While revisiting SPH, the basic steps which are followed in the present study are only outlined. Detail information may be found in \citep{liu2003smoothed} and the references therein. 

\subsection{SPH}
The conservation Equations (mass, momentum and energy) in a Lagrangian framework are given by,
\begin{equation}
\frac{d\rho}{dt} = -\rho\frac{\partial \mathbf{v}^{\beta}}{\partial \mathbf{x}^{\beta}}
\label{density}
\end{equation}
\begin{equation}
\frac{d\mathbf{v}^{\alpha}}{dt} = \frac{1}{\rho}\frac{\partial\sigma^{\alpha\beta}}{\partial \mathbf{x}^{\beta}} 
\label{moment}
\end{equation}
\begin{equation}
\frac{de}{dt} = \frac{\sigma^{\alpha\beta}}{\rho}\frac{\partial \mathbf{v}^{\beta}}{\partial \mathbf{x}^{\beta}} 
\label{energy}
\end{equation}
where, $\rho$ is the density; $v^\alpha$ and $\sigma^{\alpha \beta}$ are the components of velocity and Cauchy stress tensor respectively; $e$ is the specific energy; $\alpha$ and $\beta$ are the spatial coordinates. 

The computational domain $\bar{\Omega}=\partial\Omega\bigcup\Omega$ is discretized in to a set of particles positioned at ${\{\mathbf{x}_i}\}_{i=1}^N$, where $N$ is the total number of particles. Let ${\{m_i}\}_{i=1}^N$, ${\{\rho_i}\}_{i=1}^N$, ${\{p_i}\}_{i=1}^N$,${\{v^{\alpha}_i}\}_{i=1}^N$, ${\{{\sigma_i}^{\alpha \beta}}\}_{i=1}^N$ be the discrete values of mass, density, pressure, velocity and Cauchy stress respectively. Any continuous field variable is approximated as a functional  representation of the discrete values through a smoothed weighting function or kernel function as, 
\begin{equation}
<f(\mathbf{x})> \cong \sum_{j \in \mathbb{N}^i} f_j W(\mathbf{x}-\mathbf{x}_j,\kappa h)\frac{m_j}{\rho_j},
\label{eq1}
\end{equation}
where $W\left(\mathbf{x},\kappa h\right)$ is a bell-shaped compactly supported kernel function with centre at $\mathbf{x}$ and smoothing length $\kappa h$, $\kappa$ being the support size of the kernel in its parametric space; and $\mathbb{N}^i$ is the influence domain of $i$-th particle (i.e., set of indices of other particles which interact with $i$-th particle) defined as $\mathbb{N}^i = \lbrace j \in \mathbb{Z}^+ ~|~ |\mathbf{x}_i-\mathbf{x}_j| < \kappa h ~and~ i \neq j \rbrace$. 

SPH being a particle based method, requires the integral forms of the continuous conservation laws [Eqs. \eqref{density} - \eqref{energy}] to be discretized at each particle position using the kernel approximations [Equation \eqref{eq1}]. The discrete form of the conservation Equations are,
\begin{equation}
\frac{d\rho_i}{dt} = \sum_{j \in \mathbb{N}^i} {m_j ({v_i}^\beta-{v_j}^\beta)W_{ij,\beta}},
\label{disdensity}
\end{equation}
\begin{equation}
\frac{dv_i^{\alpha}}{dt} = \sum_{j \in \mathbb{N}^i} {m_j\left(\frac{\sigma_i^{\alpha\beta}}{\rho_i^{2}}+\frac{\sigma_j^{\alpha\beta}}{\rho_j^{2}}-\Pi_{ij}\delta^{\alpha \beta}\right) W_{ij,\beta}},
\label{dismoment}
\end{equation}
\begin{equation}
\frac{de_i}{dt} = -\frac{1}{2} \sum_{j \in \mathbb{N}^i} {m_j({v_i}^\beta-{v_j}^\beta)\left(\frac{\sigma_i^{\alpha\beta}}{\rho_i^{2}}+\frac{\sigma_j^{\alpha\beta}}{\rho_j^{2}} - \Pi_{ij} \delta^{\alpha \beta}\right) W_{ij,\beta}},
\label{disenergy}
\end{equation}
where, $W_{ij,\beta} = \frac{\partial W(\mathbf{x},\kappa h)}{\partial \mathbf{x}^{\beta}}|_{(\mathbf{x}_i-\mathbf{x}_j)}$. In Equation \ref{dismoment} and \ref{disenergy}, \textbf{$\Pi_{ij}$} is the artificial viscosity \citep{monaghan1988introduction} which is often required to stabilize the numerical algorithm in the presence of shock. The following form \cite{gray2001sph} of  artificial viscosity is used in the present study. 
\begin{equation}
\Pi_{ij}=
\begin{cases}
		\frac{-\gamma_1\overline{C}_{ij}\mu_{ij} + \gamma_2\mu^2_{ij}}{\overline{\rho}_{ij}^2} & \text{for} \, \mathbf{x}_{ij}.\mathbf{v}_{ij} < 0\\
		0 & \text{otherwise} \\
\end{cases}
\label{artvisc}
\end{equation}
where, $\mu_{ij}= \frac{h\left(\mathbf{v}_{ij}.\mathbf{x}_{ij}\right)}{X^2_{ij} + \eta h^2}$; $\overline{C}_{ij} = \frac{C_i + C_j}{2}$; $\overline{\rho}_{ij} = \frac{\rho_i + \rho_j}{2}$; $\gamma_1$ and $\gamma_2$ are parameters which control the intensity of the artificial viscosity; $\eta$ is a small number to avoid singularity when two interacting particles ($i$-th and $j$-th) are  close to each other; $C_i = \sqrt{\frac{E}{\rho_i}}$ and $C_j = \sqrt{\frac{E}{\rho_j}}$ are the wave propagation speed evaluated at $i$-th and $j$-th particles respectively; and $\mathbf{v}_{ij} = \mathbf{v}_i- \mathbf{v}_j$ and $\mathbf{x}_{ij} = \mathbf{x}_i- \mathbf{x}_j$ indicate the relative velocity and displacement of the $i-j$ particle pair. 

The particles are moved as per the XSPH Equation \citep{monaghan1989problem} as,

\begin{equation}
\frac{dx_i^\alpha}{dt} = v_i^\alpha  - \epsilon \sum_{j \in \mathbb{N}^i} \frac{m_j}{\overline{\rho}_{ij}} v_{ij}^\alpha W_{ij}, 
\label{XSPH}
\end{equation}
where $v_{ij}^\alpha=v_i^\alpha-v_j^\alpha$ and $\epsilon \in [0,1]$.  

The Cauchy stress component is written in terms of hydrostatic and deviatoric stresses as $\sigma^{\alpha\beta}= -P\delta^{\alpha\beta}+S^{\alpha\beta}$, where $P$ and $S^{\alpha\beta}$ are respectively the pressure and the components of traceless symmetric deviatoric stress tensor. Pressure is computed from the following Equation of state,
\begin{equation}
p=K\left(\frac{\rho}{\rho_0} -1\right), 
\label{eos}
\end{equation}
where, $\rho_0$ is he initial density and $K$ is the bulk modulus. The discretized Equation for the deviatoric stress rate is,
\begin{equation}
\begin{split}
\label{eq_stress}
 \dot{S}_i^{\alpha \beta}=  - \sum_{j \in \mathbb{N}^i} \frac{m_j}{\rho_j} G_i\Big( v_{ij}^\alpha W_{ij,\beta}+v_{ij}^\beta W_{ij,\alpha} - \frac{2}{3} \delta^{\alpha \beta} v_{ij}^\gamma W_{ij,\gamma} \Big) + \\
  \frac{1}{2}\sum_j \frac{m_j}{\rho_j} \Big[  S^{\alpha \gamma} \Big( v_{ij}^\beta W_{ij,\gamma}-v_{ij}^\gamma W_{ij,\beta} \Big)+ S^{\gamma \beta} \Big( v_{ij}^\alpha W_{ij,\gamma}-v_{ij}^\gamma W_{ij,\alpha} \Big) \Big] 
\end{split}
\end{equation}
where $G$ is the shear modulus.

Once the Equations are discretized over space, the resulting ODEs in time are integrated via any appropriate time integration technique. In the present study the two step predictor-corrector method with constant time step is used.

\subsection{Tensile Instability} \label{TI}
In a continuum if two material points are relatively displaced (approach to or move away from each other) from their equilibrium positions, internal force develop which opposes the motion. This internal force ($F_b$) evolves with the internal stresses satisfying the conservation of linear momentum as,
\begin{equation}
F_b = m\frac{dv^{\alpha}}{dt} = m\frac{1}{\rho}\frac{\partial\sigma^{\alpha\beta}}{\partial \mathbf{x}^{\beta}}
\label{tens1}
\end{equation}
In one dimension, the SPH discretization of Equation \eqref{tens1} takes the form (without the artificial viscosity term),
\begin{equation}
m_i\frac{dv_i}{dt} = - m_i\sum_{j \in \mathbb{N}^i} {m_j\left(\frac{\sigma_i}{\rho_i^{2}}+\frac{\sigma_j}{\rho_j^{2}}\right) W'_{ij}}.
\label{tens2}
\end{equation}
The above Equation gives the cumulative force exerted on particle $i$ from its neighbour $j\in \mathbb{N}^i$. The interaction force between any particle pair (say $i-j$) is expressed as,
\begin{equation}
F_{ij} = - m_i m_j {\left(\frac{\sigma_i}{\rho_i^{2}}+\frac{\sigma_j}{\rho_j^{2}}\right) W'_{ij}},
\label{tens3} 
\end{equation}

\begin{figure}[h!]\vspace*{4pt}
    \centering
    \begin{subfigure}[t]{0.5\textwidth}
        \centering
        \includegraphics[trim={4cm 0cm 2cm 1cm},clip,width=\textwidth]{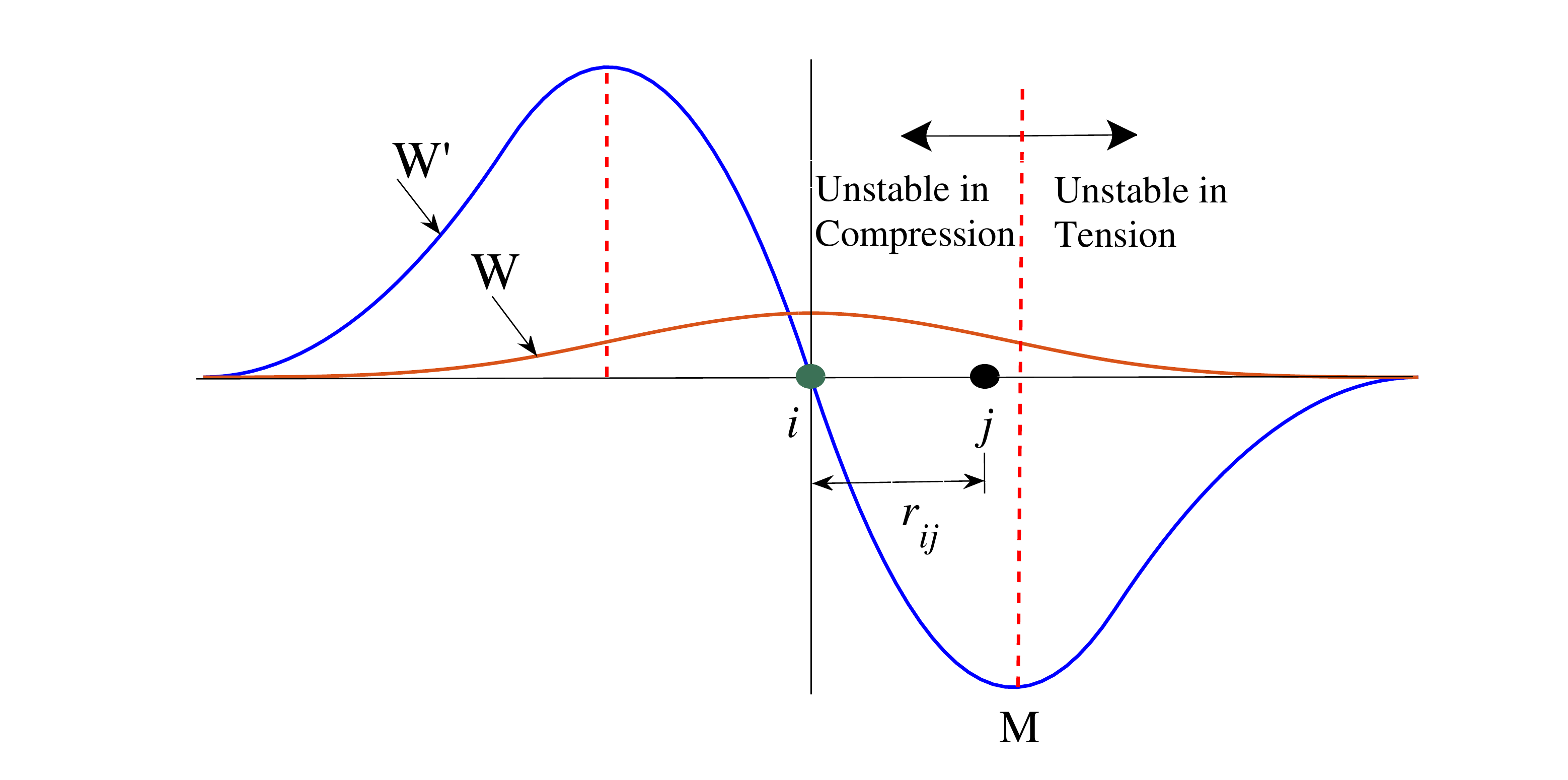}
        \caption{}\label{Instability1}
    \end{subfigure}%
    ~    
    \begin{subfigure}[t]{0.5\textwidth}
        \centering
        \includegraphics[trim={4cm 0cm 2cm 1cm},clip,width=\textwidth]{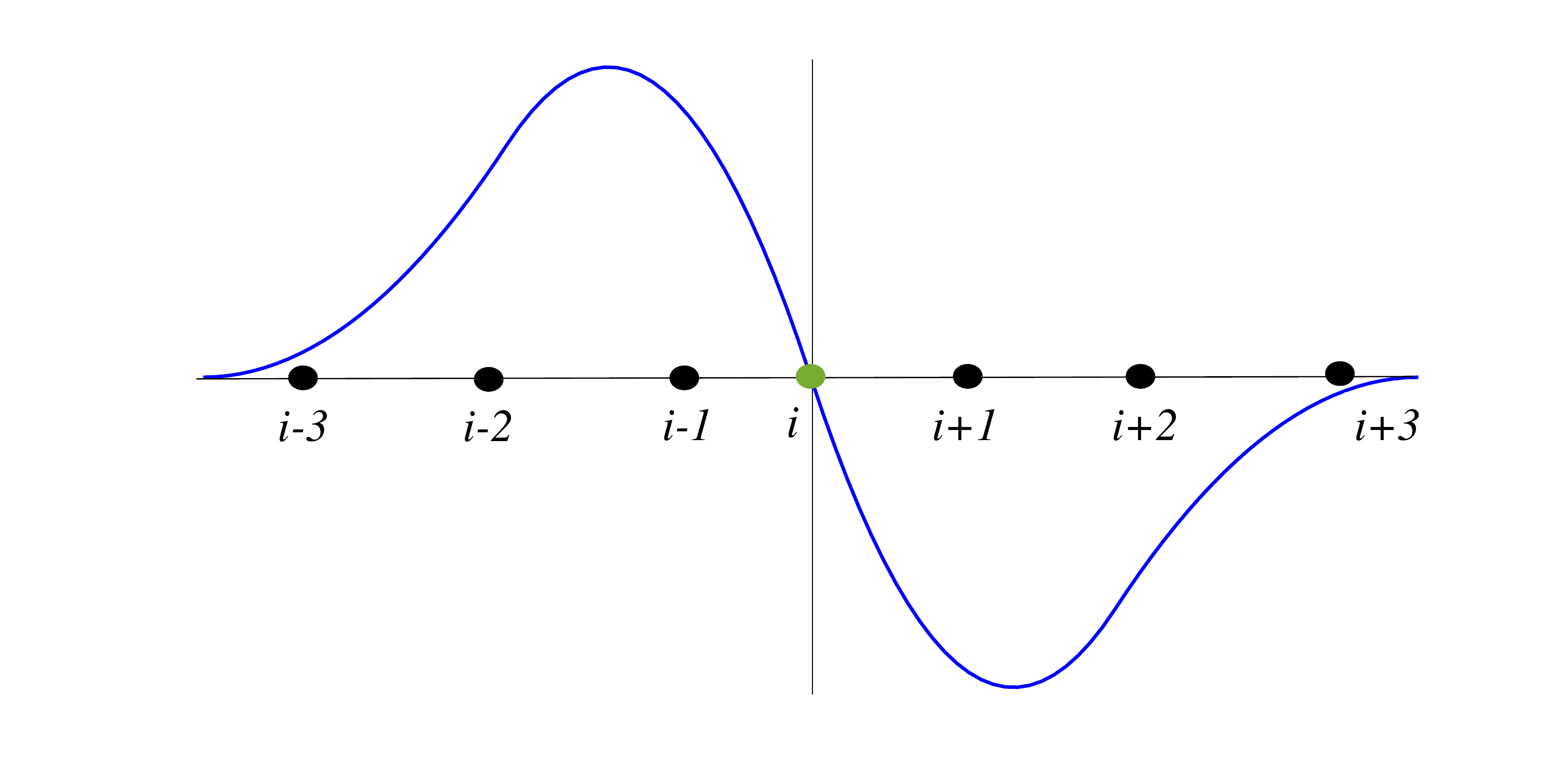}
        \caption{}\label{Instability2}
    \end{subfigure}%
     
   \caption{Geometric representation of tensile instability}\vspace*{-6pt}
    \label{Instability}
\end{figure}


From Equation \ref{tens3} it is evident that if the interacting particles are in constant state of stress (\textit{i.e} $\sigma_i = \sigma_j = \sigma$), the interaction force is guided by the first derivative ($W'$) of the kernel function. Figure \ref{Instability1} illustrates the shape of a cubic kernel (Equation \ref{BS3_st}) function ($W$) and its first derivative ($W'$) centred at the $i$-th particle i.e., $x_i$. Now consider the interaction of $i$-th paricle with any of its neighbour, say $j$-th particle. If the $j$-th particle is shifted from the equilibrium position the interaction force $F_{ij}$ between the pair ($i-j$) should increase to retain the particles ($i$ and $j$) in equilibrium state. The shape of $W'$ does not always allow such to happen. Initially starting from zero at the $i$-th particle $W'$ increases (i.e., positive $W''$) with inter-particle distance and after reaching the peak value at M it starts declining (i.e., negative $W''$) and eventually become zero at the boundary of the influence domain. Therefore, initially when $W''$ is positive the interaction force ($F_{ij}$) monotonically increases with increase in inter-particle separation. But when the $j$-th particle crosses the maxima the magnitude of the interaction force starts decreasing with further increase in inter-particle distance resulting in further separation of the $i-j$ particle pair. This situation is analogous to having a negative stiffness in the system which causes disruption in stability of the computation. Similar argument can be put in case of approaching particles. 

\begin{figure}[h!]
    \centering
    \begin{subfigure}[t]{0.33\textwidth}
        \centering
        \includegraphics[width=\textwidth]{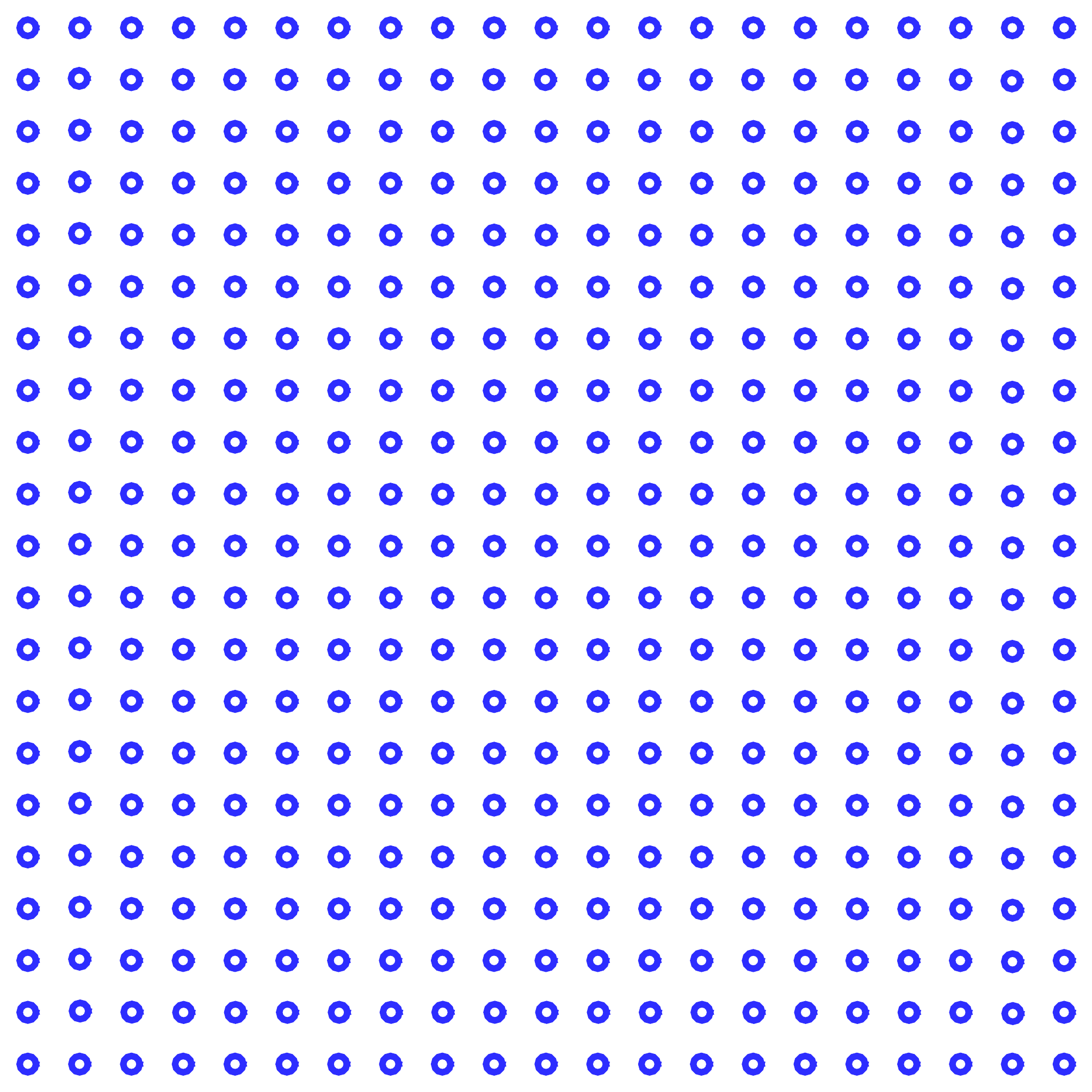}
        \caption{Initial particle distribution}\label{box1_0}
    \end{subfigure}%
    ~    
    \begin{subfigure}[t]{0.33\textwidth}
        \centering
        \includegraphics[width=\textwidth]{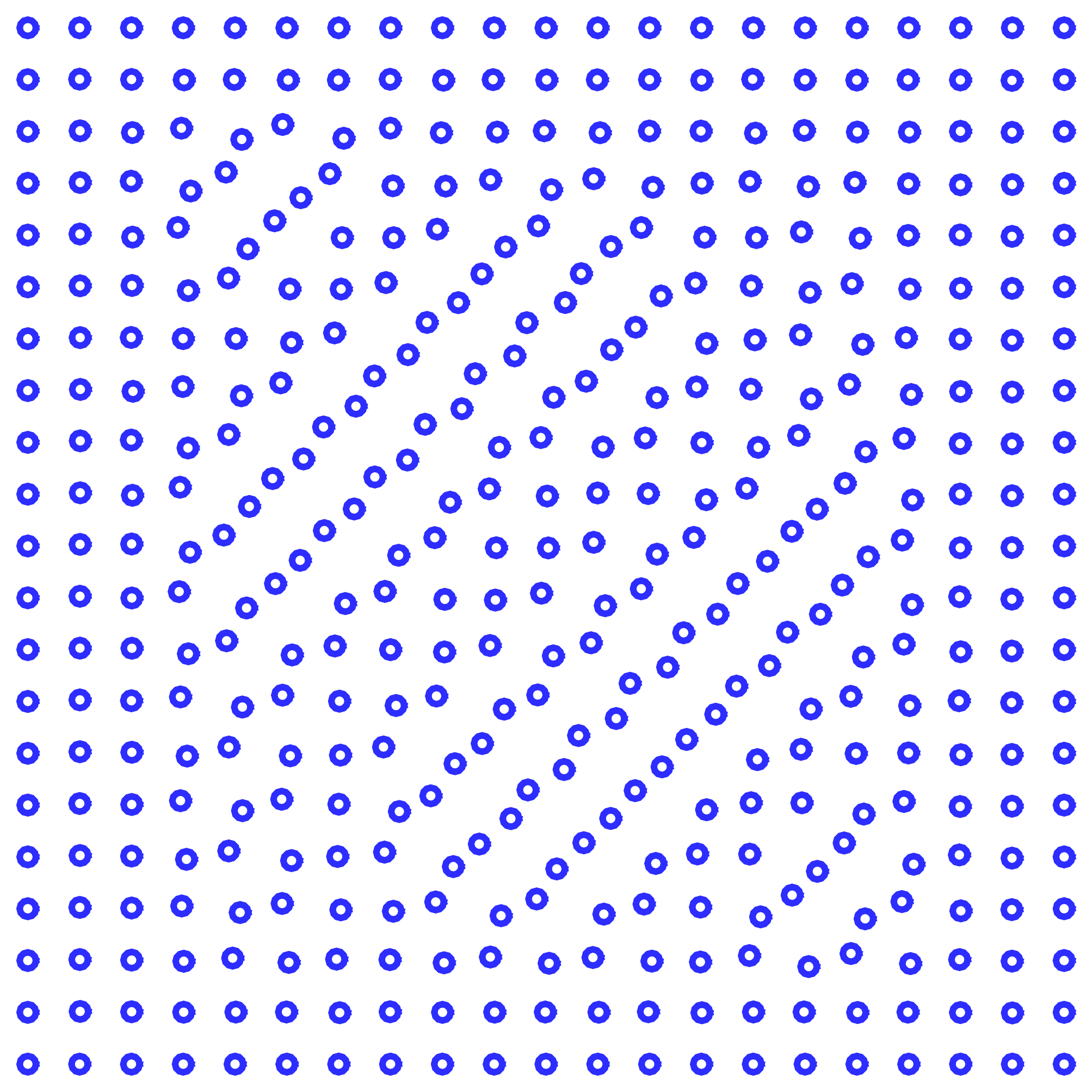}
        \caption{Configuration at 2.5$\mu$s}\label{box1_1}
    \end{subfigure}%
    ~
    \begin{subfigure}[t]{0.33\textwidth}
        \centering
        \includegraphics[width=\textwidth]{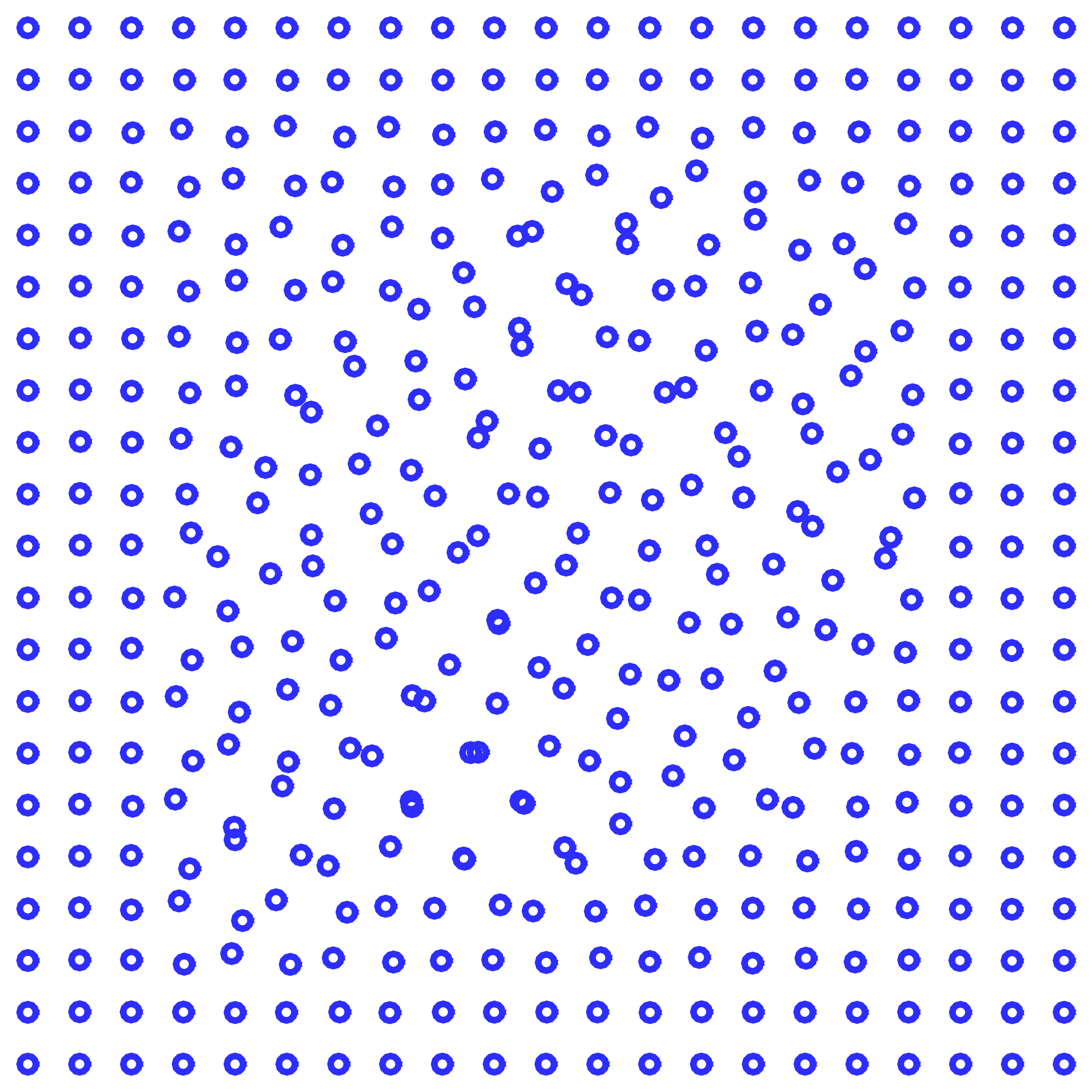}
        \caption{Configuration at 5$\mu$s}\label{box1_2}
    \end{subfigure}%
    
   \caption{Demonstration of \textit{tensile instability}}
    \label{box1}
\end{figure}

The instability manifests itself by unphysical clustering or clumping of particles, sometime causing numrical fracture. This is  illustrated via a two dimensional example \cite{swegle1995smoothed} in Figure \ref{box1}. Herein, a square grid of isotropic linear elastic particles (Figure \ref{box1_0}) with $\rho =7850 kg/m^3$, $E = 200GPa$ and $\nu = 0.3$ is considered. The boundary particles are kept fixed. All interior particles are preassigned with an uniform stress field and the central particle is perturbed with a velocity of $10^{-07}$ m/s. No artificial viscosity and artificial stress corrections are used. Particle configurations after $2.5 \mu s$ and $5.0 \mu s$ computed via SPH with standard cubic kernel are shown in Figures \ref{box1_1} and \ref{box1_2} respectively. The effect of instability is evident in the figures. 

\section{Removing Tensile Instability} 
As per the analysis by \citet{swegle1995smoothed}, the \textit{tensile instability} may be kept away from the SPH computation if the inequaqlities $\sum_{j \in \mathbb{N}^i}^{} W^{''}_{i,ij} \leq 0$ for tension and $\sum_{j \in \mathbb{N}^i}^{} W^{''}_{i,ij} \geq 0$ for compression are preserved. In the standard SPH, the kernel function ($W(x,\kappa h)$) and the smoothing length ($h$) are chosen independent of the material constitution. Therefore for a given kernel, the sign of $W^{''}(x,\kappa h)$ depends only on the distance between the interacting particles; not on their state of stress. Now the central idea of the present algorithm is to introduce this feature in the SPH formulation i.e., continuously change the sign of $W^{''}(x,\kappa h)$ at a material point depending on the state of stress at its neighbourhood such that the above stability condition is satisfied. In other words, the shape of the kernel function is adapted so as to ensure that the neighbour particles lie within the stable region of the kernel derivative (as shown in Figure \ref{Instability}). This adaptation needs to be done in such a way that, (a) the fundamental properties of kernel function are preserved, (b) no ad hoc parameters are introduced, (c) the computational efficiency of SPH code is not compromised, and (d) definition of neighbour is not changed i.e., for a given particle the size of the influence domain remains same.  

The B-spline basis functions are the building blocks of the proposed adaptive algorithm. Therefore, to start with, the construction and salient features of B-spline basis are outlined in the next sub-section. 

\subsection{B-spline as adaptive kernel in SPH} \label{Bspline}
B-Spline basis functions have inherent property of compactness and higher order continuity with minimal number of points \cite{piegl2012nurbs}. These functions also hold partition of unity characteristics. The recursive definition (Cox-eBoor algorithm) of $i$-th normalized B-spline basis functions of degree $p$ in  one dimension is,

\begin{equation}
N_{i,0}\left(\zeta\right) = 
\begin{cases}
1 & for \zeta_i\leq \zeta \leq \zeta_{i+1}\\
0 & otherwise 
\end{cases}
\label{Bspline1}
\end{equation}   

\begin{equation}
N_{i,p}\left(\zeta\right) = \frac{\zeta - \zeta_i}{\zeta_{i+p} - \zeta_i}N_{i,p-1}\left(\zeta\right) + \frac{\zeta_{i+p+1} - \zeta}{\zeta_{i+p+1} - \zeta_{i+1}}N_{i+1,p-1}\left(\zeta\right)
\label{Bspline2}
\end{equation}
where, $\Xi=\lbrace\zeta_1,\zeta_2, \zeta_3,...,\zeta_{n+p+1} | \zeta_i \in \mathbb{R}\rbrace$ is the knot vector and $n$ is the number of basis functions. B-spline basis functions are, (a) strictly positive: $N_{i,p} (\zeta) \geq 0 \forall i, p$, (b) compactly supported: $N_{i,p} (\zeta) = 0 \forall \zeta \in [\zeta_i, \zeta_{i+p+1}]$ i.e., $\left(\zeta_{i+p+1} - \zeta_i \right)$ is the support size of $N_{i,p}$, (c) partition of unity: $\sum_{j} N_{i,p}(\zeta) = 1$, and (d) differentiability: $N_{i,p} (\zeta)$ is $C^p$ continuous everywhere in the domain except at the knot locations where it is $C^{p-1}$ continuous. The above properties qualify the B-spline basis function to be a valid kernel in SPH. The most important feature which constitutes the premise of the present formulation is that for any B-spline basis function, say $N_{i,p}(\zeta)$, the shape can be modified by changing the location of intermediate knots $\lbrace \zeta_{i+1},...\zeta_{i+p} \rbrace$ and still its support (or influence domain) will remain same as $\left(\zeta_{i+p+1} - \zeta_i \right)$ if the positions of extreme knots $\lbrace \zeta_{i},\zeta_{i+p+1} \rbrace$ are unchanged. For the use in SPH computation, here $N_{1,p}$ defined over knot vector $\lbrace \zeta_1,\zeta_2, ,...,\zeta_{1+p+1}\rbrace$ is taken as the kernel function. 

\begin{figure}[h!]\vspace*{-4pt}
    \centering
    \begin{subfigure}[t]{0.33\textwidth}
        \centering
        \includegraphics[trim={1cm 2.0cm 1cm 2.0cm},clip,width=\textwidth]{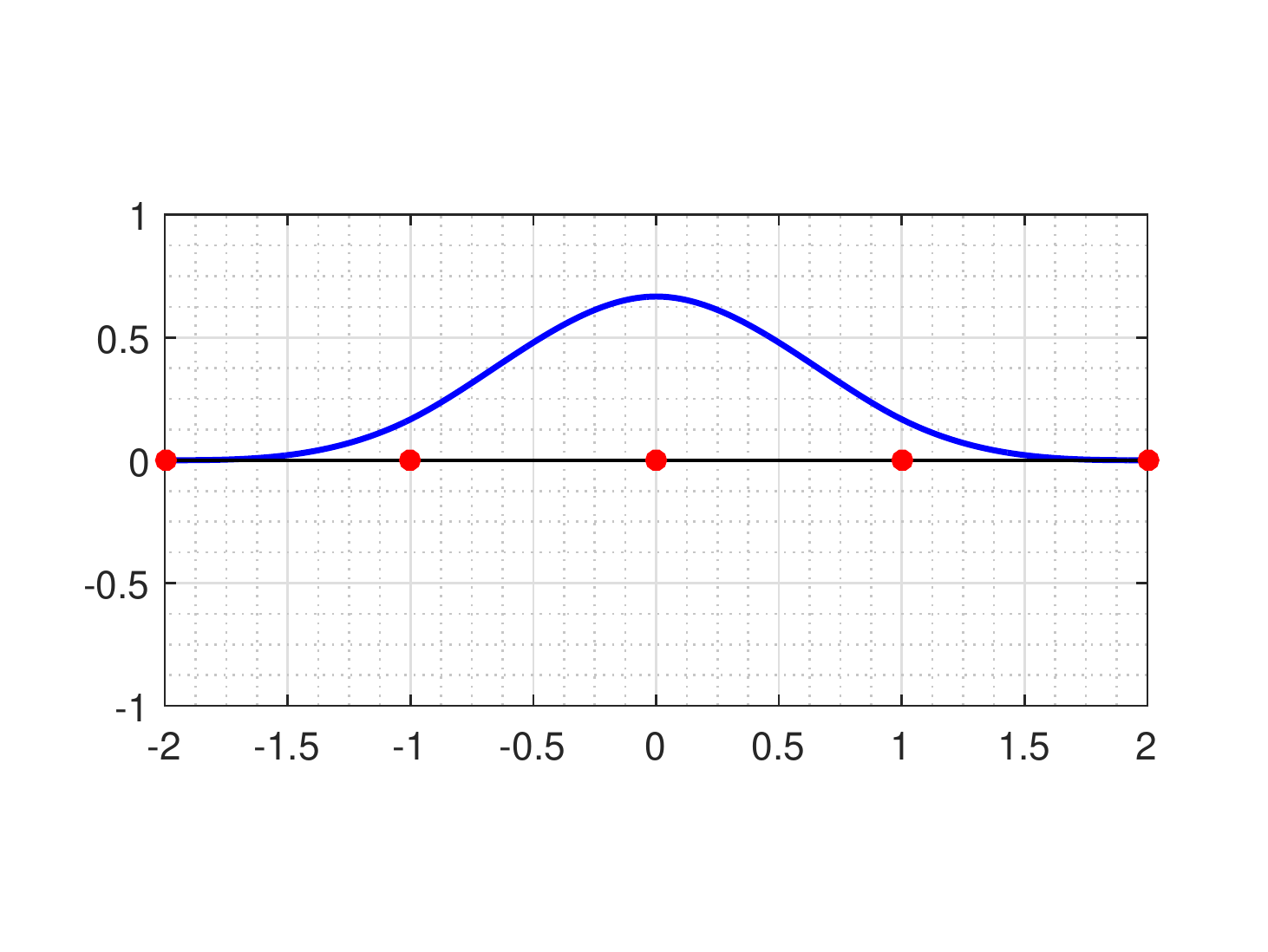}
    \end{subfigure}%
    ~    
    \begin{subfigure}[t]{0.33\textwidth}
        \centering
        \includegraphics[trim={1cm 2.0cm 1cm 2.0cm},clip,width=\textwidth]{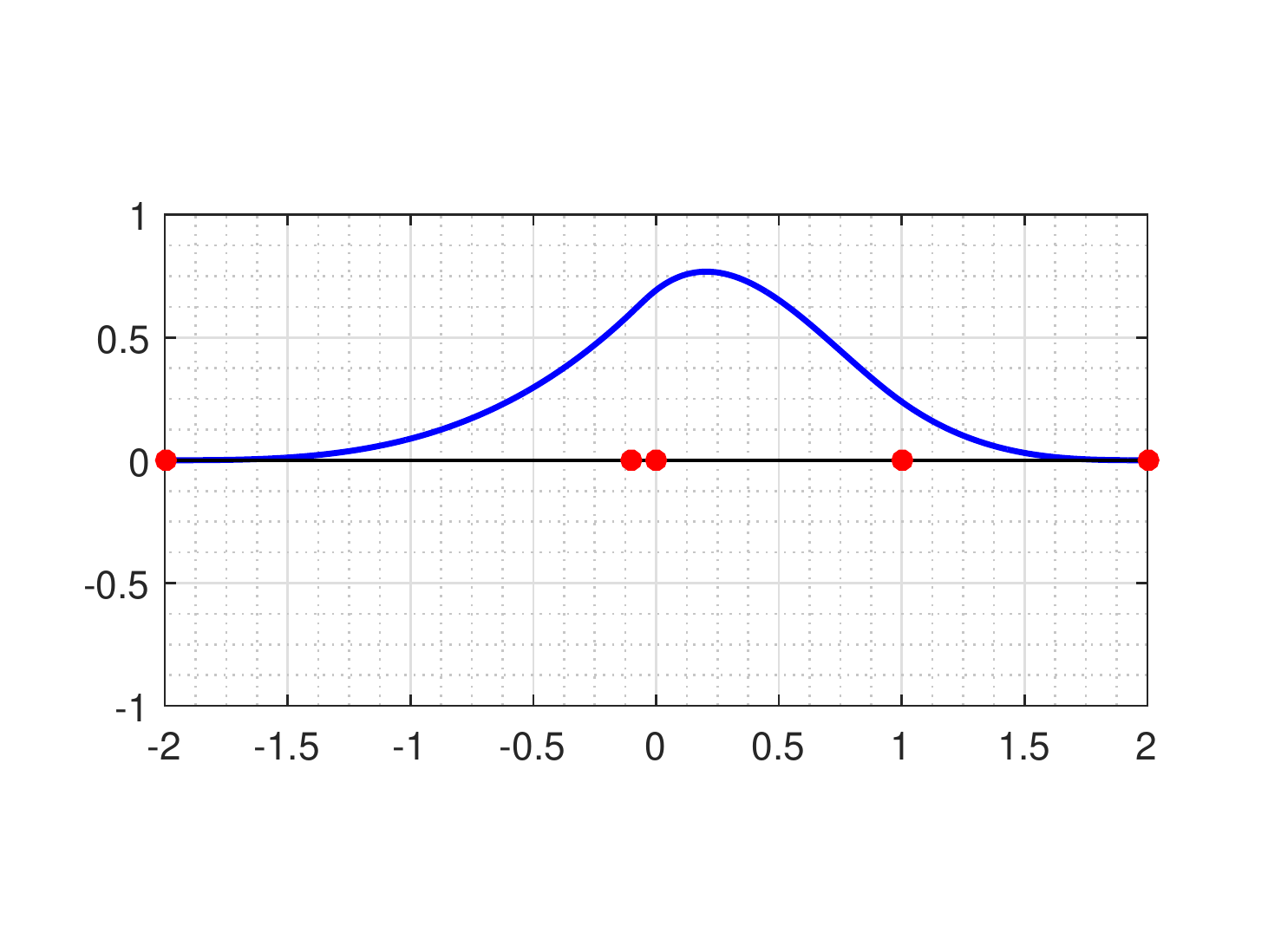}
    \end{subfigure}%
    ~
    \begin{subfigure}[t]{0.33\textwidth}
        \centering
        \includegraphics[trim={1cm 2.0cm 1cm 2.0cm},clip,width=\textwidth]{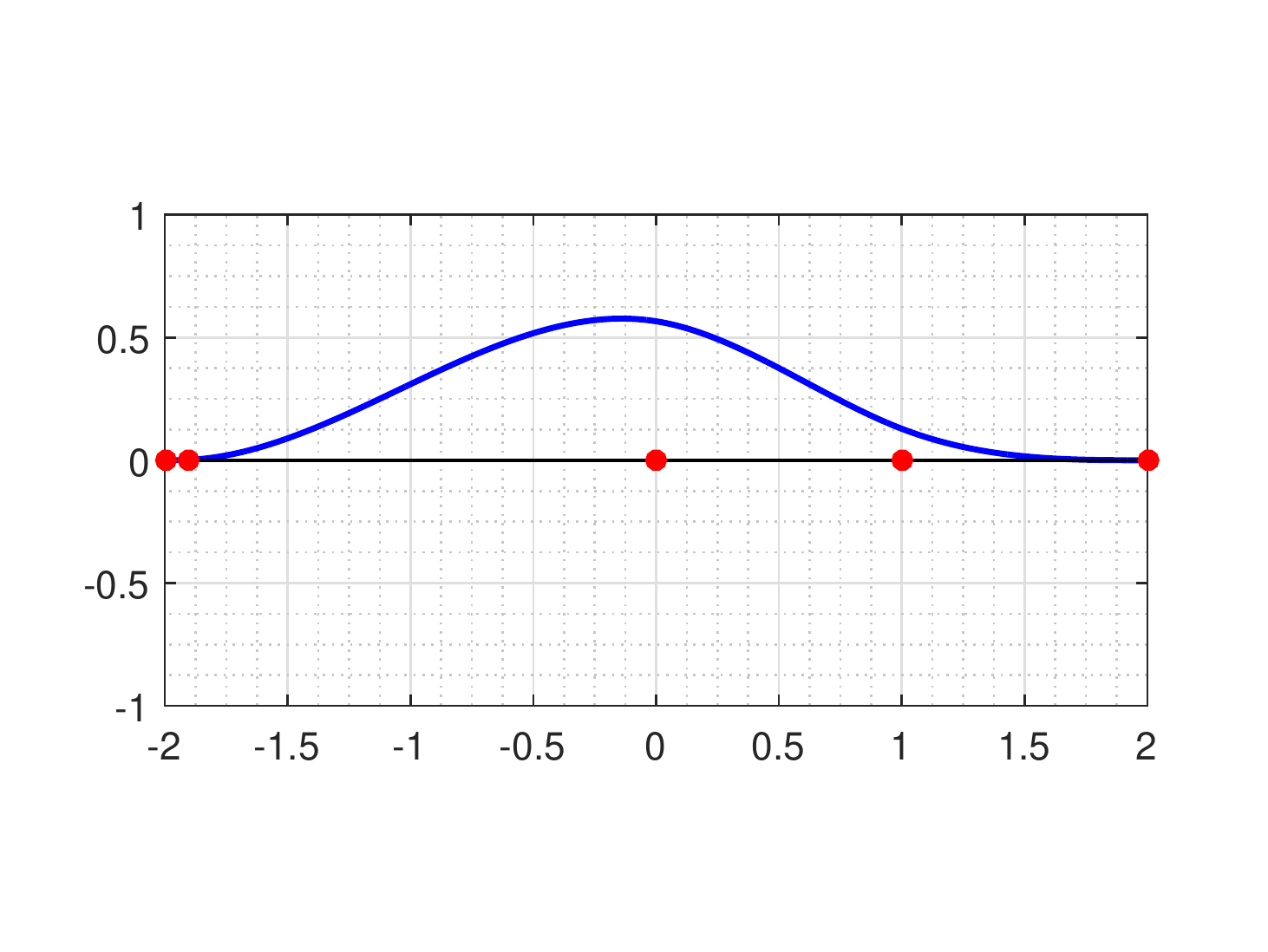}
    \end{subfigure}%
    
        \begin{subfigure}[t]{0.33\textwidth}
        \centering
        \includegraphics[trim={1cm 2.0cm 1cm 2.0cm},clip,width=\textwidth]{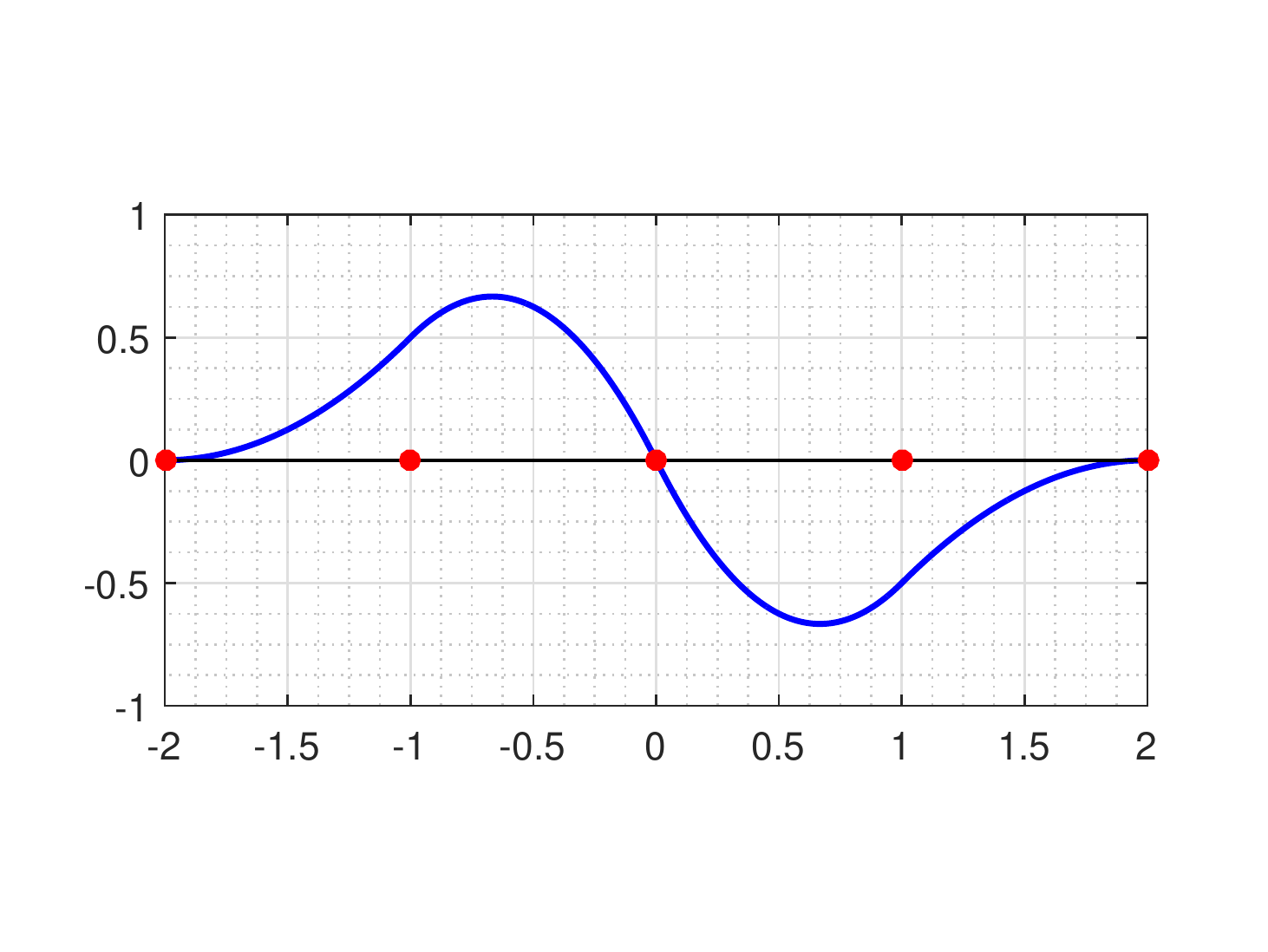}
        \caption{$a_1 = a_2 = 1.0$}
    \end{subfigure}%
    ~    
    \begin{subfigure}[t]{0.33\textwidth}
        \centering
        \includegraphics[trim={1cm 2.0cm 1cm 2.0cm},clip,width=\textwidth]{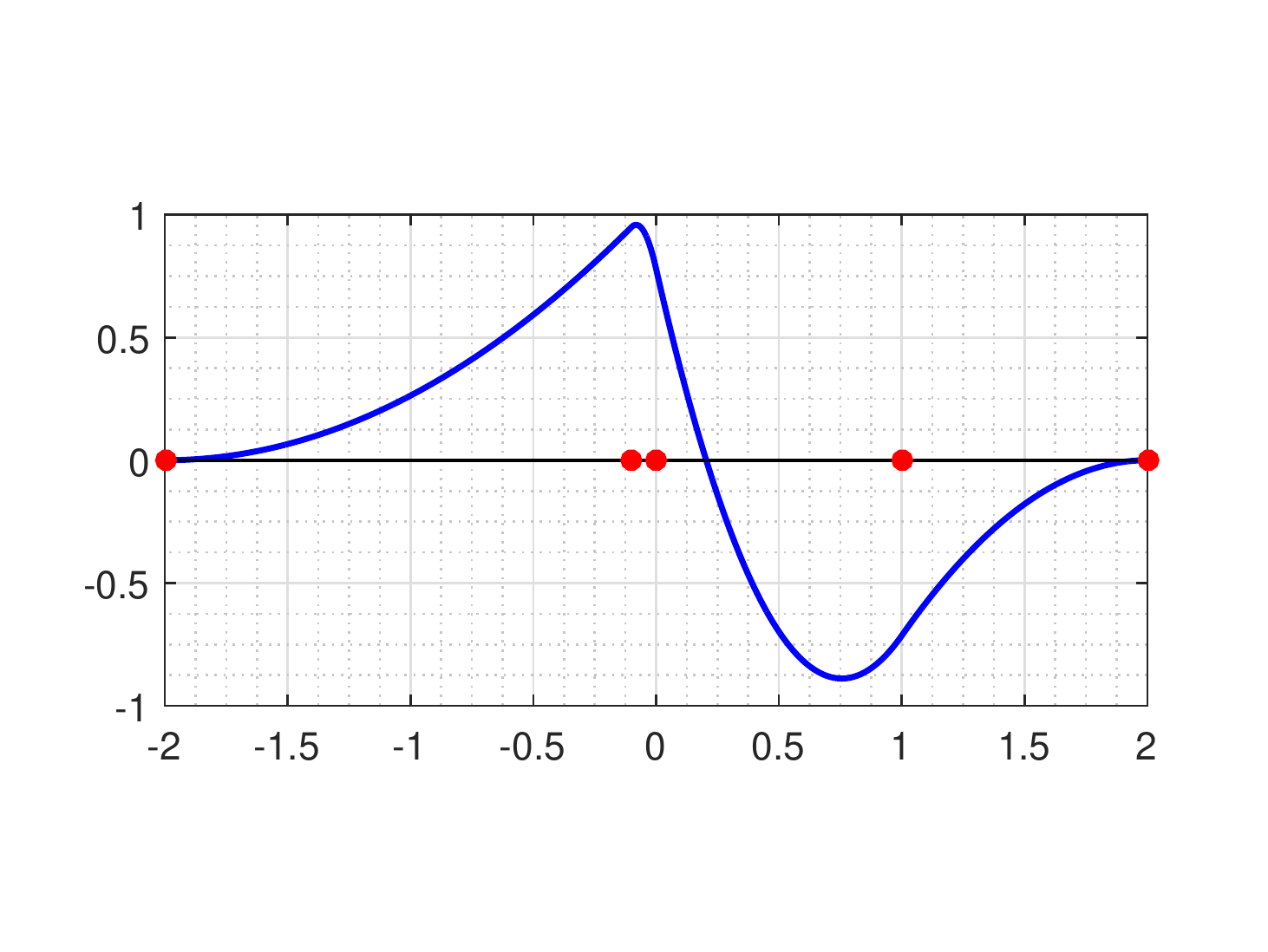}
        \caption{$a_1 = 0.1, a_2 = 1.0$}
    \end{subfigure}%
    ~
    \begin{subfigure}[t]{0.33\textwidth}
        \centering
        \includegraphics[trim={1cm 2.0cm 1cm 2.0cm},clip,width=\textwidth]{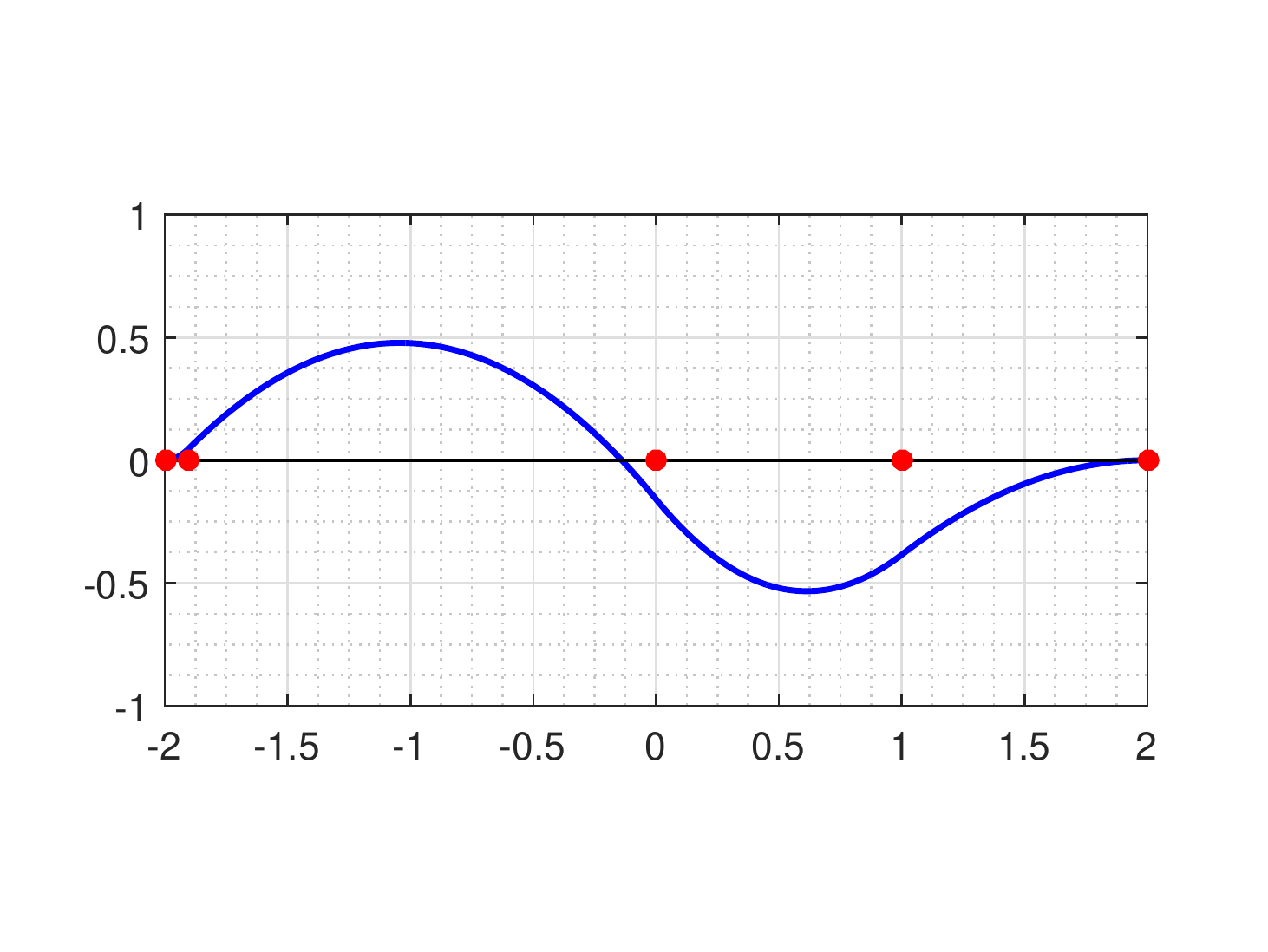}
        \caption{$a_1 = 1.9, a_2 = 1.0$}
    \end{subfigure}%
    
   \caption{Cubic B-spline kernel and its first derivative constructed over knot $\left\lbrace-b, -a_1, 0, a_2, b \right\rbrace$ for fixed $b=2$ and different $a_1$ and $a_2$. Knot positions are shown by red dots.}
    \label{BS3}
\end{figure}

Change of shape of cubic B-spline kernel function (i.e., $N_{1,3}$) with different intermediate knot positions is demonstrated in Figure \ref{BS3}. The kernel is defined over knot $\lbrace -b, -a_1, 0, a_2, b\rbrace$. Figures are shown for fixed $b=2$ and different $a_1$ and $a_2$. Though obvious, it is important to take a note (for discussion in subsequent sections) that if the knot positions are not symmetric, the kernel function will also not be symmetric. The expressions (obtained from Equation \eqref{Bspline1} and \eqref{Bspline2}) of quadratic (i.e., $N_{1,2}$) and cubic (i.e., $N_{1,3}$) B-spline kernel with a symmetric knot vectors $\lbrace -b, -a, a, b\rbrace$ and $\lbrace -b, -a, 0, a, b\rbrace$ respectively are provided in Table \ref{Table1}. This enables to form a more generalised analytical expression, where both shape (varying the value of intermediate knot $a$) and support size (varying the value of $b$) of the kernel function may be changed as per the requirement. 

\begin{table}[h!]
\caption{Expressions for parametrised B-Spline functions}
\begin{tabular*}{\hsize}{@{\extracolsep{\fill}}lll@{}}
\hline
Order 		& Analytic form 			& Knot Vector \\
\hline
Quadratic 	& $	W(q,h)=
				\begin{cases}
				\frac{ab - q^2}{a(a+b)} & for ~ 0\leq q \leq a\\
				\frac{(b-q)^2}{b^2 - a^2} & for ~ a \leq q \leq b \\
				0 & for ~ b\leq q 
				\end{cases}$
				 						& [-b,-a,a,b] \\
\hline
Cubic 		& $	W(q,h)=
				\begin{cases}
				\frac{(a+b)q^3 - 3abq^2 +a^2b^2}{a^2b(a+b)} & for ~ 0\leq q \leq a\\
				\frac{(b-q)^3}{b(b^2-a^2)} & for ~ a \leq q \leq b \\
				0 & for ~ b\leq q 
				\end{cases}$
 										& [-b,-a,0,a,b] \\
\hline
\end{tabular*}
\label{Table1}
\end{table}


\subsection{Adaptive Algorithm} 
\begin{figure}[h!]\vspace*{-4pt}
    \centering
    \begin{subfigure}[t]{0.5\textwidth}
        \centering
        \includegraphics[trim={3cm 1cm 2cm 1cm},clip,width=\textwidth]{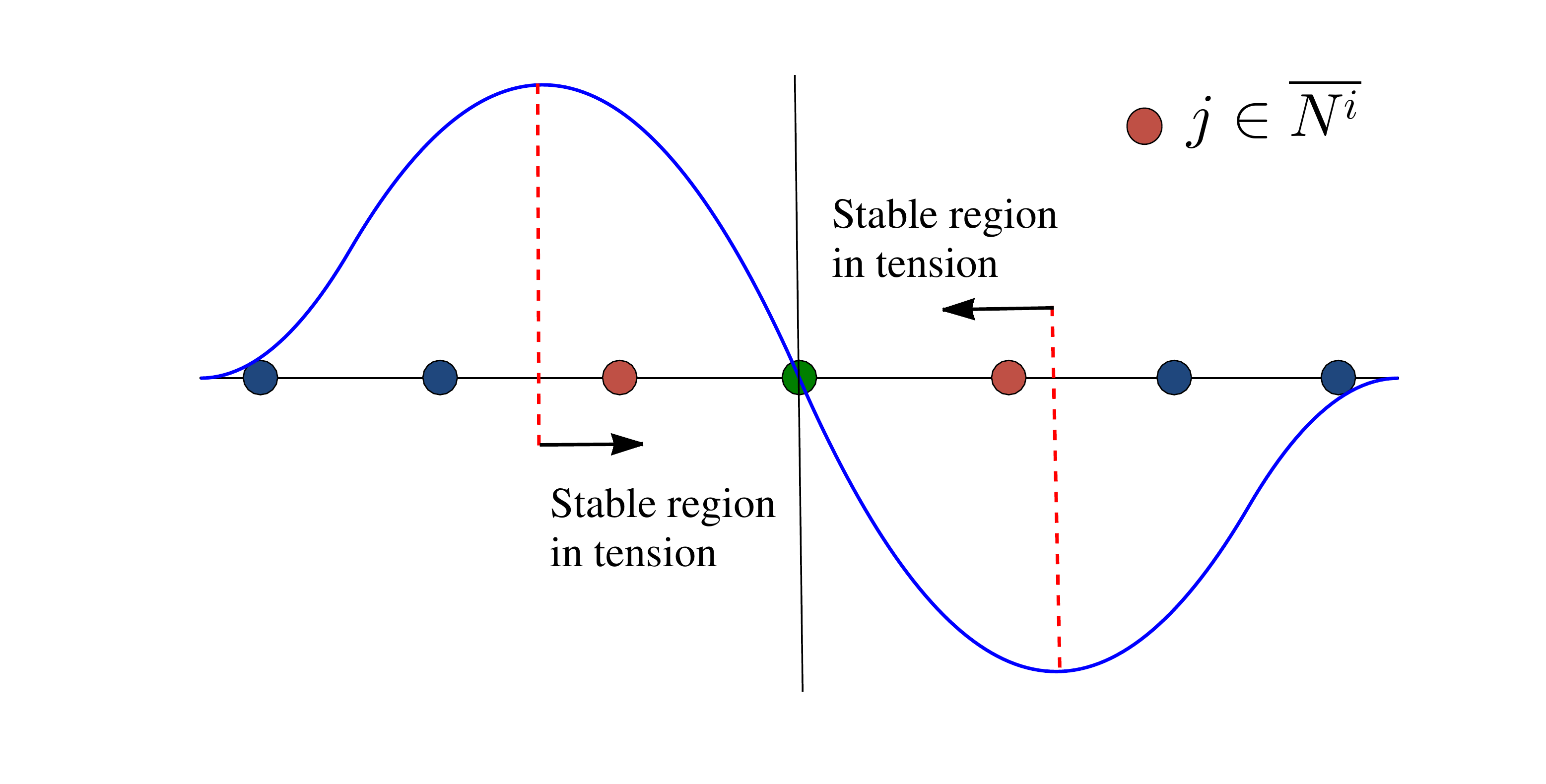}
    \end{subfigure}%
    ~    
    \begin{subfigure}[t]{0.5\textwidth}
        \centering
        \includegraphics[trim={3cm 1cm 2cm 1cm},clip,width=\textwidth]{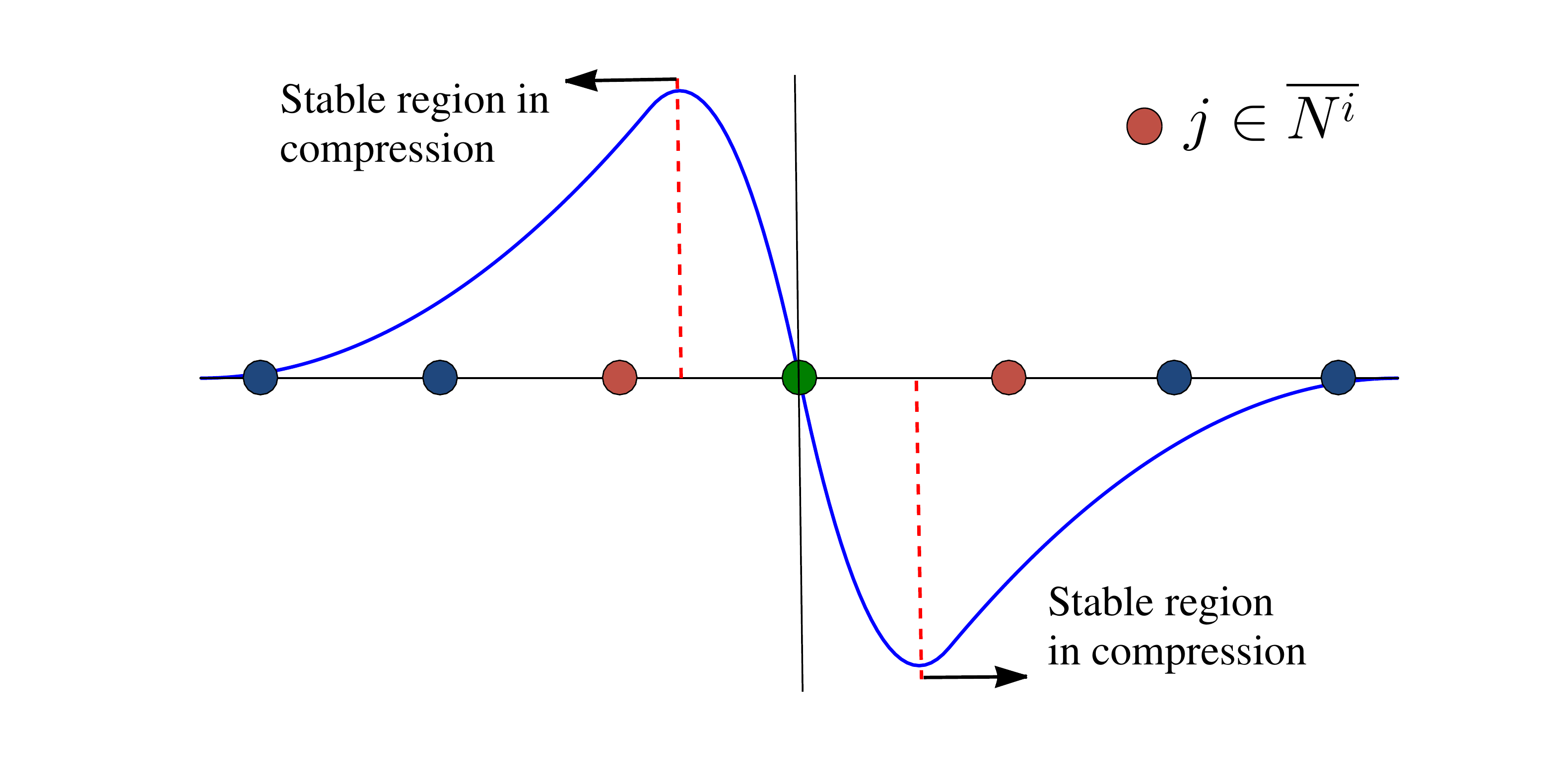}
    \end{subfigure}%
    ~
\caption{Stable region for different state of stress: a. higher $\chi$ for tension; b. lower $\chi$ for compression }\vspace*{-6pt}
\label{TInst}
\end{figure}

For a better understanding the algorithm is first discussed in one dimension. Consider an arbitrary particle (say $i$-th), with position $x_i$ and influence domain $\mathbb{N}^i = \lbrace j \in \mathbb{Z}^+ ~|~ |x_i-x_j| < bh ~and~ i \neq j \rbrace$. Let $\overline{\mathbb{N}^i} =\lbrace i-1, i+1 \rbrace \subseteq \mathbb{N}^i$ be the set of immediate neighbours as represented by red dots in Figure \ref{TInst}. The \textit{tensile instability} is known to be a short wavelength instability which starts with clumping of pair of immediate neighbours as depicted in Figure \ref{box1_1}. Though in the stability condition the summation is over all the particles within the influence domain, stability of the computation is significantly governed by the interaction of immediate neighbours. Therefore, the crucial step in the algorithm is to ensure that the immediate neighbours (particles in $\overline{\mathbb{N}^i}$) are brought into the stable zone of the kernel function. Towards achieving this, the required shift in the position of the intermediate knot is discussed in this section. 

For a given particle, the magnitude and the direction of the shift (i.e., whether towards the particle or away from the particle) depend on the relative positions and the state of stress of its neighbour particles. One approach would be to check the interaction of $i$ with all its immediate neighbour $j \in \overline{\mathbb{N}^i}$ separately and adapt the kernel such that all $j \in \overline{\mathbb{N}^i}$ come within the stable zone of $i$. For instance in the 1D case, with cubic B-spline kernel defined over knot $\lbrace -b, -a_1, 0, a_2, b\rbrace$, first check the interaction between $i-(i-1)$ and depending on whether the particles are approaching (compression) or moving away (tension) fix $a_1$. Similarly depending on the interaction of $i-(i+1)$ fix $a_2$. While this seems to be more logical, the problem with this approach is that for $a_1 \neq a_2$ the kernel no longer remains symmetric (as shown in Figure \ref{BS3}) and may affect the conservation. Moreover, implementation in higher dimension would be numerically very intensive.   

Instead of treating the each individual particle pair separately, herein the cumulative effect of all immediate neighbours i.e., $j \in \overline{\mathbb{N}^i}$ on particle $i$ is considered. The kernel function at $x_i$ ($W(x-x_i,h)$) is formed with a symmetric knot vector $\lbrace -b, a, 0, a, b\rbrace$. Now, when the neighbouring particles approach towards $i$, the density of the particle $\rho_i$ increases and vice versa. Therefore the density ratio $\rho_i/\rho_0$ ($\rho_0$ is the initial density) is taken as the criteria for knot shifting i.e., if $\rho_i/\rho_0 >1$, the intermediate knot is shifted towards particle $i$ and similarly if $\rho_i/\rho_0 <1$, the intermediate knot is moved away from $i$. 

Once the direction of the knot shifting (i.e., towards or away with respect to particle $i$) is identified, next is to determine the magnitude of the shift. The influence domain (i.e., value of $b$) is kept fixed and only the shape of the kernel over the influence domain is adapted by varying $\chi=a/b$ ratio (depending on the value of $\rho_i/\rho_0$) in order to obtain a stable region. The two extreme positions of the intermediate knot are $\chi=\epsilon$ and $\chi=1-\epsilon$, where $\epsilon \in \mathbb{R}^+$ is a small number used to avoid any repetition of knot at the centre (i.e, $x=0$) or at the boundary (i.e., $x=\pm b$) of the kernel support. For the cubic B-spline kernel, $W'$ curve attains maximum at a distance $\frac{\chi}{1 + \chi}b$. Assuming a quasi-uniform particle distribution (at least locally) in the neighbourhood of $i$ with characteristic spacing $\Delta p$, the condition so that the immediate neighbour lies at the peak of $W'$ is,
\begin{equation}
\Delta p/h = \frac{\chi}{1+\chi}b \\
\Rightarrow \chi = \frac{\Delta p/h}{b-\Delta p/h}
\end{equation}

Now the following condition is necessary in order to ensure a stable interaction of particle $i$ with its immediate neighbour $j \in \overline{\mathbb{N}^i}$.  
\begin{equation}
\chi_i = 
\begin{cases}
		\ < \frac{\Delta p/h}{b-\Delta p/h} & \text{for} \, \frac{\rho_i}{\rho_0} > 1\\
		> \frac{\Delta p/h}{b-\Delta p/h} & \text{for} \, \frac{\rho_i}{\rho_0} < 1 \\
\end{cases}
\label{CondBS3}
\end{equation}

Similar condition may be obtained for any order B-spline kernel. For instance the condition for quadratic B-spline ( $W'$ is maximum at $\chi = 1$) would be,
\begin{equation}
\chi_i = 
\begin{cases}
		\ < \frac{\Delta p/h}{b} & \text{for} \, \frac{\rho_i}{\rho_0} > 1\\
		> \frac{\Delta p/h}{b} & \text{for} \, \frac{\rho_i}{\rho_0} < 1 \\
\end{cases}
\label{CondBS2}
\end{equation}

\begin{figure}[h!]
    \centering 
    \begin{subfigure}[t]{0.5\textwidth}
        \centering
        \includegraphics[trim={1cm 0.5cm 1cm 0.5cm},clip,width=\textwidth]{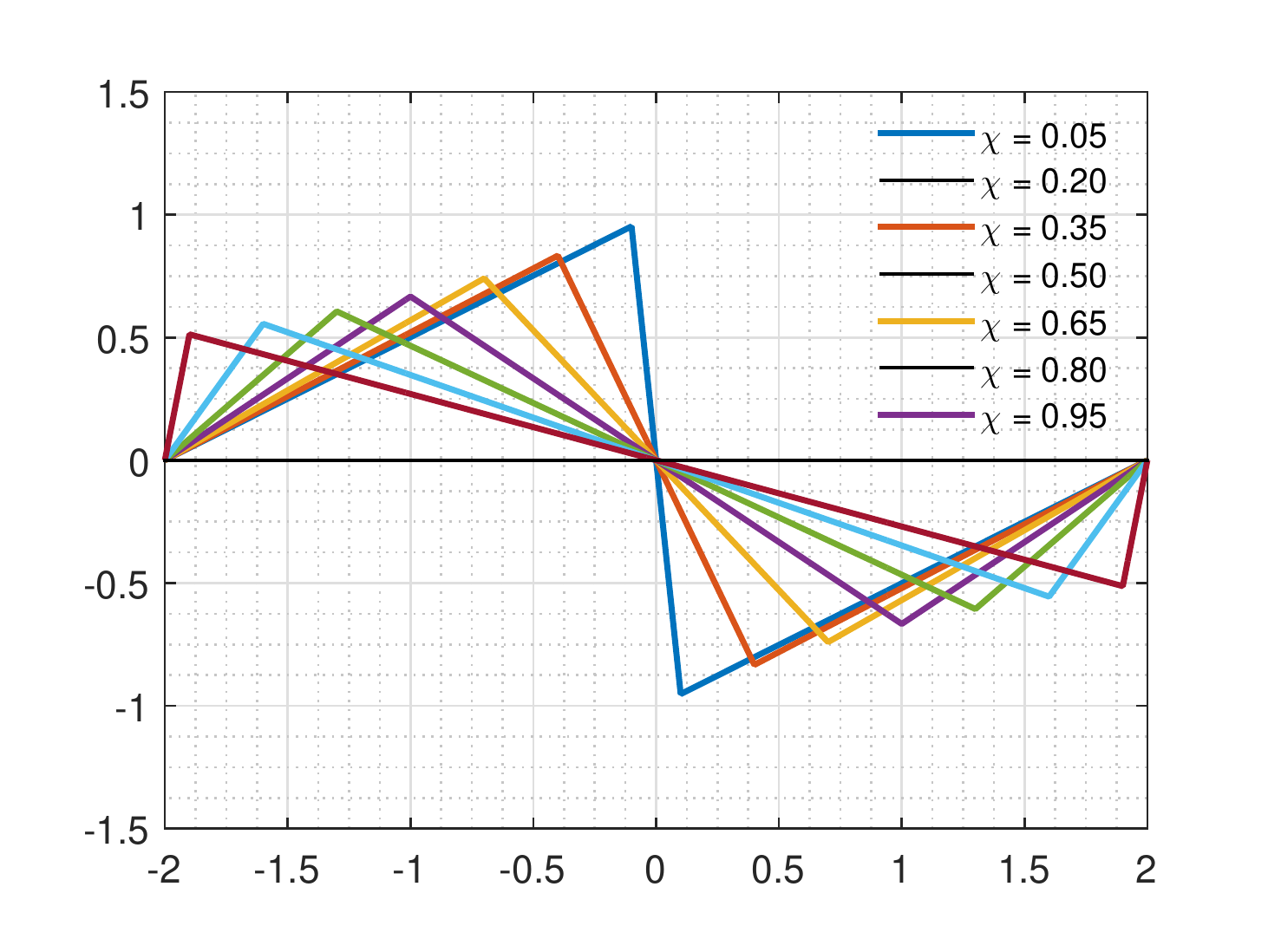}
        \caption{Quadratic on $\left\lbrace -b, -a, a, b  \right\rbrace$}
    \end{subfigure}%
    ~
    \begin{subfigure}[t]{0.5\textwidth}
        \centering
        \includegraphics[trim={1cm 0.5cm 1cm 0.5cm},clip,width=\textwidth]{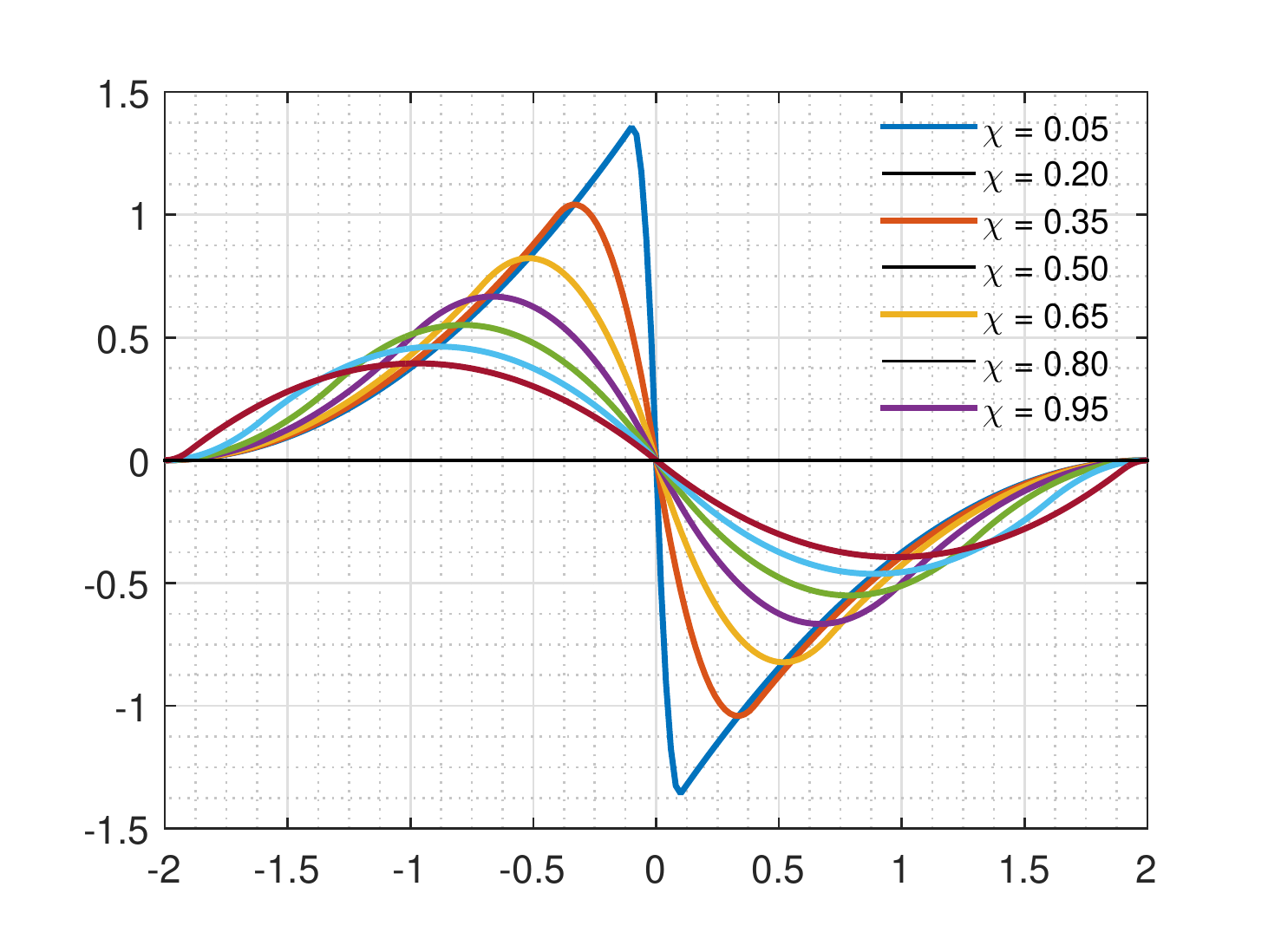}
        \caption{Cubic on $\left\lbrace -b, -a, 0, a, b  \right\rbrace$}
    \end{subfigure}%
   \caption{1st derivative of quadratic and cubic kernel for fixed $b=2$ and different $\chi=a/b$.}
    \label{Kernel_deriv}
\end{figure}

\begin{figure}[!h]
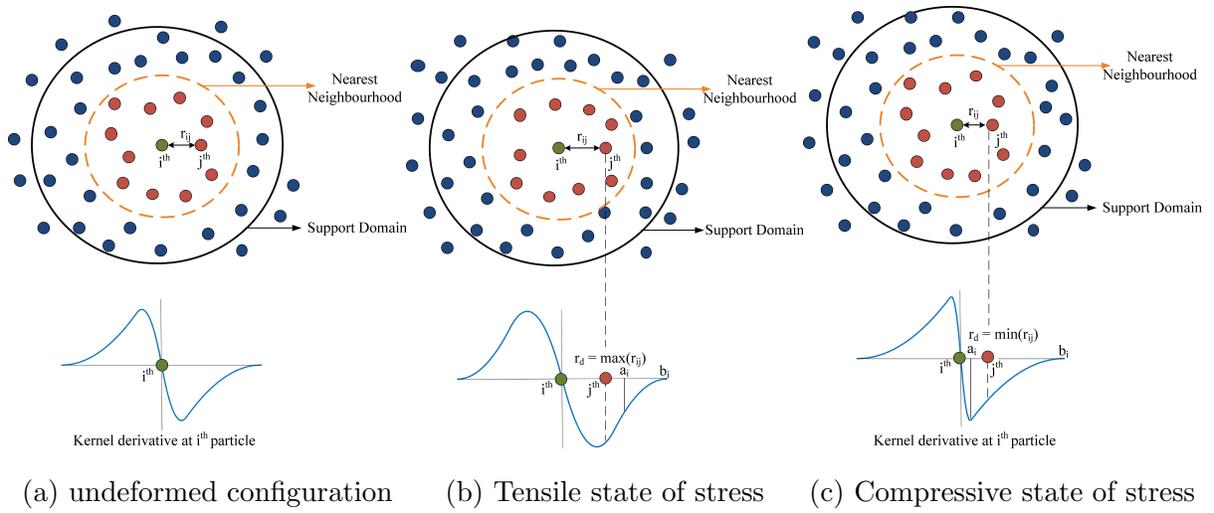

	\centering
		\begin{subfigure}[t]{0.32\textwidth}
			\centering
				\includegraphics[trim={0 0 0 0},width=\textwidth]{kernel-undeformed.pdf}\label{ker-u}
				\caption{undeformed configuration}\vspace{-6pt}
		\end{subfigure}%
		\begin{subfigure}[t]{0.32\textwidth}
			\centering
				\includegraphics[trim={0 0 0 0},width=\textwidth]{kernel-tension.pdf}\label{ker-t}
				\caption{Tensile state of stress}\vspace{-6pt}
		\end{subfigure}%
		\begin{subfigure}[t]{0.32\textwidth}
			\centering
				\includegraphics[trim={0 0 0 0},width=\textwidth]{kernel-compression.pdf}\label{ker-c}
				\caption{Compressive state of stress}\vspace{-6pt}
		\end{subfigure}%
	\caption{Adaptive Kernel pictorial illustration}\label{Kernel-pic}
\end{figure}

\begin{figure}[h!]
	\centering
		\begin{subfigure}[t]{0.33\textwidth}
			\centering
				\includegraphics[trim={0.75cm 1cm 0.75cm 1cm},width=1.0\textwidth]{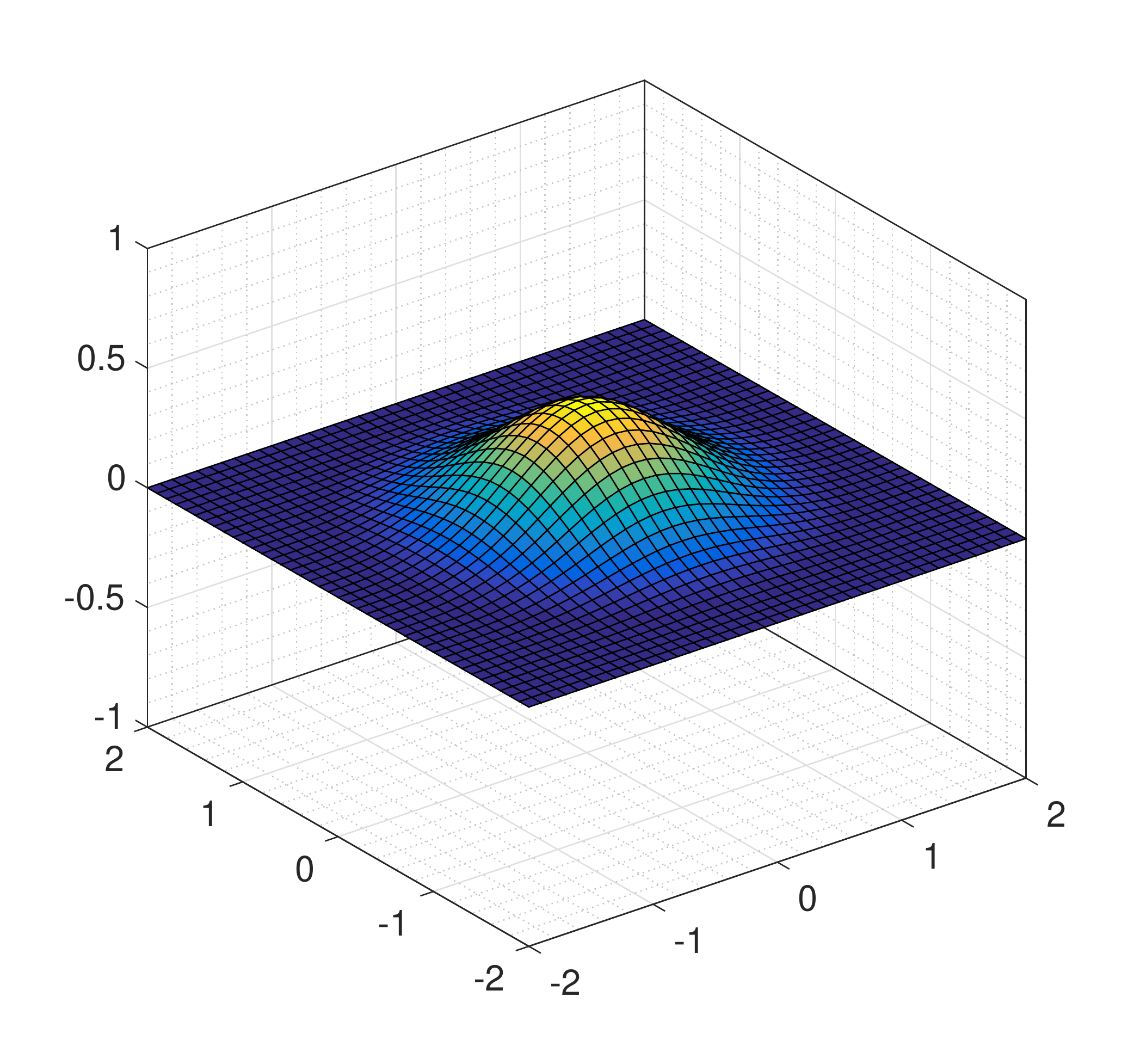}
		\end{subfigure}%
		\begin{subfigure}[t]{0.33\textwidth}
			\centering
				\includegraphics[trim={0.75cm 1cm 0.75cm 1cm},width=1.0\textwidth]{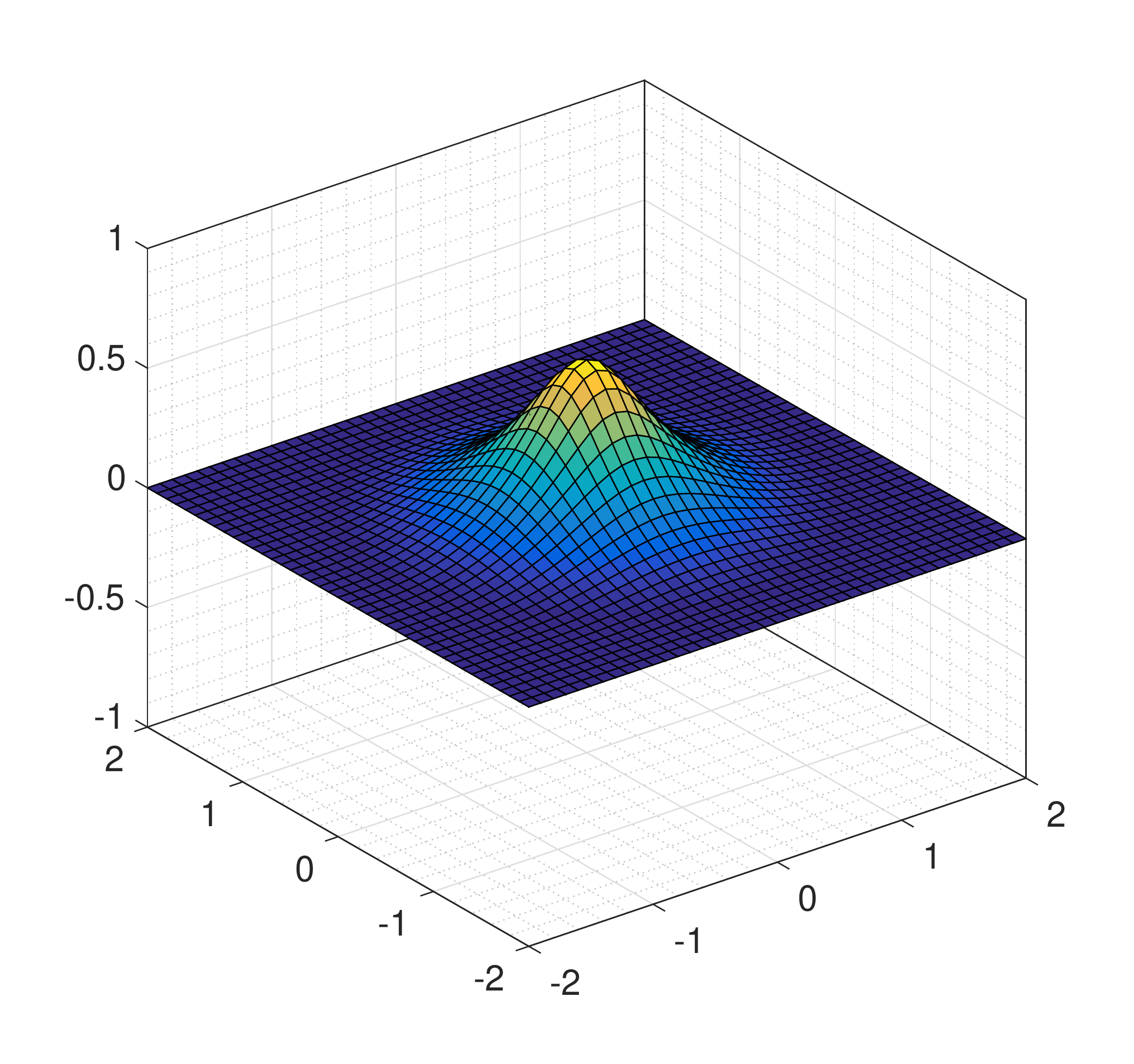}
		\end{subfigure}%
		\begin{subfigure}[t]{0.33\textwidth}
			\centering
				\includegraphics[trim={0.75cm 1cm 0.75cm 1cm},width=1.0\textwidth]{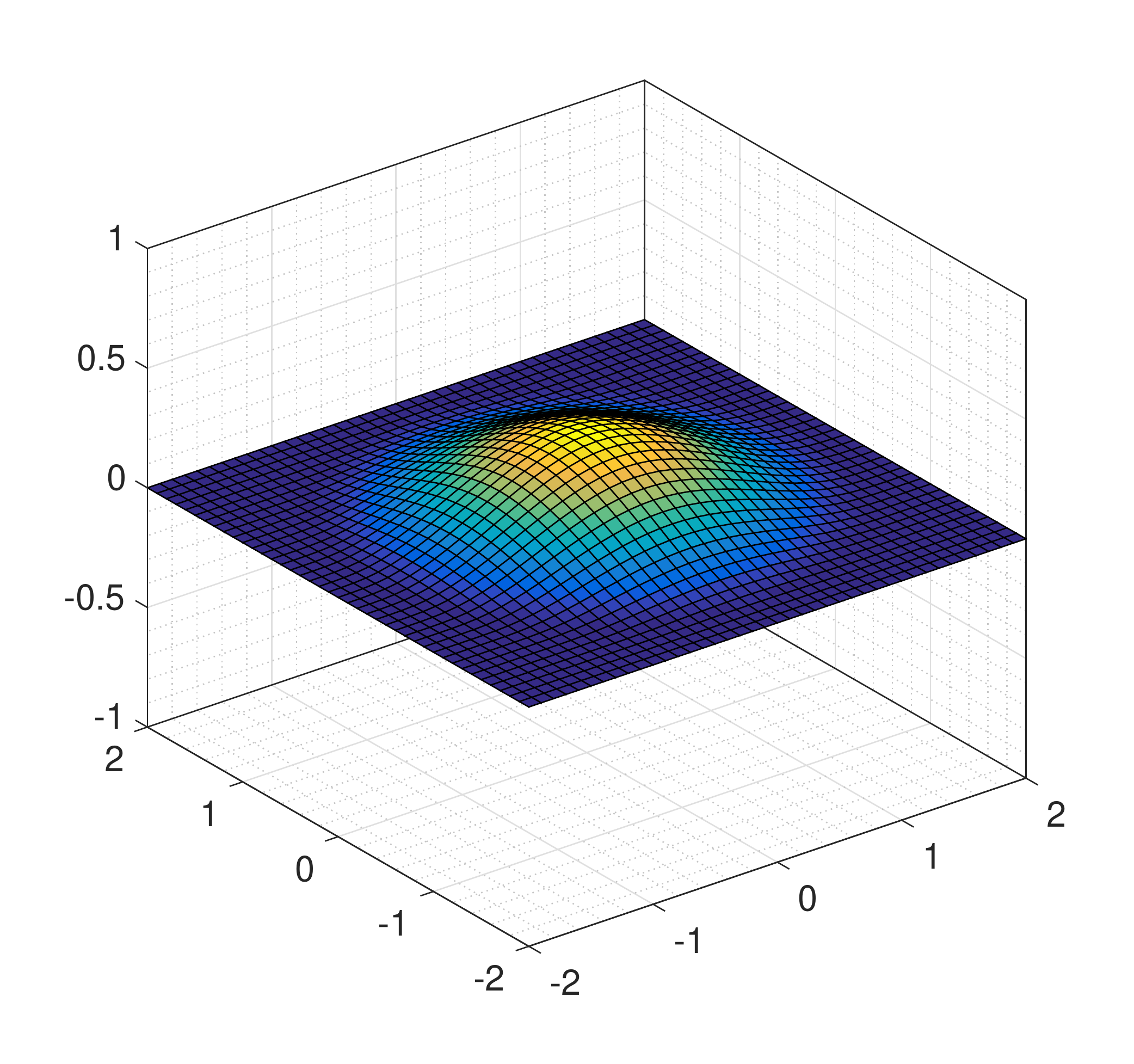}
		\end{subfigure}%
	
		\begin{subfigure}[t]{0.33\textwidth}
			\centering
				\includegraphics[trim={0.75cm 1cm 0.75cm 1cm},width=1.0\textwidth]{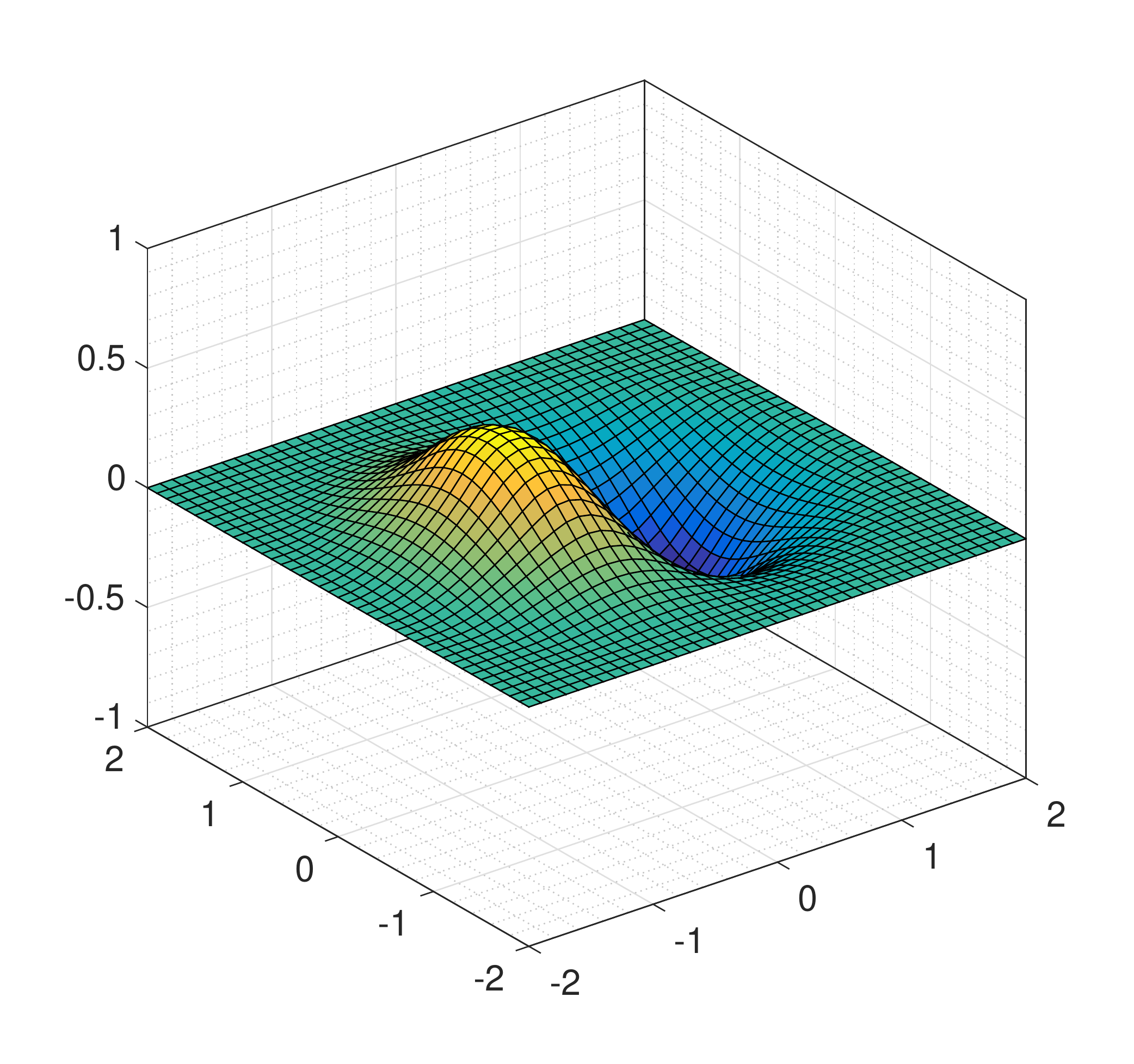}
				\caption{$\chi = 0.5$}
		\end{subfigure}%
		\begin{subfigure}[t]{0.33\textwidth}
			\centering
				\includegraphics[trim={0.75cm 1cm 0.75cm 1cm},width=1.0\textwidth]{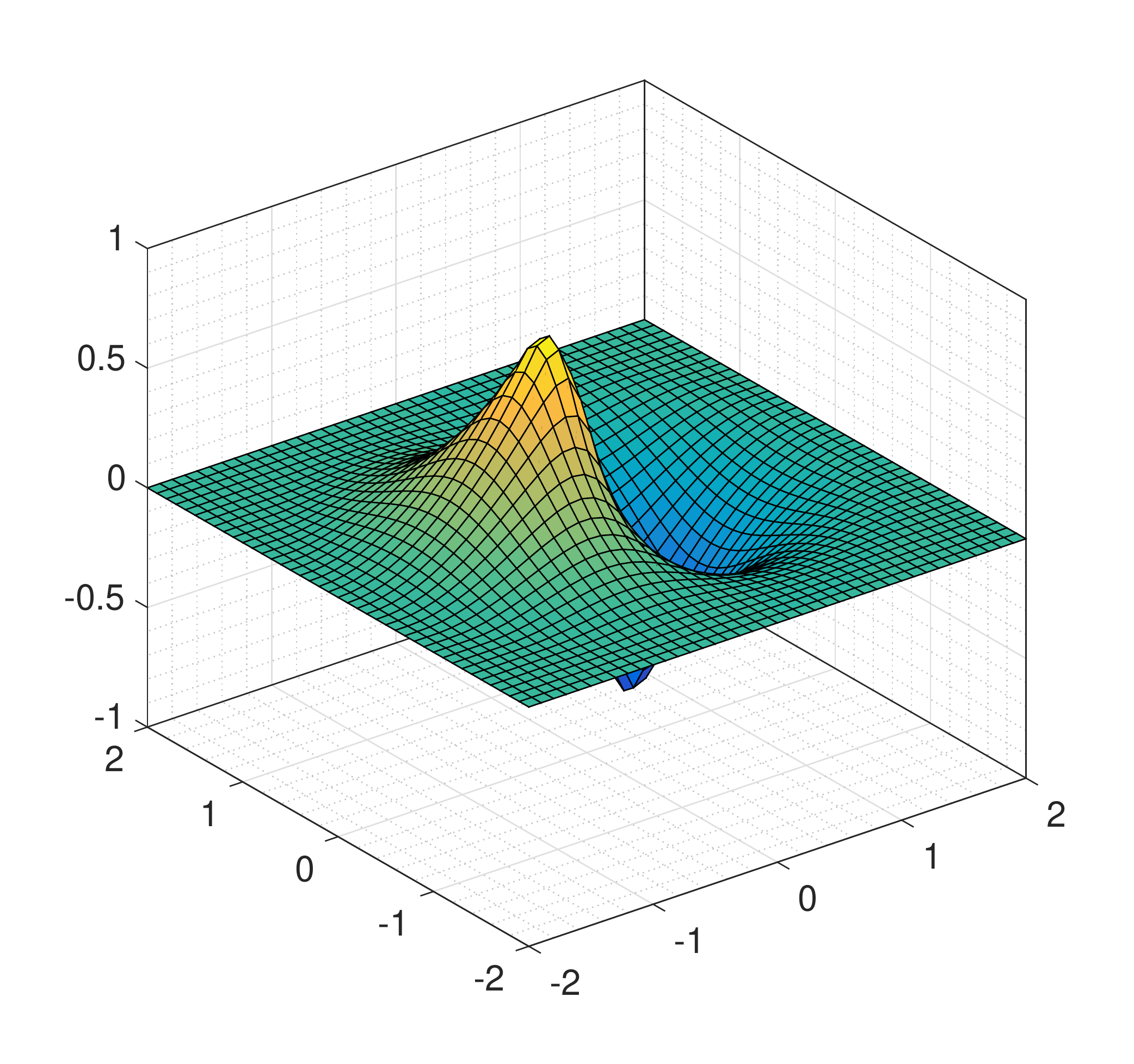}
				\caption{$\chi = 0.2$}
		\end{subfigure}%
		\begin{subfigure}[t]{0.33\textwidth}
			\centering
				\includegraphics[trim={0.75cm 1cm 0.75cm 1cm},width=0.8\textwidth]{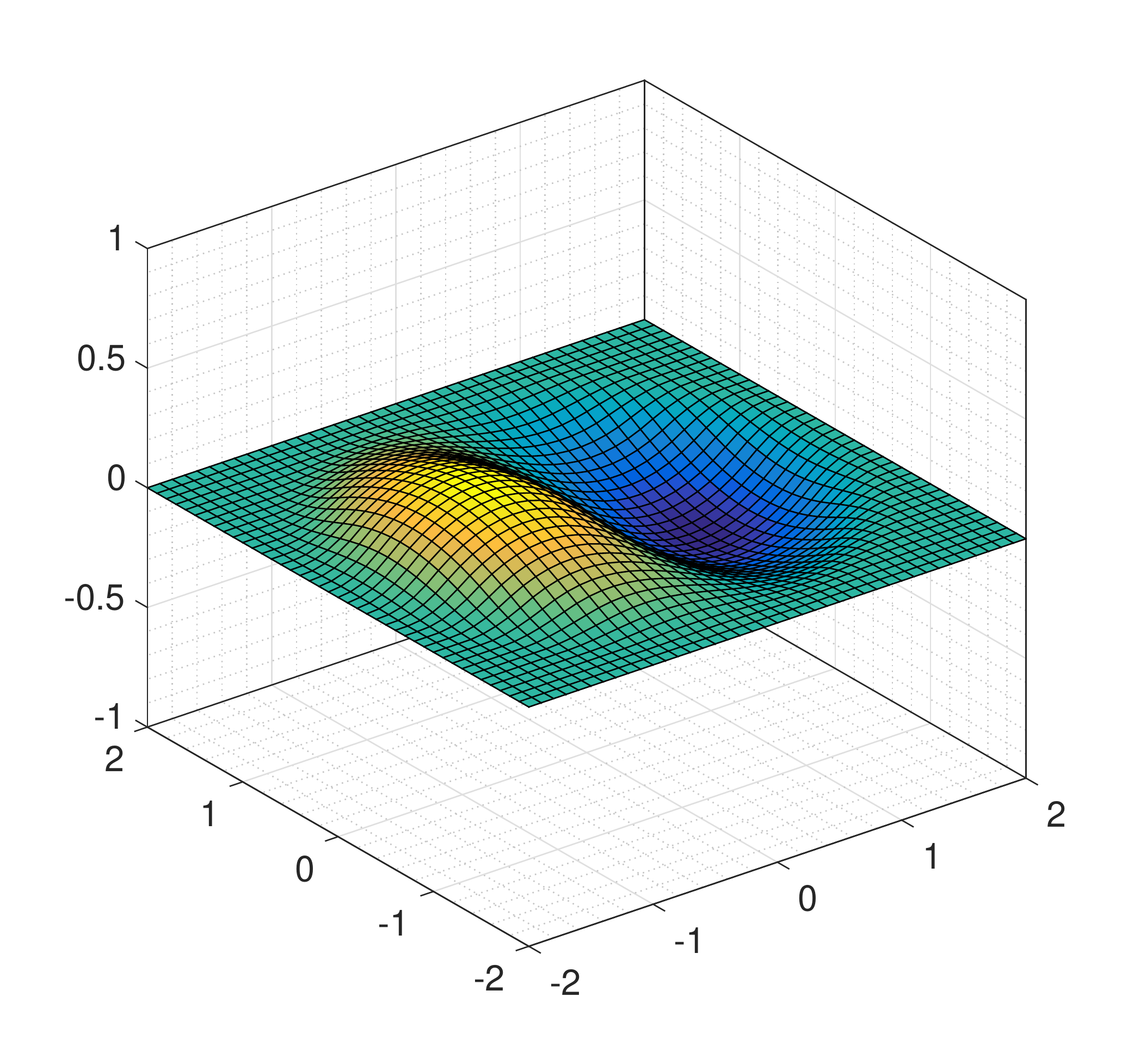}
				\caption{$\chi = 0.8$}
		\end{subfigure}%
	\caption{Cubic B-spline kernel and its first derivative in 2D for fixed $b=2$ and different $\chi = a/b$}\label{BS3D}
\end{figure}

The first derivative of quadratic and cubic B-spline kernel for different value so $\chi$ are plotted in Figure \ref{Kernel_deriv}. The similar concept may be extended to higher dimension. The algorithm in 2D is demonstrated in Figure \ref{Kernel-pic}. Cubic B-spline kernel and its first derivative in 2D for different $\chi$ are shown in Figure \ref{BS3D}. 


Any value of $\chi_i$ satisfying the above condition ensures a stable interaction between immediate neighbours. However, this may not always guarantee a stable computation as the interaction with other neighbours (i.e., $j \in \mathbb{N}^i \setminus  \overline{\mathbb{N}^i}$) may also contribute to the instability. In the next section a more general condition is obtained through dispersion analysis.

\section{Dispersion Analysis}\label{disp_an}
Any standard SPH method is associated with its inherent dispersion that arises due to the functional approximation over the support domain. In this current exercise, the additional dispersion introduced in the system for using adaptive kernel approximation is investigated. Let us consider a stable configuration of particles ${\{\overline{x}_i}\}_{i \in \mathbb{Z}^+}$  with characteristic spacing $\Delta p$ in a one dimensional domain $\overline{\Omega} \subset \mathbb{R}$. The system is provided with an infinitesimal perturbation from its stable state and subsequently the wave speed of the perturbation is calculated. Here onwords any variable ($f$) described in the stable state is noted as $\overline{f}$ and in the perturbed state as $f$ respectively for the ease of understanding. For the given stable configuration, assuming a uniform density $\overline{\rho}$, the mass associated with any particle may be obtained as $m = \overline{\rho} \Delta p$. This results in a uniform state of hydrostatic stress $\overline{\sigma}_i = \overline{\sigma} = K [\frac{\overline{\rho}}{\rho_0} - 1]$ at each particle (i.e., $\forall i \in \mathbb{Z}^+$ ), where $\rho_0$ is the initial density and $K$ is the bulk modulus. Ignoring the artificial viscosity term $\Pi_{ij}$ (as it does not have any effect in removing tensile instability \cite{swegle1994analysis}), the  governing Equations of continuity and motion at any particle in the stable state are described as per the conservation of mass and the momentum as,

\begin{equation}
\frac{d\overline{\rho}}{dt} = - \sum_{j \in \mathbb{N}^i } \overline{\rho} \, \Delta p \left[{\overline{v}_j - \overline{v}_i} \right] \triangledown \overline{W}_{ij}
\label{disp5}
\end{equation}
\begin{equation}
\frac{d{\overline{v}}_i}{dt} = - \sum_{j \in \mathbb{N}^i} \overline{\rho} \,\Delta p \left[2\frac{{\overline{\sigma}}_i}{\overline{\rho}^2} \right] \triangledown \overline{W}_{ij}
\label{disp4}
\end{equation}
where $\overline{W}_{ij} = W(\overline{x}_{ij}) = W(\overline{x}_i-\overline{x}_j)$ is the B-spline kernel function described in the stable configuration. The displacement $x_i$ at any $i^{th}$ particle is updated as,
\begin{equation}
\frac{d x_i}{dt} =  v_i
\end{equation}

Now when the system is provided with an infinitesimal perturbation ($\delta x$) it also induces a similar perturbation $\delta v$ in velocity and $\delta \rho$ in the assigned density at the particles. The configuration such obtained is referred to as the perturbed configuration $\Omega \subset \mathbb{R}$. Let us describe the position (${{x}_i} = \overline{{x}}_i + \delta x_i$), velocity (${{v}_i} = \overline{{v}}_i + \delta v_i$) and density (${{\rho}_i} = \overline{{\rho}} + \delta \rho_i$) at any arbitrary $i$-th particle in the perturbed configuration ($\Omega \subset \mathbb{R}$) as,
\begin{equation}
{{x}_i} = \overline{{x}_i} + Xe^{i(k\overline{x}_i - \omega t)}
\label{disp1}
\end{equation}
\begin{equation}
{{v}_i} = \overline{{v}_i} + Ve^{i(k\overline{x}_i - \omega t)}
\label{disp2}
\end{equation}
\begin{equation}
{\rho_i} = \overline{{\rho}_i} + D e^{i(k\overline{x}_i - \omega t)}
\label{disp3}
\end{equation}
where, $X, V, D$ are the respective coefficients for displacement velocity and density variation. $\omega$ and $k$ are the frequency and wave number of the oscillation. The governing conservation Equations in the perturbed configuration can be written as,
\begin{equation}
\frac{d\rho_i}{dt} = - \sum_{j \in \mathbb{N}^i}  \rho_j \Delta p \left( v_j - v_i \right) \triangledown W_{ij}
\label{disp7}
\end{equation}
\begin{equation}
\frac{d{{v}}_i}{dt} = - \sum_{j \in \mathbb{N}^i} \rho_j \Delta p \left(\frac{{\sigma}_i}{{\rho_i}^2} +\frac{{\sigma}_j}{{\rho_j}^2} \right) \triangledown W_{ij}
\label{disp6}
\end{equation}
where $W_{ij} = W(x_i -x_j) = W(\overline{x}_{ij} + \delta x_{ij})$ with $\delta x_{ij} = \delta x_{i} - \delta x_{j}$, is the kernel function described in the perturbed configuration. $W_{ij}$ can be expanded by Taylor series as,
\begin{equation}
W_{ij} = \overline{W}_{ij} + \delta x_{ij} \triangledown \overline{W}_{ij}
\label{disp7a}
\end{equation}
Similarly, $\triangledown W_{ij}$ can be expressed as,
\begin{equation}
\triangledown W_{ij} = \triangledown \overline{W}_{ij} + \delta x_{ij} \triangledown^2 \overline{W}_{ij}
\label{disp7b}
\end{equation}

As $\delta {x_{ij} < < 1}$, the higher order terms of $\delta x_{ij}$ are neglected in Equations \ref{disp7a} and \ref{disp7b}. The state of stress at the $i^{th}$ particle in perturbed configuration may be written as,  
\begin{equation}
\sigma_i = K[\frac{\rho_i}{\rho_0} -1],
\label{disp8}
\end{equation}
which further may be expanded as, 
\begin{equation}
\frac{{\sigma}_i}{\rho_i^2} = \frac{K}{\left( \overline{\rho} + \delta \rho_i \right)^2 }[\frac{\overline{\rho} + \delta \rho_i}{\rho_0} - 1]
\label{disp9}
\end{equation}
\begin{equation}
\Rightarrow \frac{{\sigma}_i}{\rho_i^2} = \frac{K}{\overline{\rho}^2 }[\frac{\overline{\rho}}{\rho_0} - 1 + \frac{\delta \rho_i}{\rho_0}][1 - 2\frac{\delta \rho_i}{\overline{\rho}}]
\label{disp10}
\end{equation}
\begin{equation}
\Rightarrow  \frac{{\sigma}_i}{\rho_i^2} = \frac{K}{\overline{\rho}^2}[\frac{\overline{\rho}}{\rho_0} - 1] + \frac{K}{\overline{\rho}^3 }[2 - \frac{\overline{\rho}}{\rho_0}]\delta \rho_i
\label{disp11}
\end{equation}
\begin{equation}
\Rightarrow  \frac{{\sigma}_i}{\rho_i^2} = \frac{\overline{{\sigma}}}{\overline{\rho}^2} + B\delta \rho_i
\label{disp12}
\end{equation}
where $B = \frac{K}{\overline{\rho}^3 }[2 - \frac{\overline{\rho}}{\rho_0}]$. Now when Equations \ref{disp1} - \ref{disp3} are replaced in the governing momentum balance Equation \ref{disp6} we can obtain,


\begin{equation}
\frac{d{v}_i}{dt} = - \sum_{j \in \mathbb{N}^i} \overline{\rho} [1 + \frac{\delta \rho_j}{\overline{\rho}}] \Delta p \left[ 2\frac{\overline{\sigma}}{\overline{\rho}^2} + B \lbrace \delta \rho_i + \delta \rho_j \rbrace \right]  \left[ \triangledown \overline{W}_{ij} + \delta x_{ij} \triangledown^2 \overline{W}_{ij} \right]
\label{disp13}
\end{equation}
Here also the higher order terms are neglected assuming the perturbations to be infinitesimal. The RHS of the Equation \ref{disp13} can be expanded as,
\begin{equation}
\begin{split}
\frac{d{\overline{v}}_i}{dt} + \frac{d}{dt} \delta v_i = - \sum_{j \in \mathbb{N}^i} \Delta p \left[2 \frac{\overline{\sigma}}{\overline{\rho}} \right] \triangledown \overline{W}_{ij} - \sum_{j \in \mathbb{N}^i} 2 \Delta p \frac{\overline{\sigma}}{\overline{\rho}} \delta x_{ij} \triangledown^2 \overline{W}_{ij} \\
- \sum_{j \in \mathbb{N}^i} 2 \Delta p \frac{\overline{\sigma}}{\overline{\rho}^2} \delta \rho_j \triangledown \overline{W}_{ij} - \sum_{j \in \mathbb{N}^i} \overline{\rho} \Delta p B \lbrace \delta \rho_i + \delta \rho_j \rbrace \triangledown \overline{W}_{ij}
\label{disp14}
\end{split}
\end{equation}
The first term in the RHS refers to the acceleration at any $i^{th}$ particle in the stable configuration \textit{i.e.} $\frac{d{\overline{v}}_i}{dt} = - \sum_{j \in \mathbb{N}^i} \Delta p \left[2 \frac{\overline{\sigma}}{\overline{\rho}} \right] \triangledown \overline{W}_{ij}$ (Equation \ref{disp6}). Putting the values of $\delta \rho_i$ and $\delta \rho_j$ the expression can be re framed as,
\begin{equation}
\begin{split}
-\omega^2 \delta x_{i} = - \sum_{j \in \mathbb{N}^i} 2 \Delta p \frac{\overline{\sigma}}{\overline{\rho}} \triangledown^2 \overline{W}_{ij} \delta x_{ij} - \sum_{j \in \mathbb{N}^i} 2 \Delta p \frac{\overline{\sigma}}{\overline{\rho}^2} D e^{i(k\overline{x}_j - \omega t)} \triangledown \overline{W}_{ij} \\
- \sum_{j \in \mathbb{N}^i} \overline{\rho} \, \Delta p B D \left\lbrace e^{i(k\overline{x}_i - \omega t)} + e^{i(k\overline{x}_j - \omega t)} \right\rbrace \triangledown \overline{W}_{ij}
\label{disp15}
\end{split}
\end{equation}

The expression of the frequency of oscillation ($\omega$) can be obtained by dividing Equation \ref{disp15} by $\delta x_i$, where $\frac{\delta x_{ij}}{\delta x_{i}} = 1 - \frac{\delta x_j}{\delta x_i} = 1 - e^{ik(\overline{x_j} - \overline{x_i})}$. As the final expression is valid everywhere within the domain (i.e., $\forall i \in \mathbb{Z}^+$ ), the particle index $i$ is omitted here onwards so that $\overline{x_{ij}}$ and $\overline{W_{ij}}$ are written as $\overline{r_j}$ and $\overline{W_{j}}$. $\overline{r_j}$ is the distance of the neighbourhood particle in the unperturbed state. Replacing $\frac{D}{X}$ by $\sum_{j=1}^{\mathbb{N}^i} \textit{i} \, \overline{\rho} \, \sin(k\overline{r_{j}}) \triangledown \overline{W}_{j}$ (as obtained from conservation of mass) (where $i^2 = -1$) the $\omega^2$ can be expressed as,
\begin{equation}
\omega^2 = 2\frac{{\overline{\sigma}}}{\overline{\rho}} \Delta p \sum_{j \in \mathbb{N}^i} \left[1- \text{cos}(k\overline{r}_{j}) \right] \triangledown^2 \overline{W}_{j} + \frac{K}{{\rho_0}^2} \lbrace \sum_{j \in \mathbb{N}^i}\Delta p \, \text{sin}(k\overline{r}_{j}) \triangledown \overline{W}_{j} \rbrace^2
\label{disp16}
\end{equation}

Expressing the kernel function and its derivatives in the parametric form , the frequency of the wave speed can be finally written as,

when, $0 < \frac{\overline{r_j}}{h} < a$
\begin{equation}
\begin{split}
\omega^2 = \frac{2\overline{{\sigma}}}{\overline{\rho}} \Delta p \frac{60}{\pi a^2 b^2 (a^2 +ab +b^2) h^3}\sum_{j \in \overline{\mathbb{N}^i}} \left[1- cos(k\overline{r_{j}}) \right] \lbrace (a+b)\frac{\overline{r_{j}}}{h} - ab \rbrace \\
+ \frac{K}{\overline{\rho}} (2- \frac{\overline{\rho}}{\rho_0}) \left\lbrace \sum_{j \in \overline{\mathbb{N}^i}} \Delta p \sin(k\overline{r}_{j}) \frac{30[(a+b)(\frac{\overline{r}_{j}}{h})^2 - 2ab\frac{\overline{r}_{j}}{h}]}{\pi a^2b^2 h^2 (a^2+ab+b^2)} \right\rbrace^2
\label{disp17a}
\end{split}
\end{equation}

when, $a < \frac{\overline{r_j}}{h} < b$
\begin{equation}
\begin{split}
\omega^2 = \frac{2\overline{{\sigma}}}{\overline{\rho}} \Delta p \frac{60}{\pi b (b-a) (a^2 +ab +b^2) h^3}\sum_{j \in \overline{\mathbb{N}^i}} \left[1- cos(k\overline{r_{j}}) \right] \lbrace b - \frac{\overline{r_{j}}}{h} \rbrace \\
+ \frac{K}{\overline{\rho}} \left\lbrace \sum_{j \in \overline{\mathbb{N}^i}} \Delta p \sin(k\overline{r}_{j}) \frac{30\lbrace b - \frac{\overline{r}_{j}}{h} \rbrace^2}{\pi b (b-a) h^2 (a^2+ab+b^2)} \right\rbrace^2
\label{disp17b}
\end{split}
\end{equation}

\begin{figure}[h!]
    \centering
    \begin{subfigure}[h!]{0.5\textwidth}
        \centering
        \includegraphics[width=1.0\textwidth]{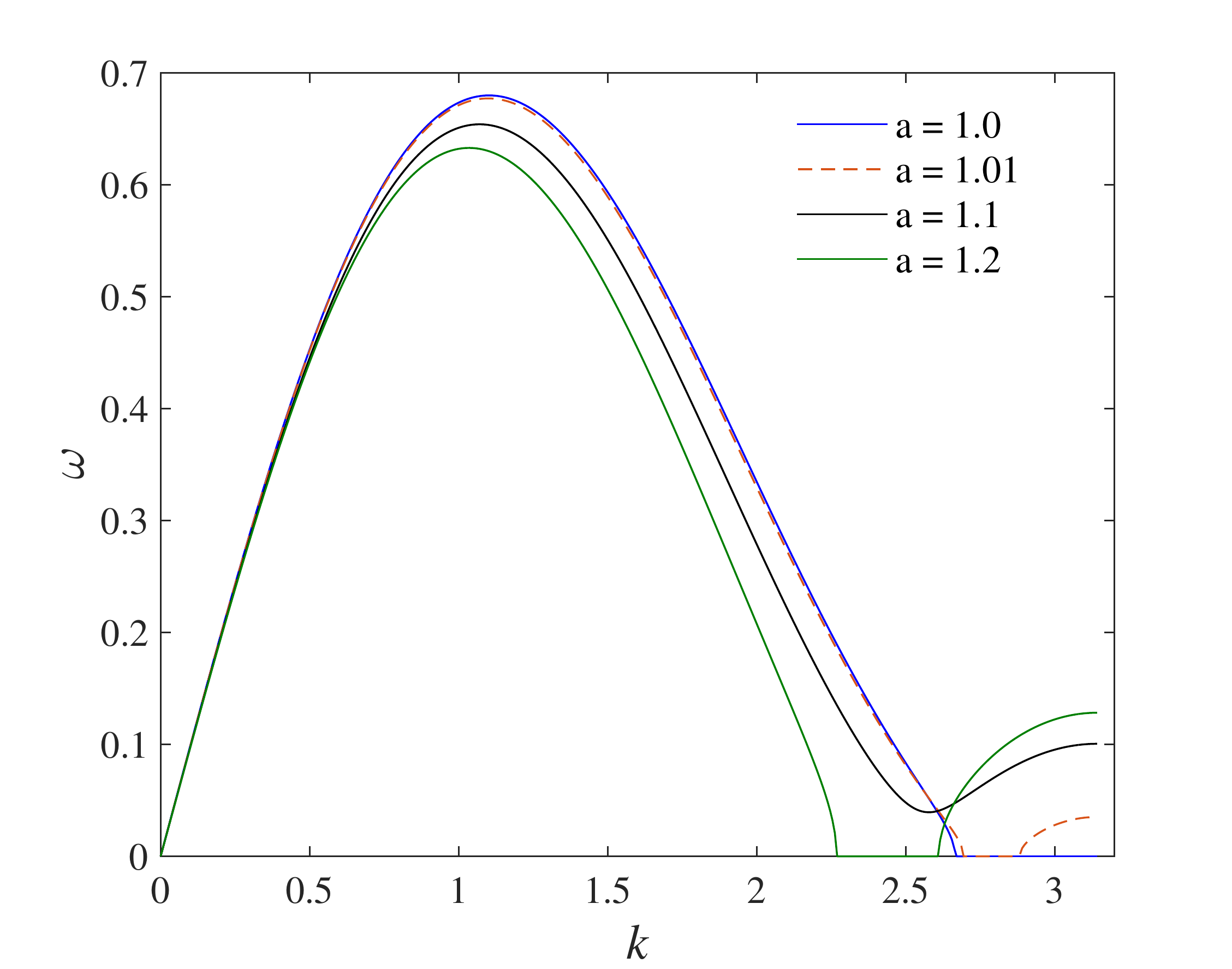}
        \caption{$\overline{\rho} = 0.96 \rho$}\label{fig: disp_fig_t}
    \end{subfigure}%
    ~
    \begin{subfigure}[h!]{0.5\textwidth}
        \centering
        \includegraphics[width=1.0\textwidth]{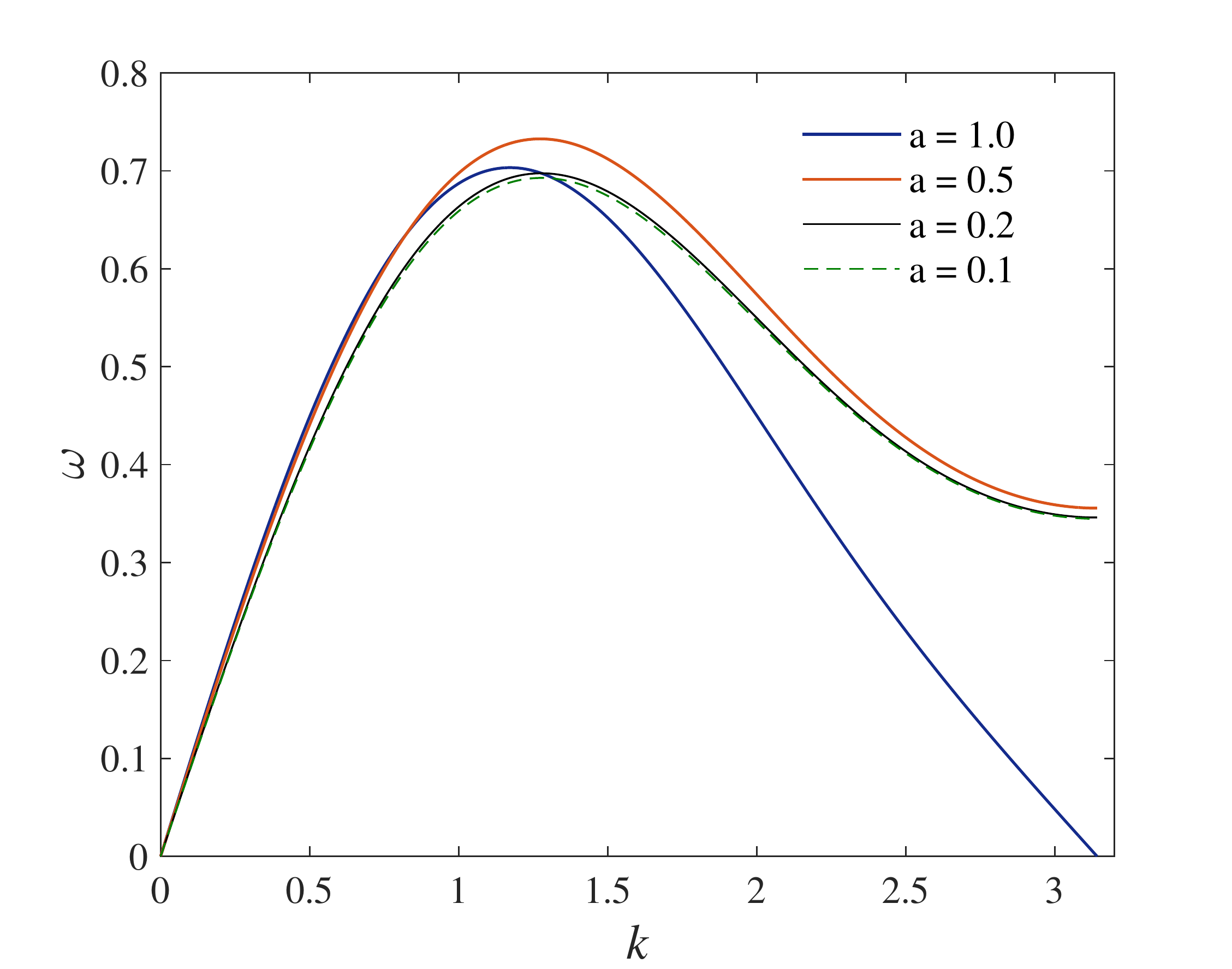}
        \caption{$\overline{\rho} = 1.05 \rho$}\label{fig: disp_fig_c}
    \end{subfigure}%
   \caption{Dispersion analysis of a 1D wave propagation problem with adaptive kernel}\vspace*{-6pt}
    \label{disp_fig}
\end{figure}

\begin{figure}[h!]
    \centering
        \includegraphics[width=0.6\textwidth]{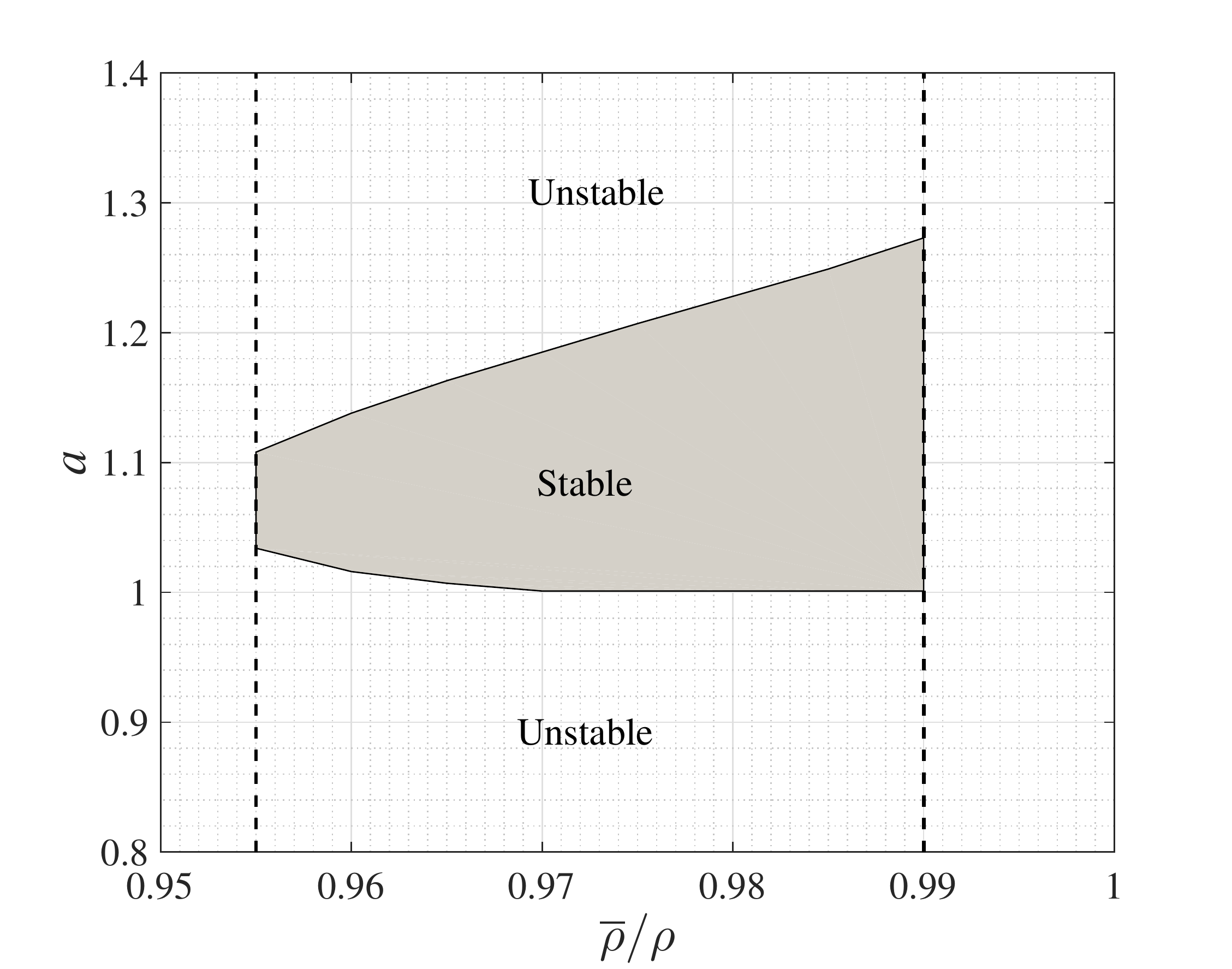}
        \caption{Stable zone for intermediate knot a for different $\overline{\rho}/\rho$ and $h=1.5\Delta p$}\label{fig: a_zone}
\end{figure}

Mathematically for any value of $k$ that results in $\omega < 0$, the system reflects zero energy modes or instability.  So the objective of removing instability clearly points out to ensure $\omega > 0$ for any wave number($k$) of the perturbing wave. Interestingly for any $k\overline{r_{j}} = n\pi, \forall n \in \mathbb{Z}^+ \rightarrow \omega = 0$. The smallest wave length that can be captured with a characteristic inter particle distance $\Delta p$ is $\lambda = 2\Delta p$, which in other form suffices to a wave number $k = \pi/\Delta p$, \textit{i.e.} $n = 1$. 

The dispersions of the numerical solution with cubic B-spline kernel for different values of $a$ are plotted in Figure \ref{disp_fig}. For this exercise $\Delta p = 1$, $h = 1.5\Delta p$, $\overline{\rho} = 0.96 \rho$ for tension and $\overline{\rho} = 1.05 \rho$ for compression are taken. As can be seen, appropriate values of $a$ ensure positive shift in the $\omega$ vs $k$ curve near the zero energy modes, which demonstrates the ability of the method, to remove the instability form the system. However, it is to be noted that for tension (Figure \ref{fig: disp_fig_t}) any value of $a$ satisfying Equation \ref{TInst} may not remove instability. Rather, there is a range within which the value of $a$ needs to be chosen for a stable computation. For a given kernel function this range depends on $h/\Delta p$ and $\overline{\rho} /\rho$ ratios. This is illustrated in Figure \ref{fig: a_zone}.   

Continuous update of the intermediate knot \textit{a} provides a way out of \textit{tensile instability} in SPH computation. The construction of B-spline basis function (section \ref{Bspline}) ensures that for any value of $a \in \left(0,b\right)$ the resulting basis satisfies all the essential properties (such as differentiability, compactness, positivity, symmetry, unique maxima, partition of unity \textit{etc.}) of a kernel and therefore does not compromise the premise of the kernel approximation in Equation \ref{eq1}. In the next section, the efficacy of the algorithm is demonstrated through some numerical example. 

\section{Numerical Examples}
\text{Tensile instability} in SPH has been identified and discussed in the literature through various examples. Some test cases among them are revisited in this section and the efficacy of the proposed adaptive formulation in removing  \textit{tensile instability} is demonstrated. In the example problems instability may occur either in tension or compression or both. The adaptive algorithm is implemented with the cubic B-spline kernel as given in Table \ref{Table1}. For a better compression of the relative advantage of the proposed algorithm, simulations are also performed with standard cubic kernel given in Equation \ref{BS3_st}.

\begin{equation}\label{BS3_st}
    W(q, h)=\alpha_d 
\begin{cases}
    1-\frac{3}{2} q^2 +\frac{3}{4} q^3,& \text{if } 0\le q\le 1\\
    \frac{1}{4}(2-q)^3,              & \text{if } 1\le q\le 2\\
    0                                 & \text{otherwise}
\end{cases}
\end{equation}
where, $\alpha_d=\frac{2}{3h}$ in 1D, $\alpha_d=\frac{10}{7\pi h^2}$ in 2D, $\alpha_d=\frac{1}{\pi h^3}$ in 3D and $q=\frac{|\mathbf{x}_i-\mathbf{x}_j|}{h}$ is the normalised position vector associated with a particle pair. 

In order to achieve the first order consistency \citep{chen1999corrective, islam2018consistency} the first derivative of the kernel function is modified as $\hat{W}_{ij,\beta}=B^{\beta \alpha}W_{ij,\alpha}$ where $\textbf{B}$ is a symmetric re-normalization matrix obtained as $\mathbf{B}^{-1}_i = -\sum_{j\in N^i} \frac{m_j}{\rho_j} \mathbf{x}_{ij}  \otimes \mathbf{\bigtriangledown} W_{ij}$. In all the simulations artificial viscosity (Equation \ref{artvisc}) and XSPH correction (Equation \ref{XSPH}) are used. This is to be noted that the purpose of the present study is to remove the \text{tensile instability} only by changing the shape of the kernel function and therefore no artificial stress correction is used. In all the example, the smoothing length ($h$) is taken as 1.5$\Delta p$. The adaptive algorithm is implemented with $b = 2$, and the intermediate knot $a_i$ updated as,

\begin{equation}\label{a_value}
    a_i= 
\begin{cases}
    1.1\frac{r_d}{h} ,& \text{if } \frac{\rho_i}{\rho_0} < 1\\
    0.2                                 & \text{otherwise}
\end{cases}
\end{equation}
where, for $i$-th particle, $r_d=max\left\lbrace r_{ij}, j\in \overline{\mathbb{N}^i} \right\rbrace$ as defined in Figure \ref{Kernel-pic}. The above values of $a_i$ are taken as per the dispersion analysis given in section \ref{disp_an}. For $b=2$ and $h=1.5\Delta p$, $a_i = 1.1r_d/h$ yields a stable computation in tension (Figure \ref{fig: a_zone}). For compression, Figure \ref{fig: disp_fig_c} suggests that any value of $a_i<1$ (for $b=2$) ensures stability and thus $a_i = 0.2$ is uniformly taken in compression. In order to preserve the symmetry in the formulation, when two particles (say $i$-th and $j$-th) interact the kernel function is constructed with an average intermediate knot $a_{ij}=(a_j+a_j)/2$.     

\subsection{2D Stability Test}
\begin{figure}[h!]
    \centering
    \begin{subfigure}[h!]{0.3\textwidth}
        \centering
        \includegraphics[width=\textwidth]{box_200.pdf}
        \caption{Standard SPH (t=100$\mu$s)}\label{P1-2}
    \end{subfigure}%
    ~ 
    \begin{subfigure}[h!]{0.3\textwidth}
        \centering
        \includegraphics[width=\textwidth]{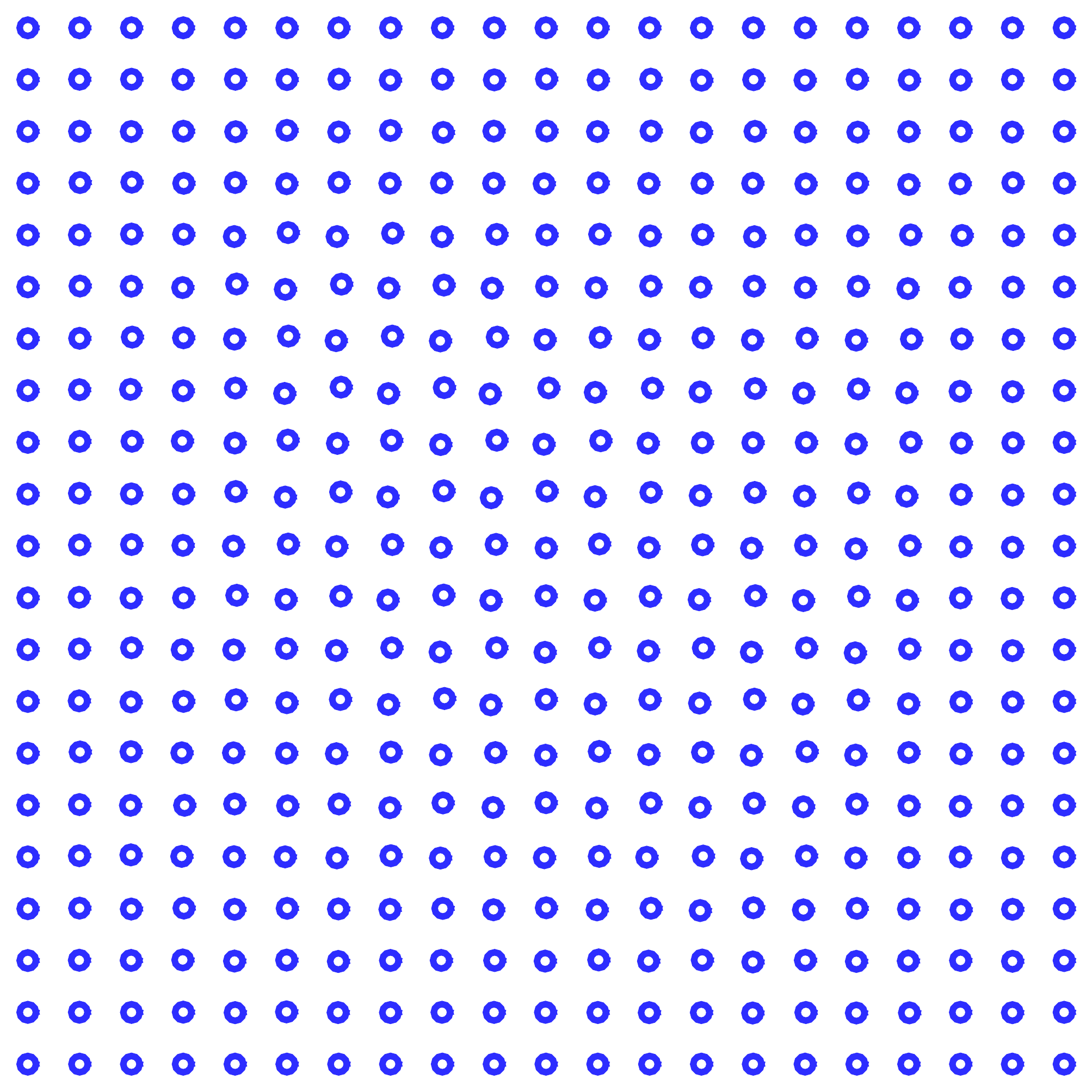}
        \caption{Adaptive SPH (t=100 $\mu$s)}\label{P1-3}
    \end{subfigure}%
    ~
    \begin{subfigure}[h!]{0.3\textwidth}
        \centering
        \includegraphics[width=\textwidth]{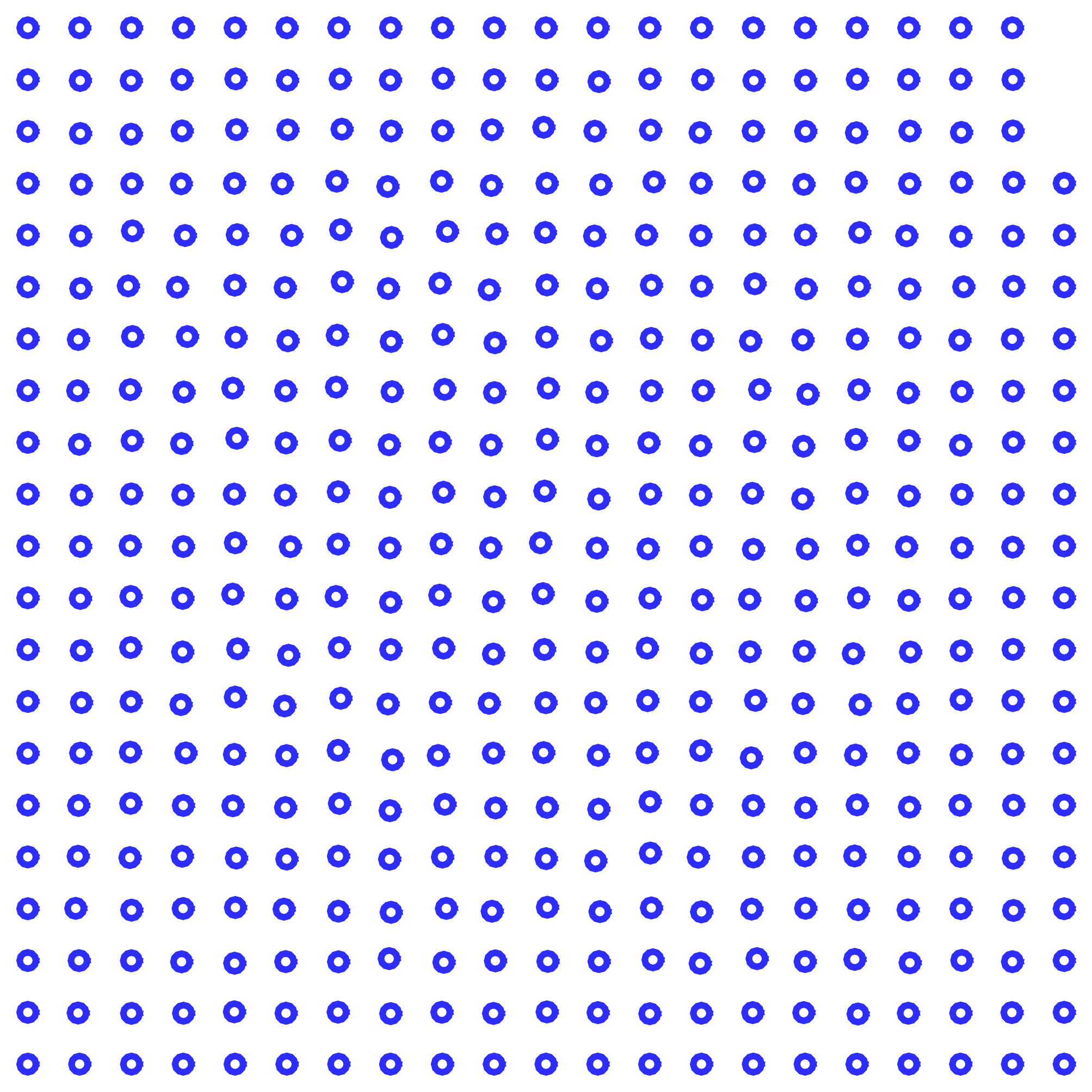}
        \caption{Adaptive SPH (t=1 ms)}\label{P1-4}
    \end{subfigure}%
   \caption{2D stability test: comparison of particle configurations at different time instants}\vspace*{-6pt}
    \label{Box}
\end{figure}

To start with, the 2D stability test considered in section \ref{TI} to illustrate the \textit{tensile instability} is considered once again. Geometry, material parameters and initial conditions are kept same as before. Simulations are performed with adaptive kernel and the computed configurations at 100 $\mu$s are compared in Figure \ref{Box}. It can be readily seen that the instability is fairly removed from the solution. In order to ensure that the computation remains stable, the simulation is continued for a long time. As evident in Figure \ref{P1-4}, there is no indication of instability even after 1 $ms$.   

\subsection{Wave propagation in a bar}
\begin{figure}[h!]\vspace*{4pt}
\centerline{\includegraphics[trim={0 0 0 0},clip,width=0.75\textwidth]{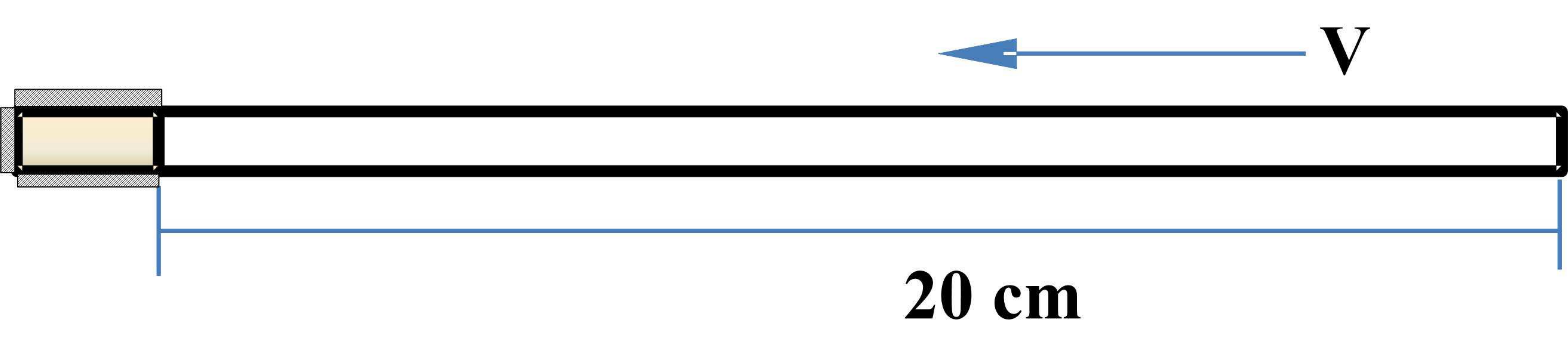}}
\caption{Schematic diagram of an axial bar under compression, L = 20 cm , B = 1 cm, V = 1 m/s}\vspace*{-6pt}
\label{P2}
\end{figure}

\begin{figure}[h!]
	\centering
		\begin{subfigure}[t]{0.5\textwidth}
			\centering
				\includegraphics[trim={0 5cm 3cm 0},width=1.0\textwidth]{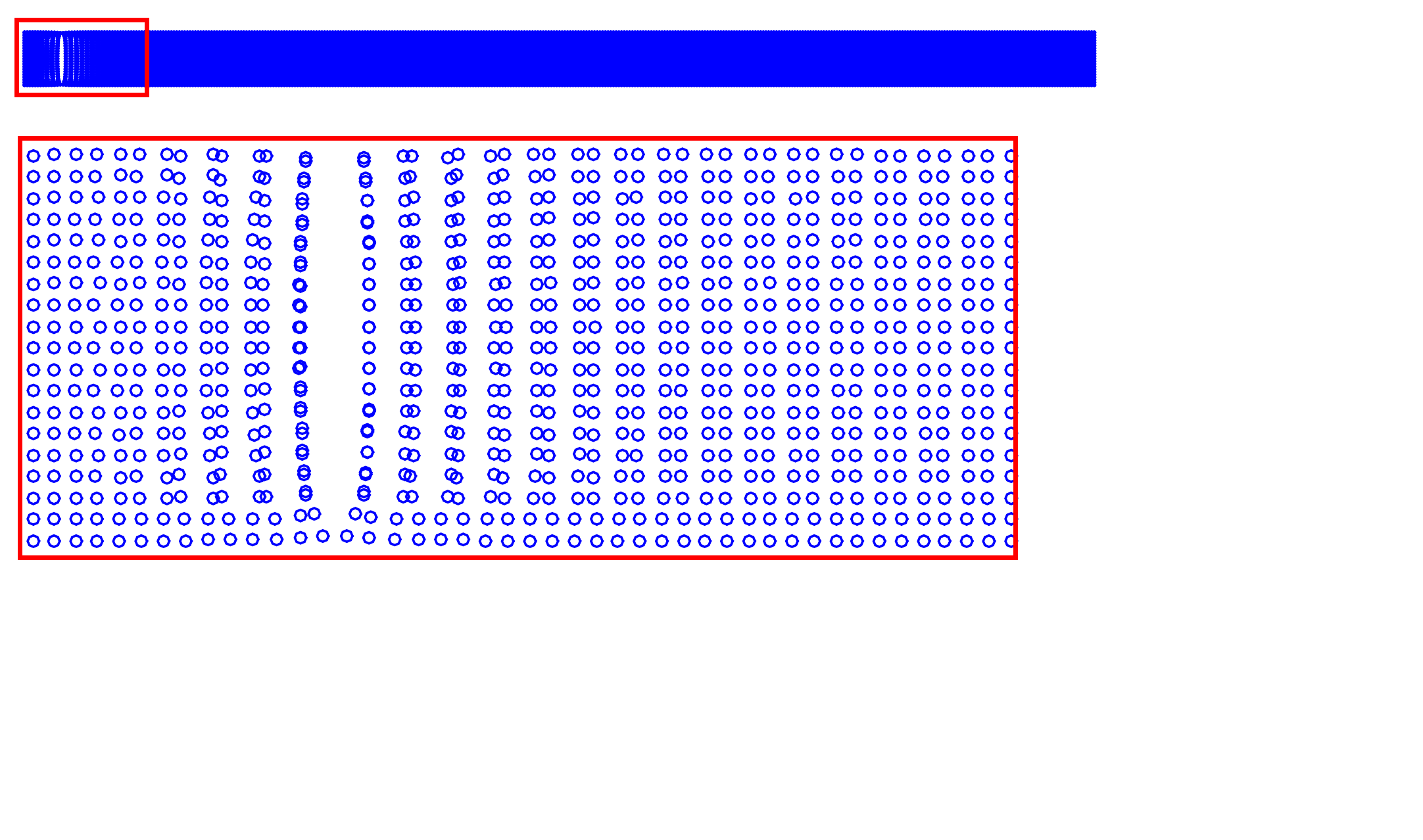}\\
				\caption{$t = 7.5 ms$}
		\end{subfigure}%
		\begin{subfigure}[t]{0.5\textwidth}
			\centering
				\includegraphics[trim={0 5cm 3cm 0},width=1.0\textwidth]{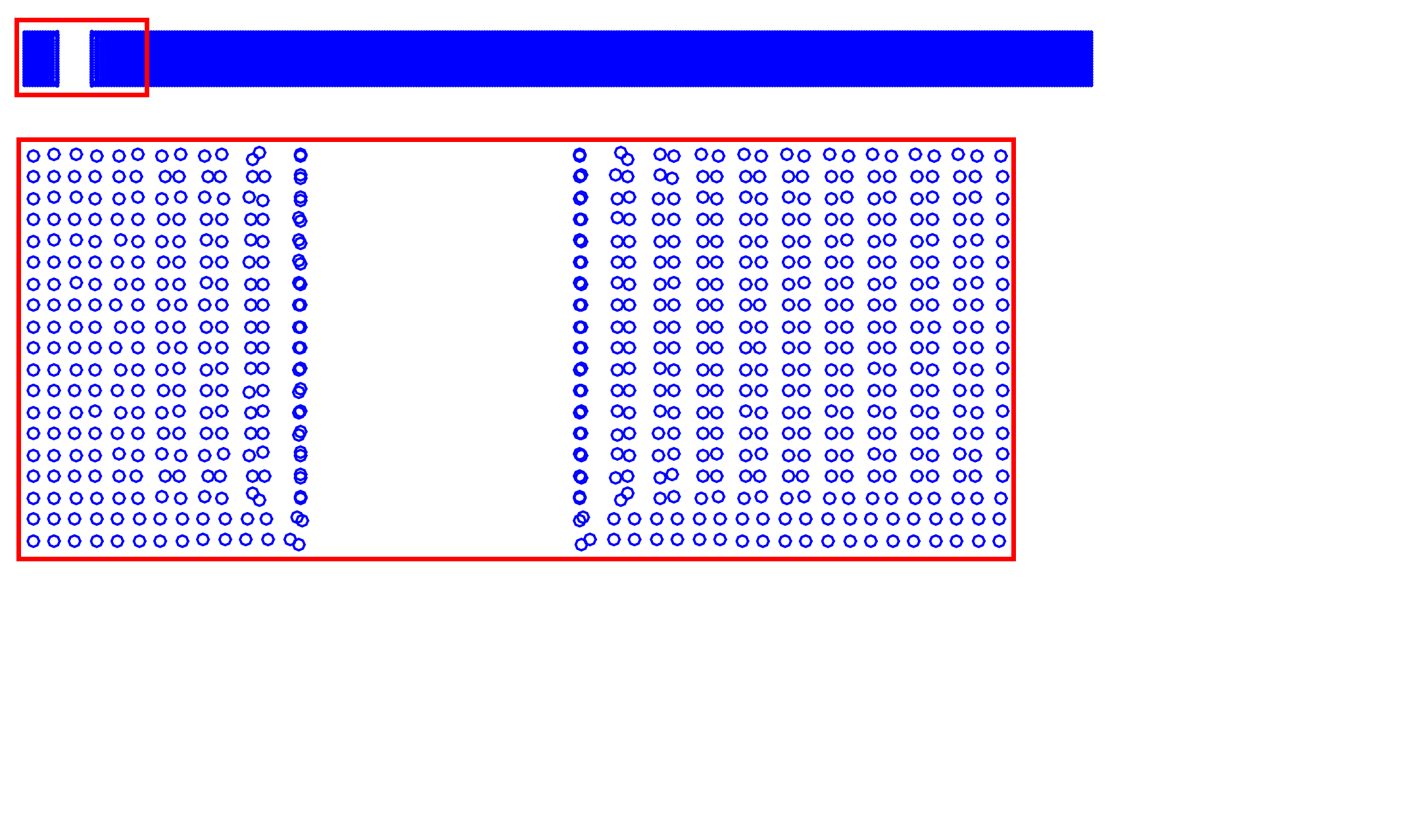}\\
				\caption{$t = 10 ms$}
		\end{subfigure}\\
	\caption{Deformed configurations at different time instants obtained via standard SPH}\label{P2_st}
\end{figure}

\begin{figure}[h!]
	\centering
		\begin{subfigure}[t]{0.5\textwidth}
			\centering
				\includegraphics[trim={0 5cm 3cm 0},width=1.0\textwidth]{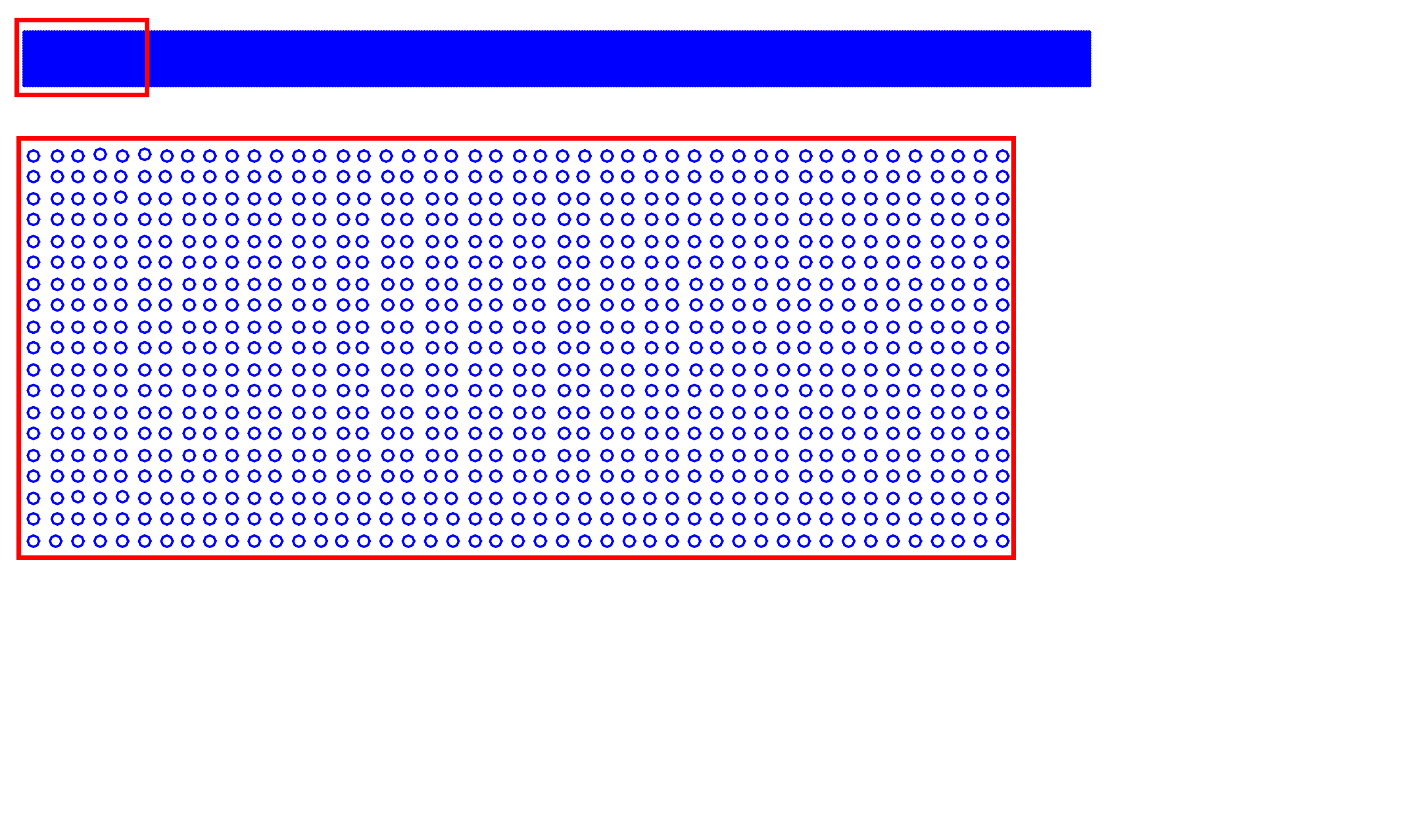}\\
				\caption{$t = 7.5 ms$}
		\end{subfigure}%
		\begin{subfigure}[t]{0.5\textwidth}
			\centering
				\includegraphics[trim={0 5cm 3cm 0},width=1.0\textwidth]{P2-10ms-adp.pdf}\\
				\caption{$t = 10 ms$}
		\end{subfigure}\\
	\caption{Deformed configurations at different time instants SPH with adaptive kernel}\label{P2_adp}
\end{figure}

Next an elastic cantilever rod of length 20cm and depth 1cm subjected to an uniform compressive velocity of $v_0 = 1 m/s$ throughout its length as depicted in Figure \ref{P2} is considered. Material properties and other computational data are provided in Table \ref{bar_value}. As the bar undergoes compression, the compressive wave traverses towards the support and is reflected back as tensile wave. With further progression of the tensile wave, instability that starts to generate in the standard SPH computation, manifests itself in the form of unphysical numerical fracture as shown in Figure \ref{P2_st}. However when the computation is performed with the adaptive kernel no instability is observed (Figure \ref{P2_adp}). 
The analytical expression for the the longitudinal displacement at a given point is,
\begin{equation}\label{PT1}
u(x,t) = \sum_{j=1}^{\infty} \frac{8(-1)^{k+1} v_0 L}{\pi^2 (2k-1)^2 c} \text{sin} \left( (2k-1) \frac{c\pi t}{2L} \right) \text{cos} \left( (2k-1) \frac{\pi x}{2L} \right), 
\end{equation}
where, $c=\sqrt{\frac{E}{\rho}}$ is the speed of elastic wave in the medium. 

\begin{table}[hbtp!]
\centering
\caption{Parameters for wave propagation in a bar}\label{bar_value}
\begin{tabular}{cccccccc}
\hline
                           & \multicolumn{3}{c}{Material Proerties}      & \multicolumn{2}{c}{Discretization} & \multicolumn{2}{c}{Simulation}                                              \\
\multirow{2}{*} & $\rho$     & $E$   & \multirow{2}{*}{$\nu$} & $\Delta p$          & $\Delta t$          & \multirow{2}{*}{$\gamma_1$} & \multirow{2}{*}{$\gamma_2$}         \\
                           & ($kg/m^3$) & (MPa) &                        & (mm)                & (sec)         &                            &                             \\ \hline
                   & 2000       & 10  & 0                    & 0.5                & $2\times 10^{-6}$         & 1.0                        & 1.0                                      \\ \hline
\end{tabular}
\end{table}

\begin{figure}[h!]\vspace*{4pt}
\centerline{\includegraphics[trim={0 0 0 0},clip,width=0.60\textwidth]{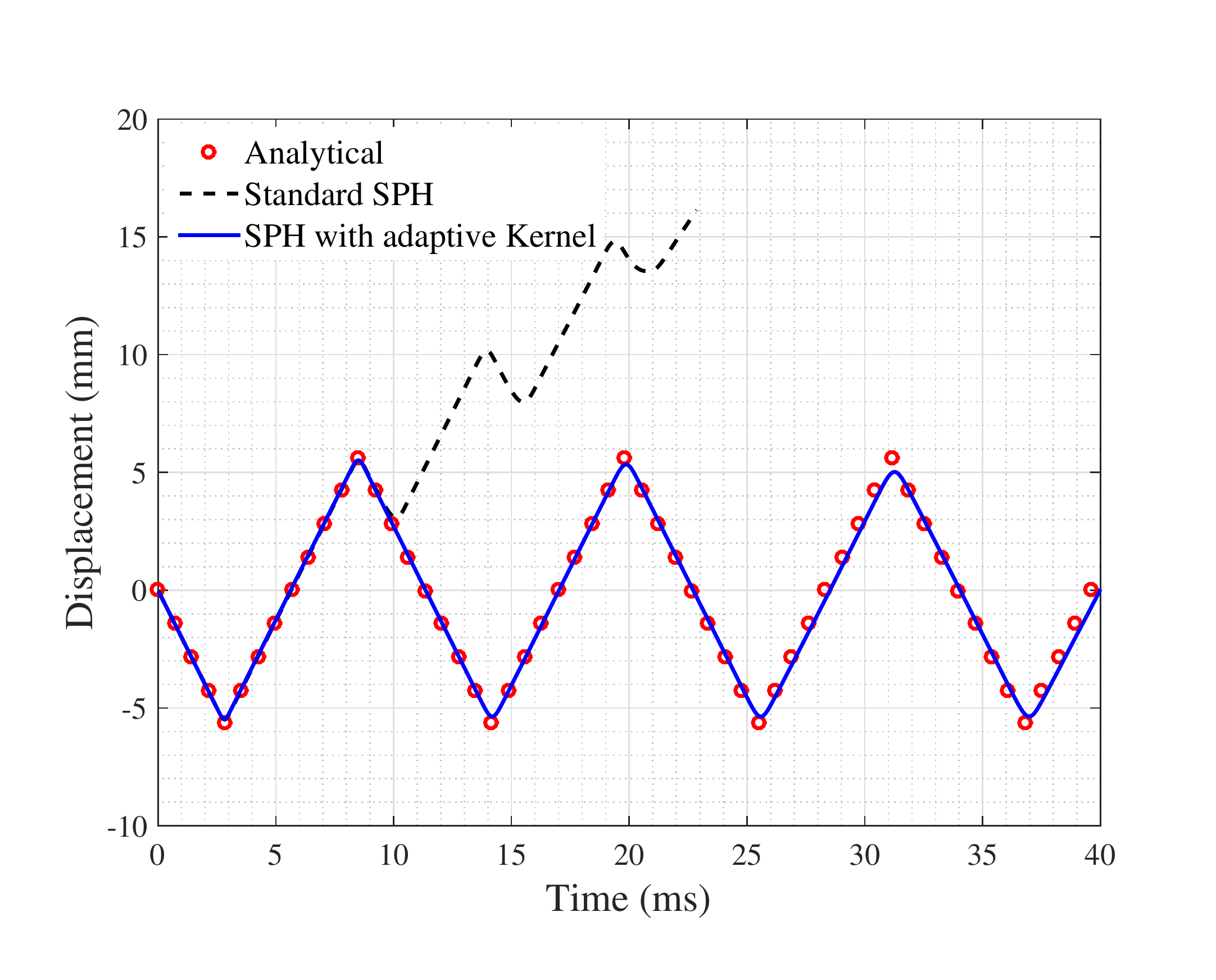}}
\caption{Comparision of displacement time histories at the tip of the bar}\vspace*{-6pt}
\label{P2-disp-plot}
\end{figure}

The computed longitudinal displacement time history at the tip of the bar is compared with that of obtained from Equation \ref{PT1} in Figure \ref{P2-disp-plot}. Displacement time history obtained via the standard SPH is also shown in the same figure. The effect of instability in the standard SPH is evident. On the other hand, the predicted response via SPH with adaptive kernel shows very good agreement with the analytical solution.  

\subsection{Transverse oscillation of a plate}  
\begin{figure}[h!]\vspace*{4pt}
\centerline{\includegraphics[trim={0 0 0 0},clip,width=0.75\textwidth]{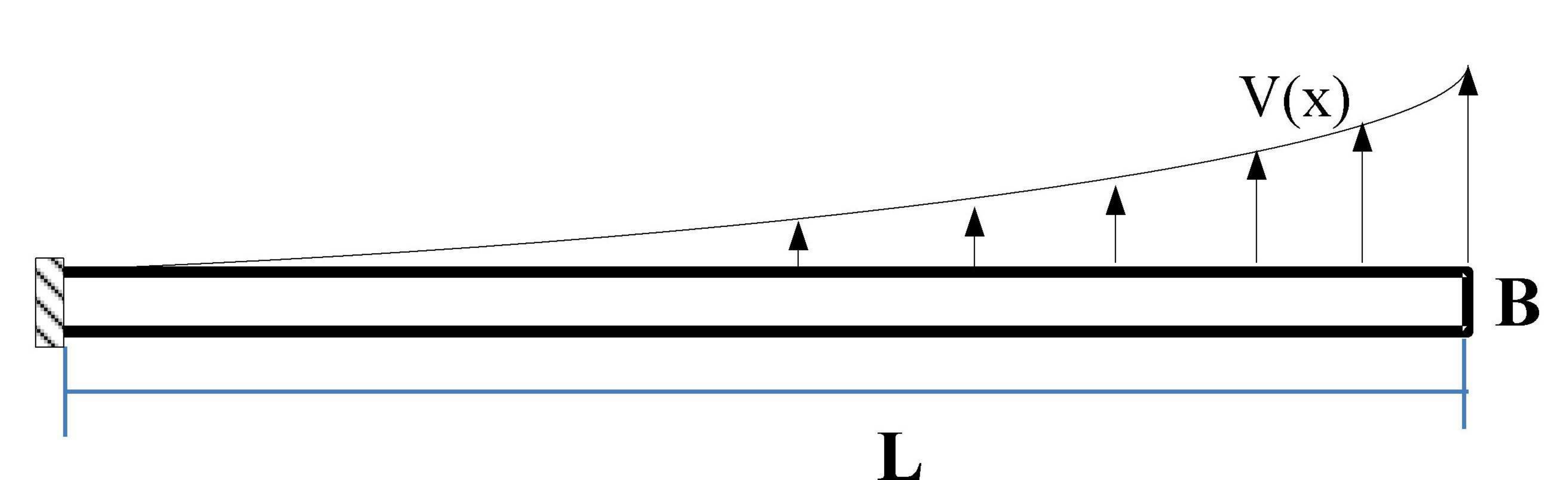}}
\caption{Cantilever plate under oscillation}\vspace*{-6pt}
\label{P3}
\end{figure}

Next a thin linear elastic cantilever plate of length $L$ and thickness $B$, oscillating in two dimensional mode as shown in Figure \ref{P3} is considered. The frequency of the oscillation is given by,  

\begin{equation}\label{beam_omega}
\omega^2 = \frac{EB^2k^4}{12\rho(1-\nu^2)}, 
\end{equation}
where, $\rho, E, \nu $ are the density, Elastic modulus and Poison ratio respectively. $k$ is the wave number determined from the condition $\text{cos}(kL)\text{cosh}(kL) = -1$. For the first mode of vibration $kL = 1.875$. The plate is set to motion by giving initial transverse velocity $v_y$ according to the first mode as,

\begin{equation}
\frac{v_y}{c} = V_f\frac{M\left[\text{cos}(kx) - \text{cosh}(kx)\right] - N\left[\text{sin}(kx) - \text{sinh}(kx)\right]}{Q}
\end{equation}\label{beam_velocity}    
where, $c$ is the speed of sound wave in the medium. $M, ~ N$ and  $Q$ are constants given as,

\begin{equation}
M = \text{sin}(kL) + \text{sinh}(kL)
\end{equation}
\begin{equation}
N = \text{cos}(kL) + \text{cosh}(kL)
\end{equation}
\begin{equation}
Q = 2 \left[\text{cos}(kL)\text{sinh}(kL) - \text{sin}(kL)\text{cosh}(kL) \right]
\end{equation}
In equation \ref{beam_velocity} $V_f$ is the transverse velocity at the free end. The material properties are taken as $\rho= 7850 kg/m^3, E = 210 GPa, \nu =0.3$. 
 
Length($L$) and thickness ($B$) of the plate are considered as 200mm and 20mm respectively, and the initial tip velocity $V_f$ is given as 0.02 m/s. In order to check the effect of adaptive kernel on the convergence behaviour, simulations are performed with different inter particle spacing ($\Delta p$). For all the cases the artificial viscosity parameters are taken as  $\gamma_1 = 1.0, \gamma_2 = 1.0$. The computed time periods for different $\Delta p$ are given in table \ref{P3_period}. For $\Delta p$ = 0.5mm the computed time period ($T_{c}$) is very close to its theoretical value ($T_{t} = 2\pi/\omega$, $\omega$ from Equation \ref{beam_omega}) and therefore results for $\Delta p$ = 0.5mm are considered for subsequent demonstration. Time step ($\Delta t$) is taken as $5\times 10^{-8}$sec. 

\begin{table}[h!]
\caption{Period of Vibration at the tip of the plate}
\begin{tabular*}{\hsize}{@{\extracolsep{\fill}}llll@{}}
\hline
$\Delta p$ (mm)	& $T_c$ (ms) & $T_{t}$(ms) \\
\hline
$4.0$	& 2.69	& 2.28	\\
$2.0$	& 2.69	& 2.28	\\
$1.0$	& 2.55	& 2.28	\\
$0.5$	& 2.35	& 2.28	\\
\hline
\end{tabular*}
\label{P3_period}
\end{table}

\begin{figure}[h!]\vspace*{4pt}
	\centering
		\begin{subfigure}[t]{0.5\textwidth}
			\centering
				\includegraphics[trim={0cm 6.5cm 5cm 0cm},width=1.0\textwidth]{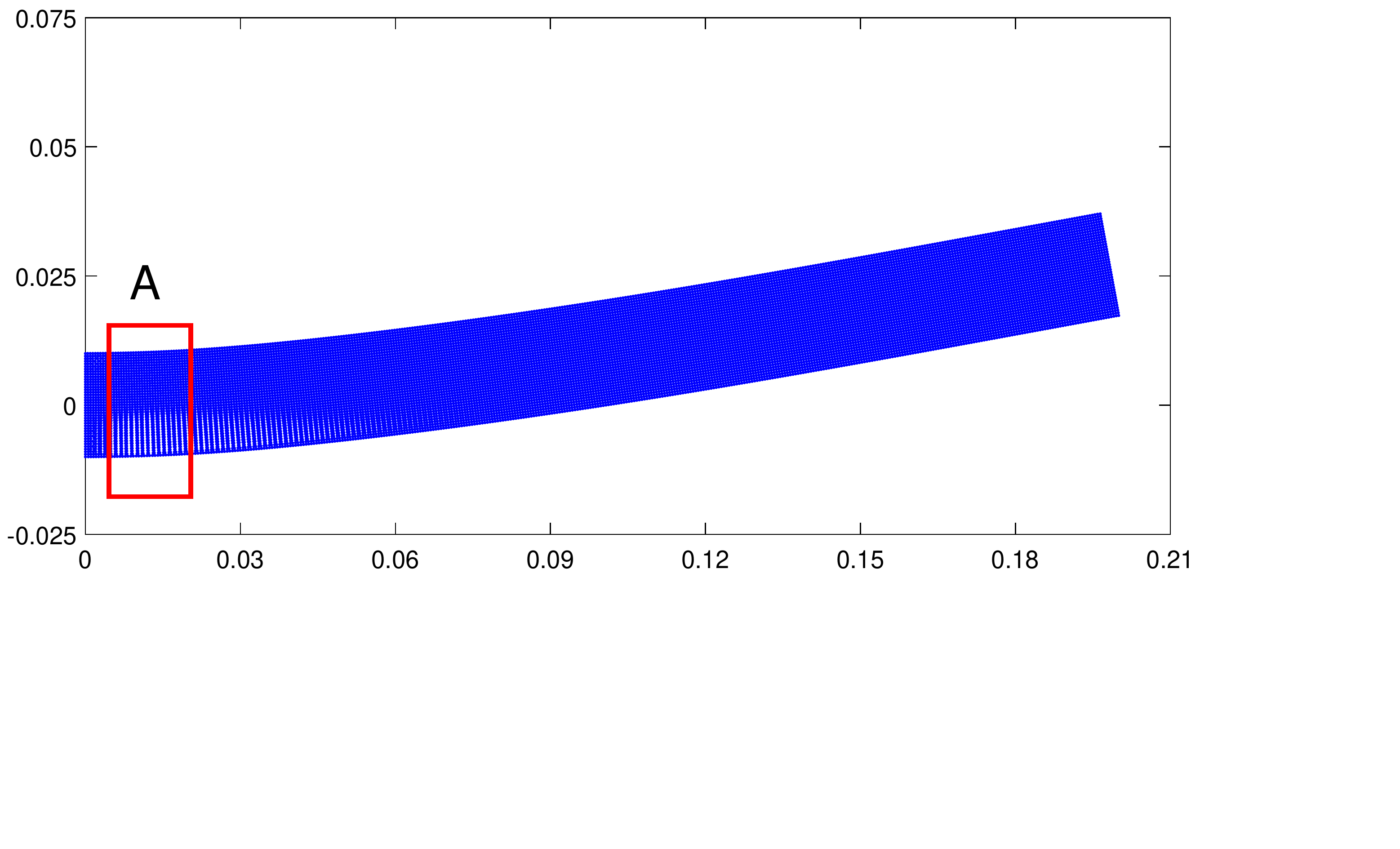}
				\caption{$t = T/8$}
		\end{subfigure}%
		~
		\begin{subfigure}[t]{0.5\textwidth}
			\centering
				\includegraphics[trim={0cm 6.5cm 5cm 0cm},width=1.0\textwidth]{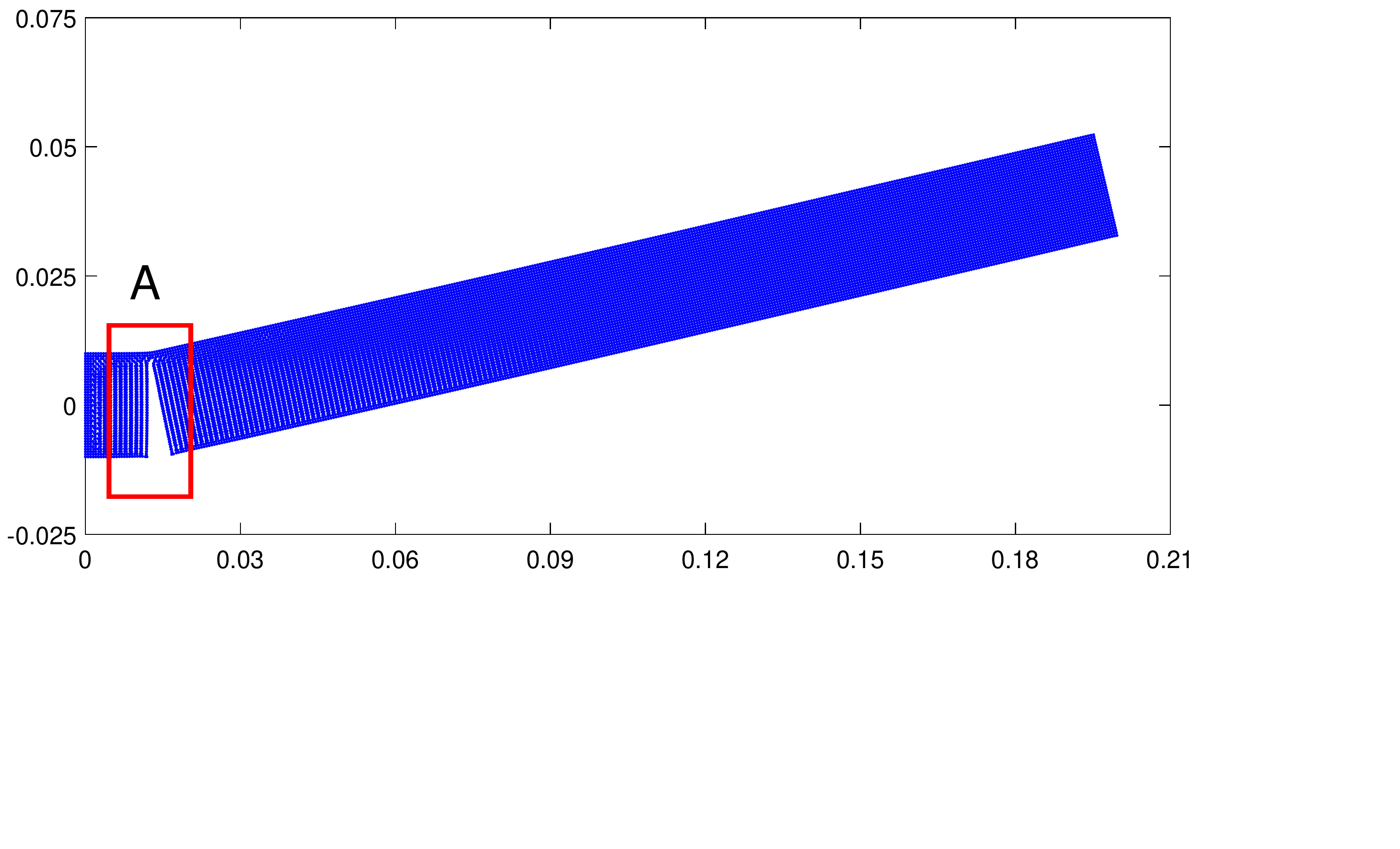}
				\caption{$t = T/4$}
		\end{subfigure} \\
		
		\begin{subfigure}[t]{0.4\textwidth}
			\centering
				\includegraphics[trim={1cm 2cm 16cm 0cm},width=1.0\textwidth]{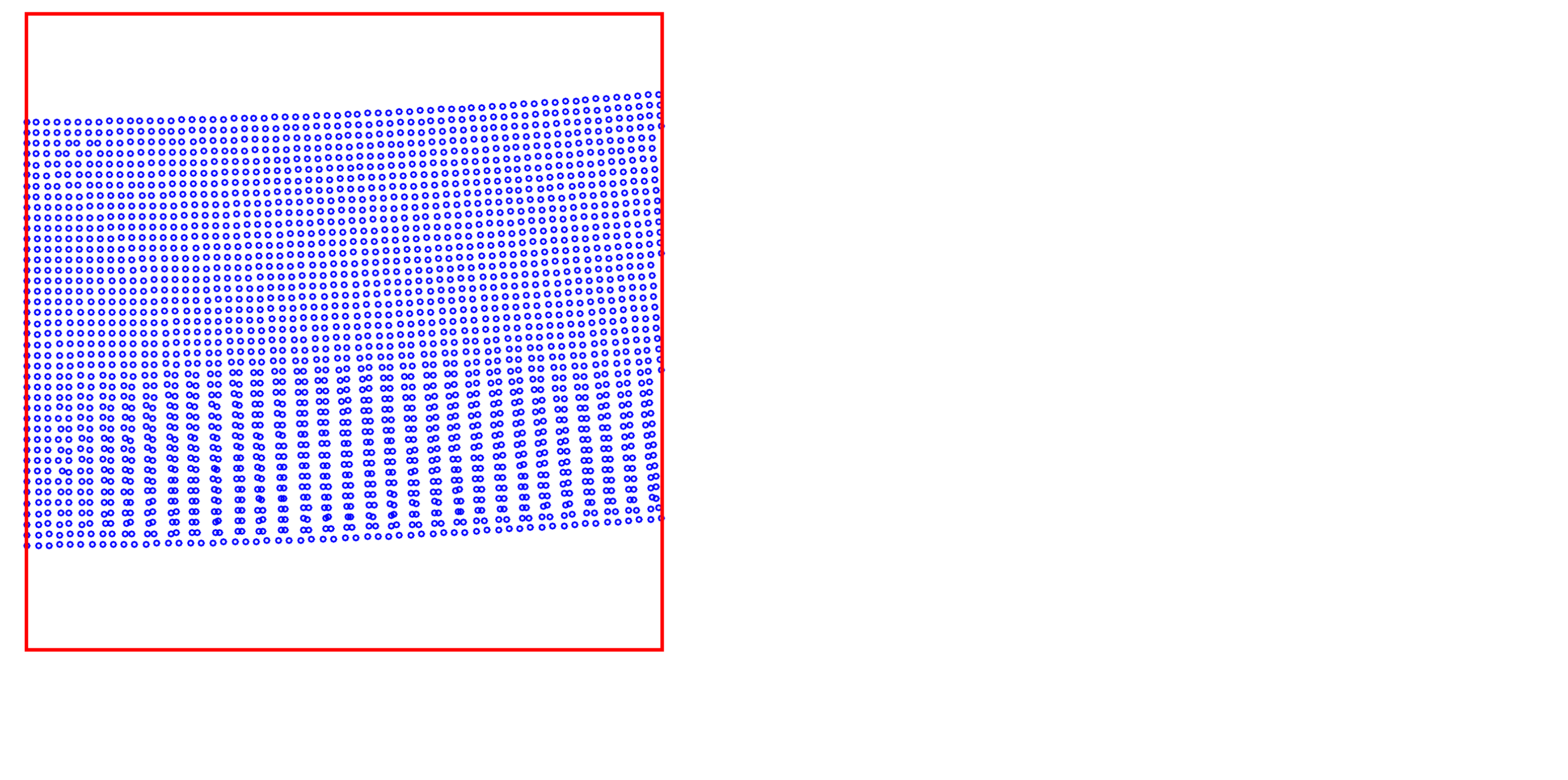}
				\caption{$t = T/8$}
		\end{subfigure}%
		~
		\begin{subfigure}[t]{0.4\textwidth}
			\centering
				\includegraphics[trim={1cm 2cm 16cm 0cm},width=1.0\textwidth]{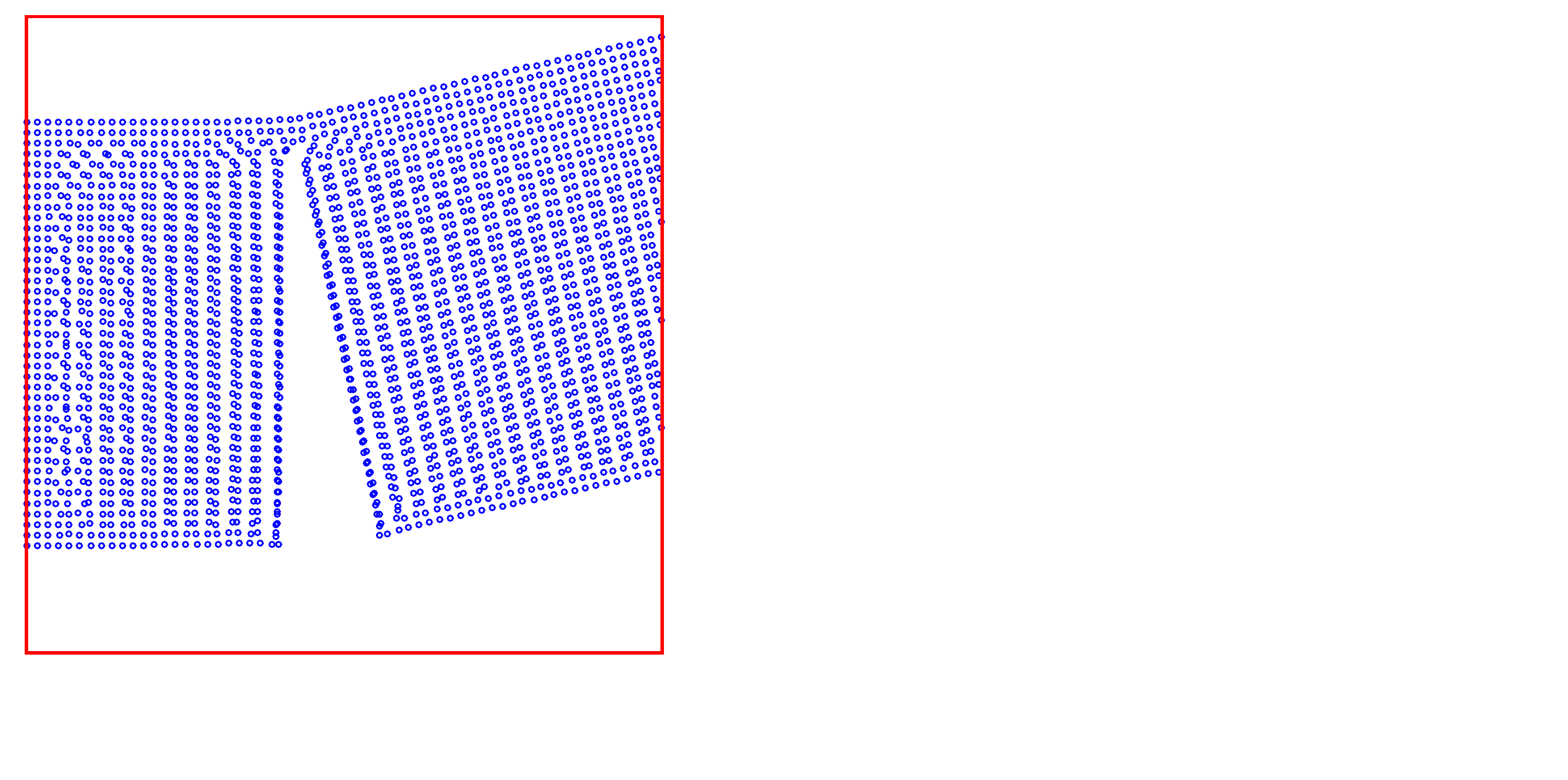}
				\caption{$t = T/4$}
		\end{subfigure}%
	\caption{Displacement profiles at different time instants using standard SPH, T = 2.35 ms}\label{P3_st}\vspace*{6pt}
\end{figure}

\begin{figure}[h!]\vspace*{4pt}
	\centering
		\begin{subfigure}[t]{0.5\textwidth}
			\centering
				\includegraphics[trim={0cm 6.5cm 5cm 0cm},width=1.0\textwidth]{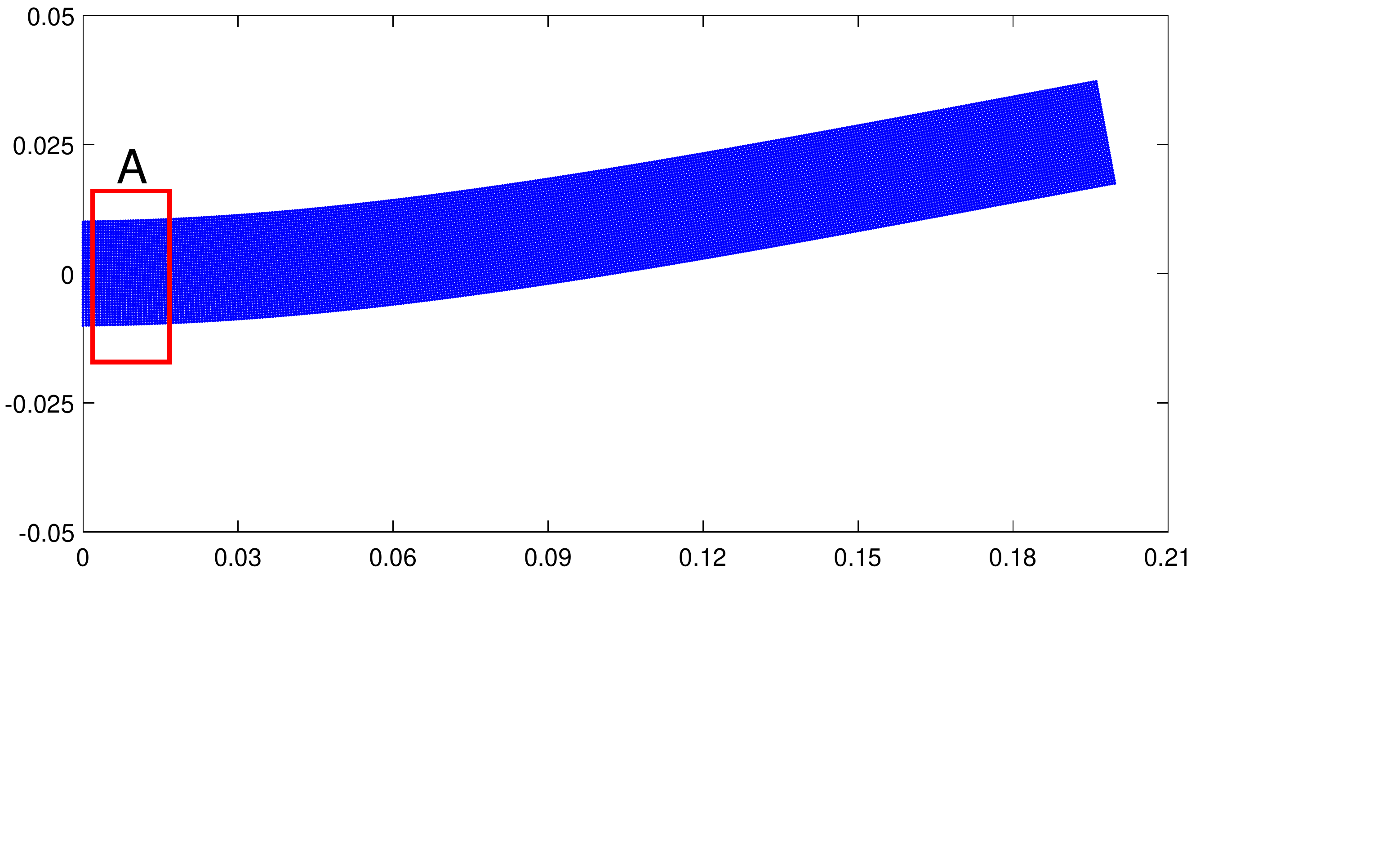}
				\caption{$t = T/8$}
		\end{subfigure}%
		~
		\begin{subfigure}[t]{0.5\textwidth}
			\centering
				\includegraphics[trim={0cm 6.5cm 5cm 0cm},width=1.0\textwidth]{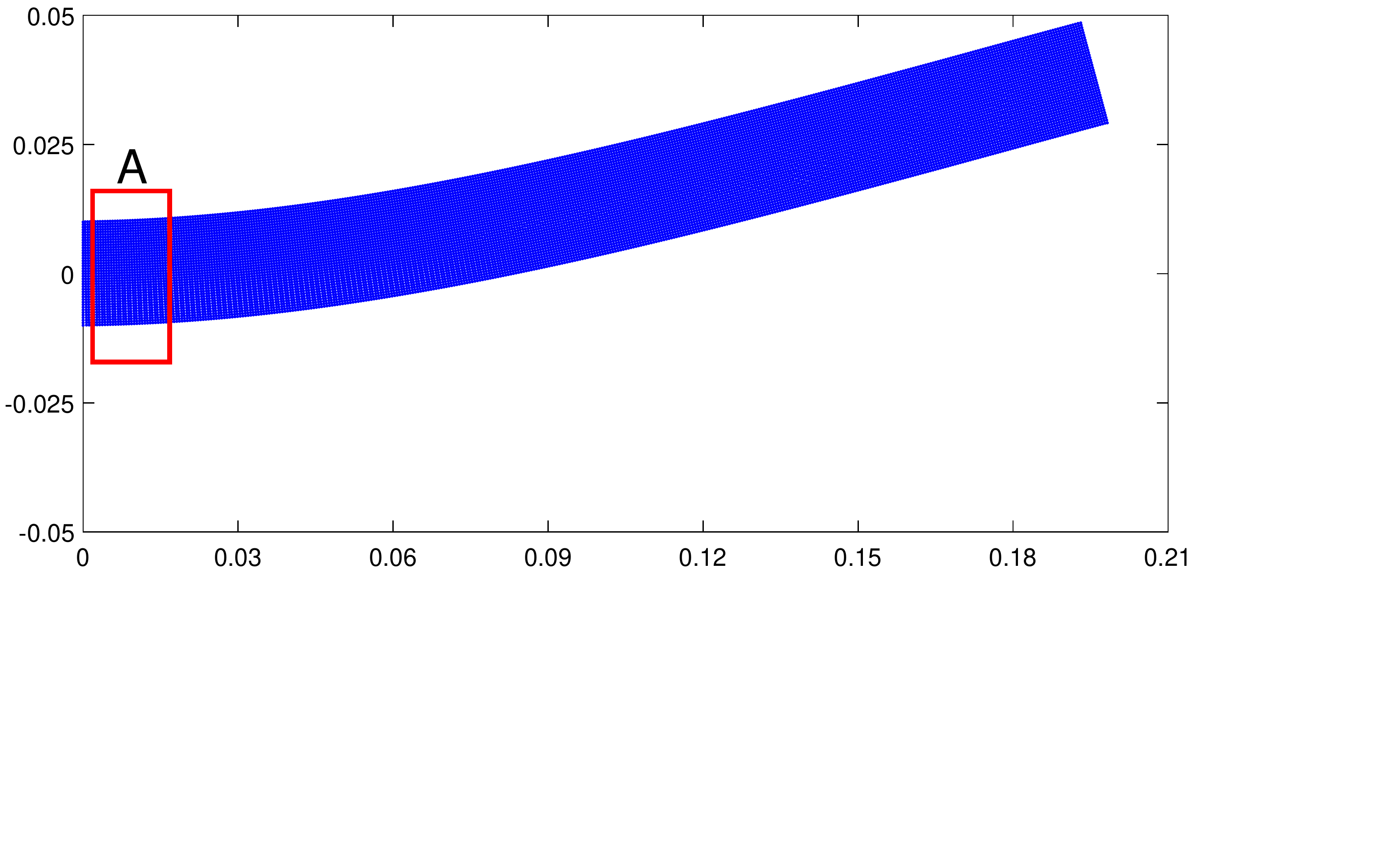}
				\caption{$t = T/4$}
		\end{subfigure} \\
		
		\begin{subfigure}[t]{0.5\textwidth}
			\centering
				\includegraphics[trim={0cm 6.5cm 5cm 0cm},width=1.0\textwidth]{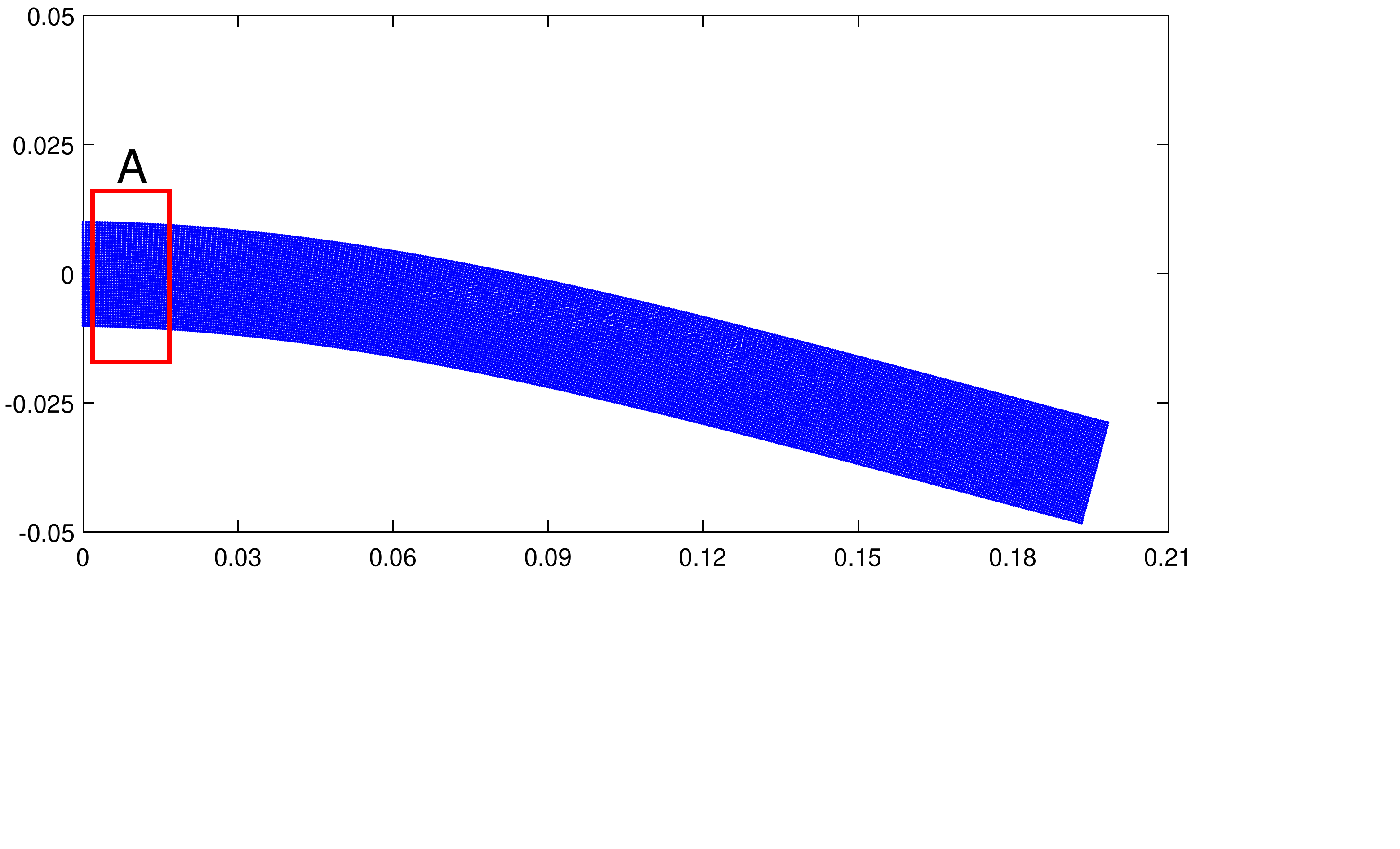}
				\caption{$t = 3T/8$}
		\end{subfigure}%
		~
		\begin{subfigure}[t]{0.5\textwidth}
			\centering
				\includegraphics[trim={0cm 6.5cm 5cm 0cm},width=1.0\textwidth]{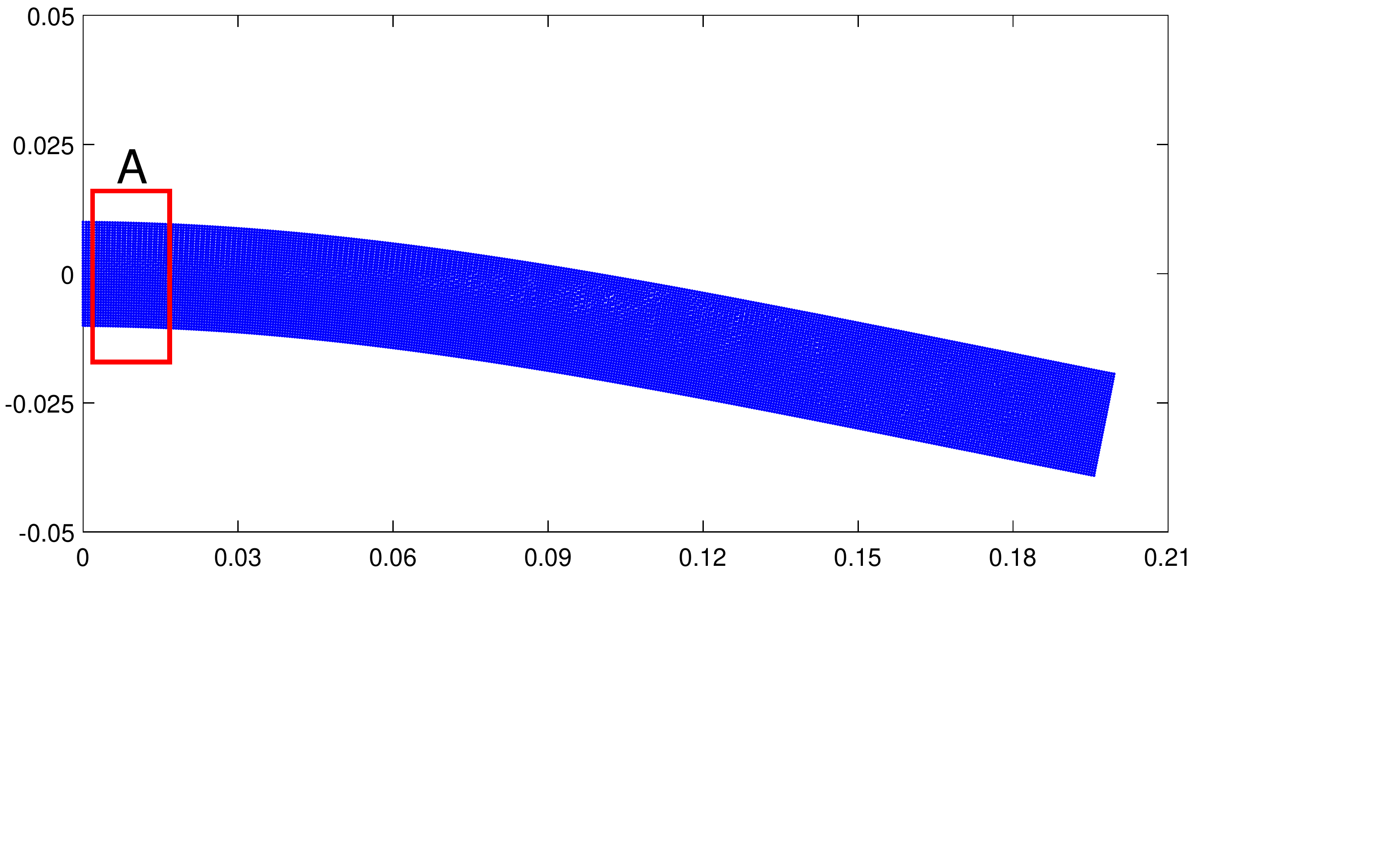}
				\caption{$t = 7T/8$}
		\end{subfigure}
		
		\begin{subfigure}[t]{0.25\textwidth}
			\centering
				\includegraphics[trim={1cm 2cm 16cm 0cm},width=1.0\textwidth]{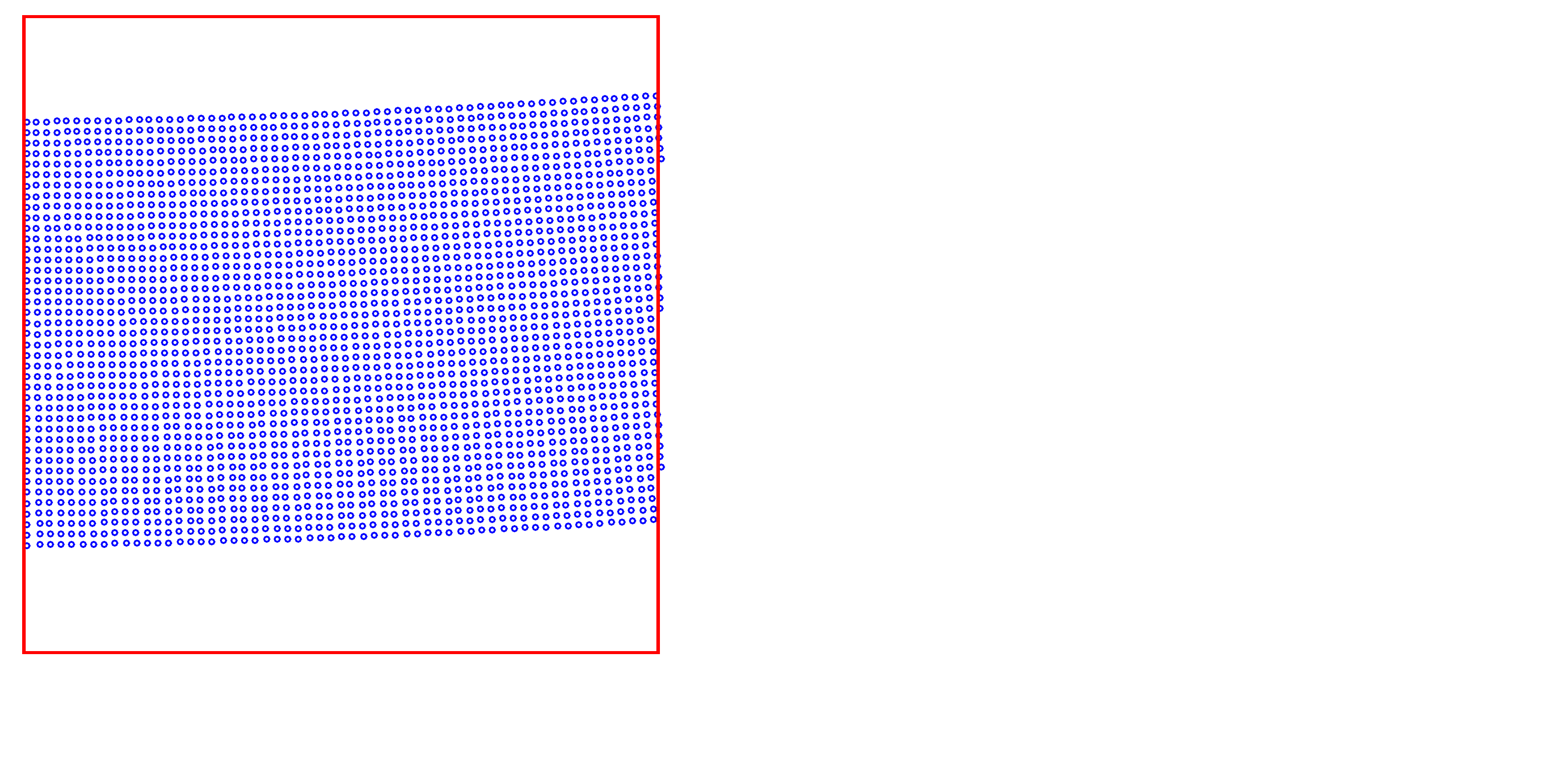}
				\caption{$t = T/8$}
		\end{subfigure}%
		~
		\begin{subfigure}[t]{0.25\textwidth}
			\centering
				\includegraphics[trim={1cm 2cm 16cm 0cm},width=1.0\textwidth]{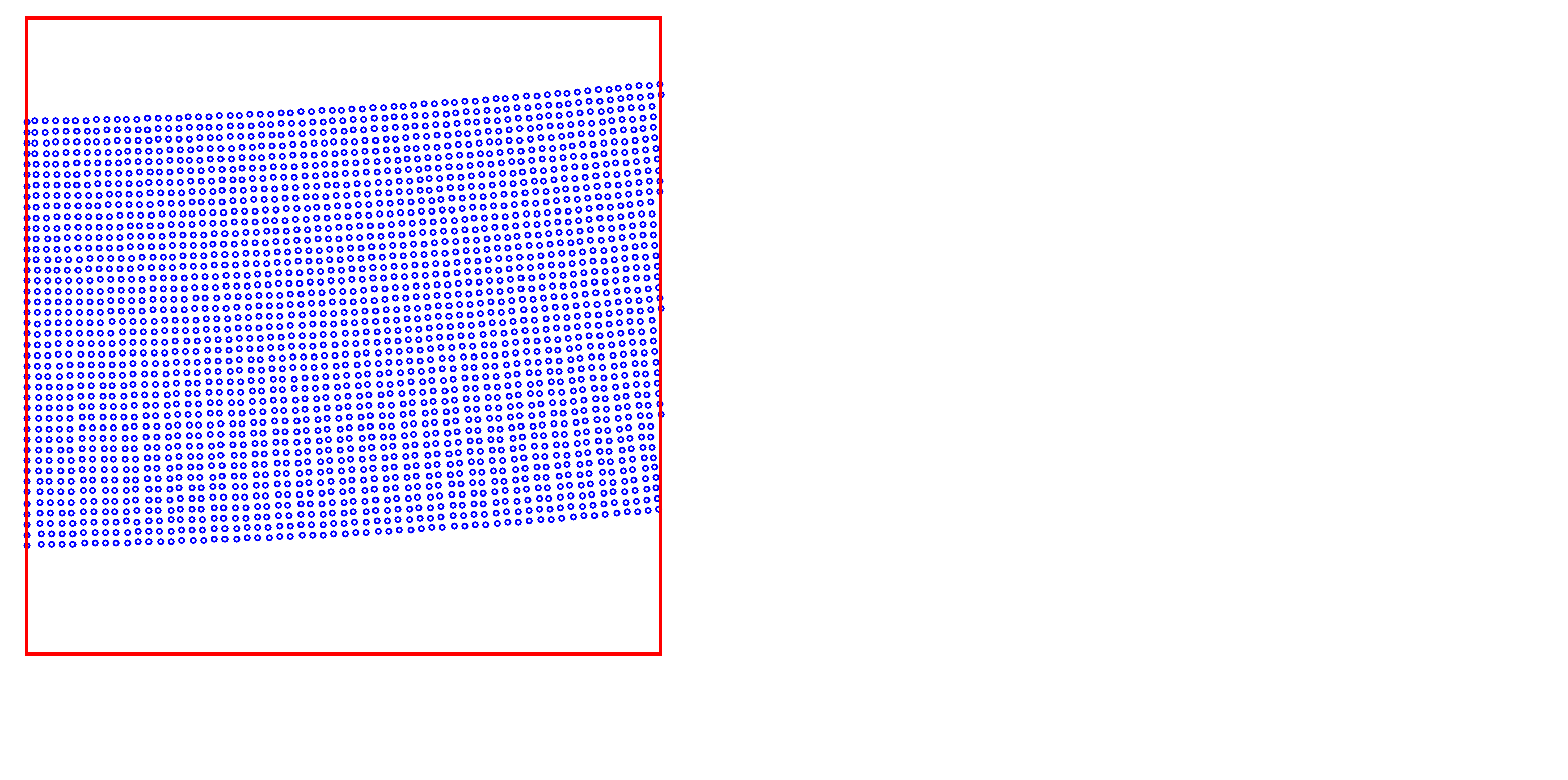}
				\caption{$t = T/4$}
		\end{subfigure}%
		~
		\begin{subfigure}[t]{0.25\textwidth}
			\centering
				\includegraphics[trim={1cm 2cm 16cm 0cm},width=1.0\textwidth]{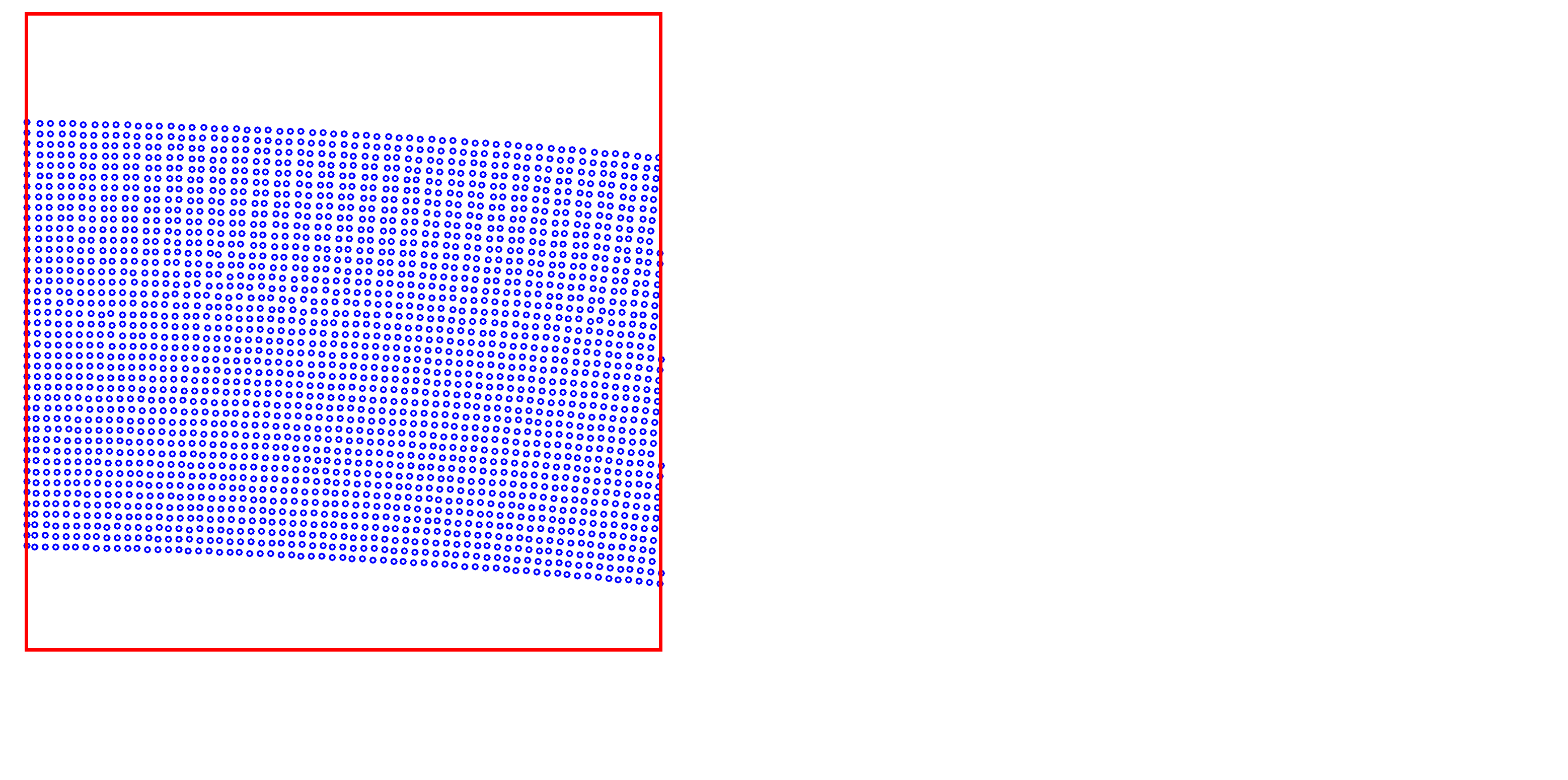}
				\caption{$t = 3T/8$}
		\end{subfigure}%
		~
		\begin{subfigure}[t]{0.25\textwidth}
			\centering
				\includegraphics[trim={1cm 2cm 16cm 0cm},width=1.0\textwidth]{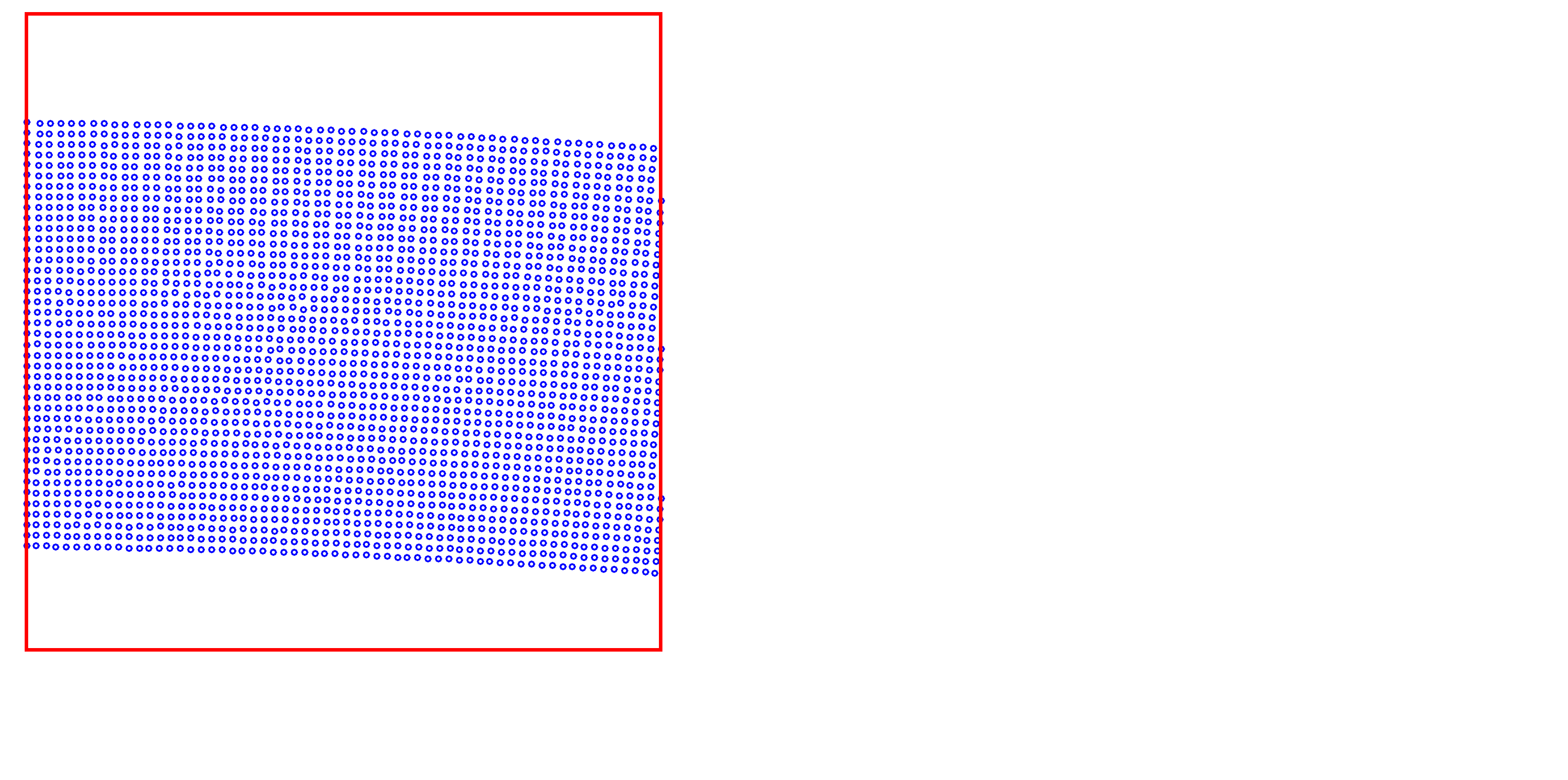}
				\caption{$t = 7T/8$}
		\end{subfigure}
	\caption{Displacement profiles at different time instants obtained via SPH with adaptive kernel, T = 2.35 ms; (e) - (h) are the enlarged view A at different time instants}\label{P3_adp}\vspace*{6pt}
\end{figure}

Deformed configurations of the plate at different time instants obtained via the standard cubic kernel and the adaptive kernel are shown in Figure \ref{P3_st} and \ref{P3_adp} respectively. As evident from the figures, while standard cubic kernel (Figure \ref{P3_st}) shows instability near the fixed end, no prominent instability is observed in case of adaptive kernel (Figure \ref{P3_adp}). 

\begin{figure}[h!]\vspace*{4pt}
\centerline{\includegraphics[trim={1cm 0 5cm 0},clip,width=\textwidth]{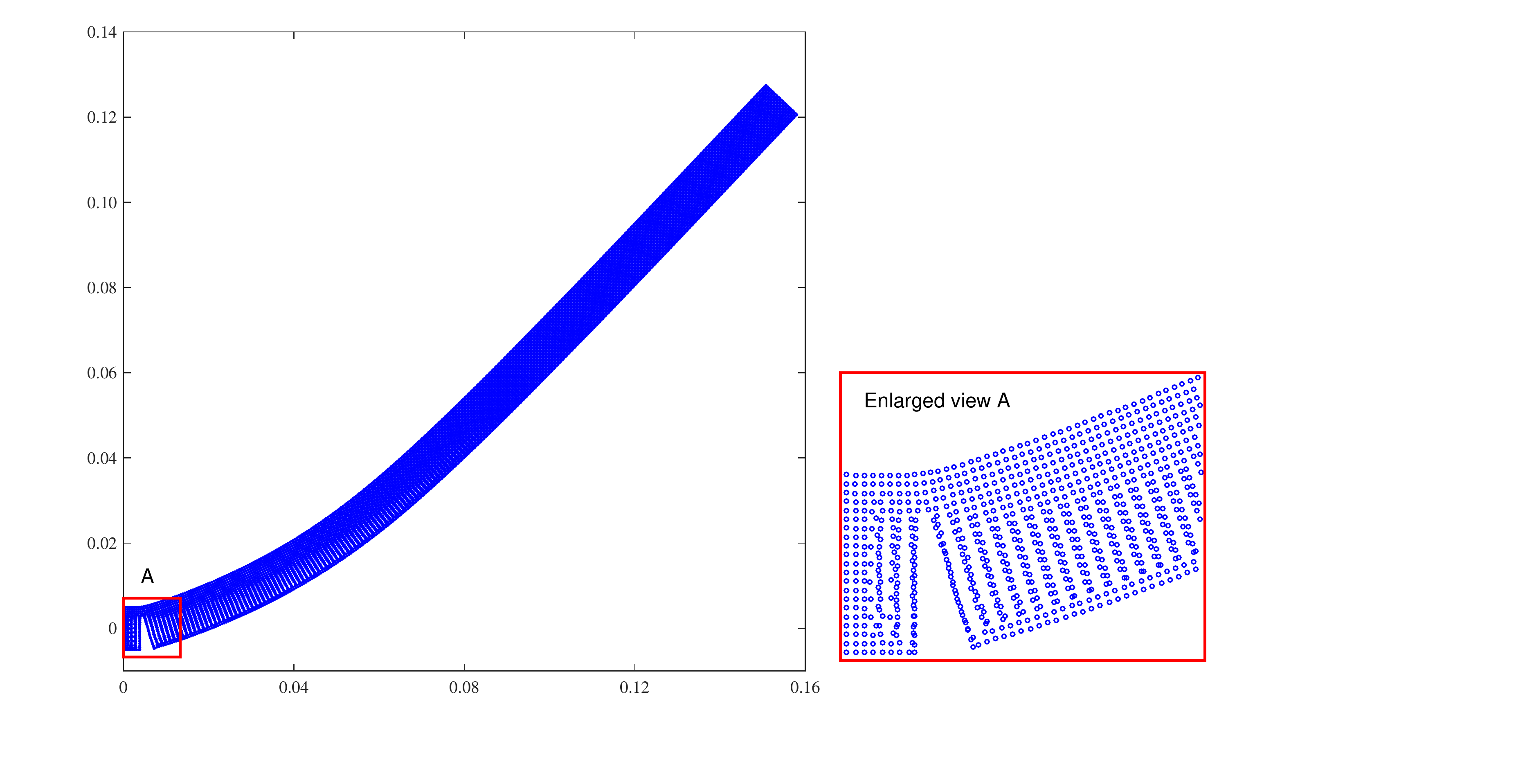}}
\caption{Deformed configuration of the plat (L = 200mm , B = 20mm) obtained via standard SPH}\vspace*{-6pt}
\label{long_beam_st}
\end{figure}

\begin{figure}[h!]\vspace*{4pt}
\centerline{\includegraphics[trim={1cm 0 5cm 0},clip,width=\textwidth]{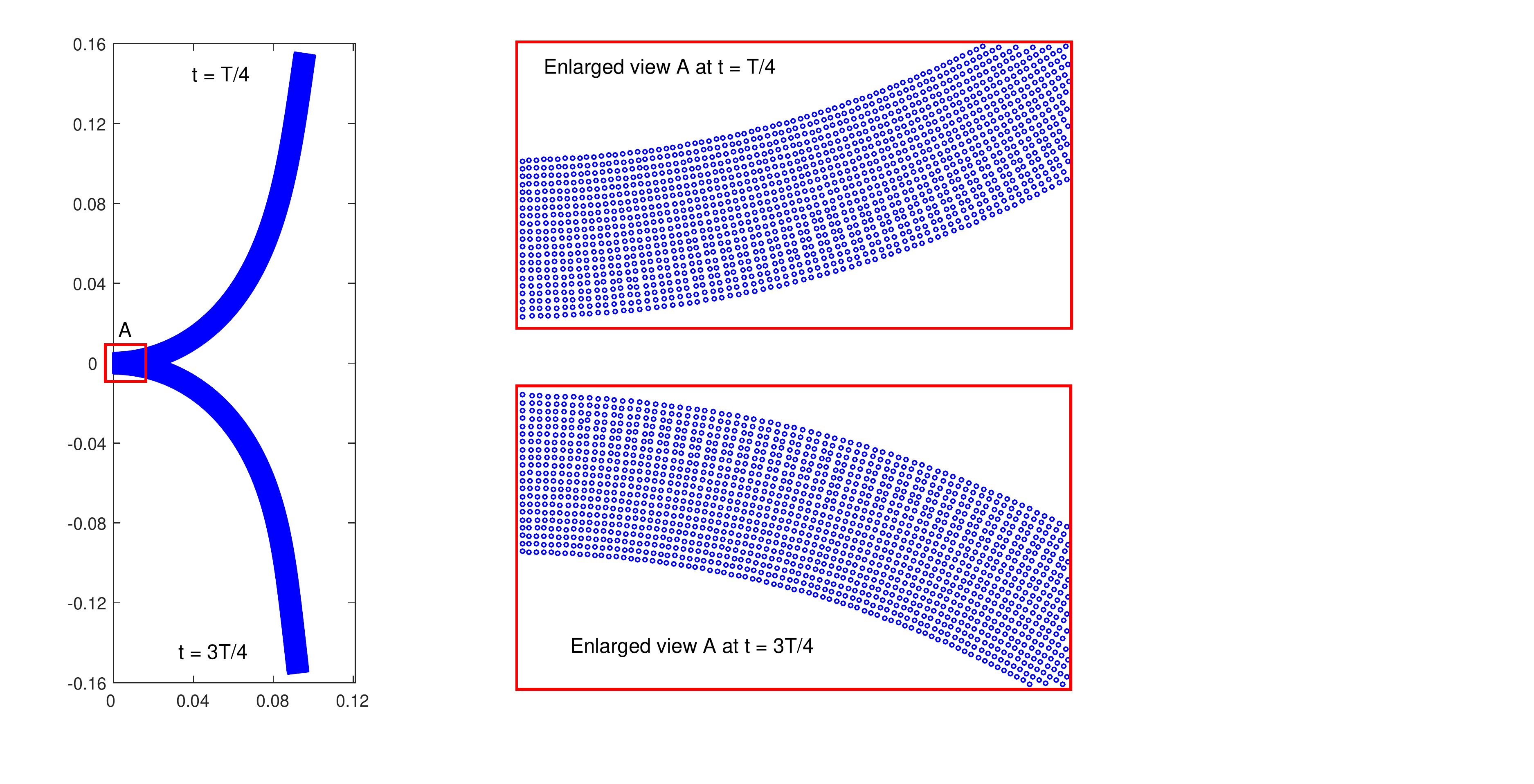}}
\caption{Deformed configurations of the plat (L = 200mm , B = 20mm) obtained via SPH with adaptive kernel, T = 4.62 ms}\vspace*{-6pt}
\label{beam_long_adb}
\end{figure}

\begin{figure}[h!]\vspace*{4pt}
\centerline{\includegraphics[trim={0 0 0 0},clip,width=0.7\textwidth]{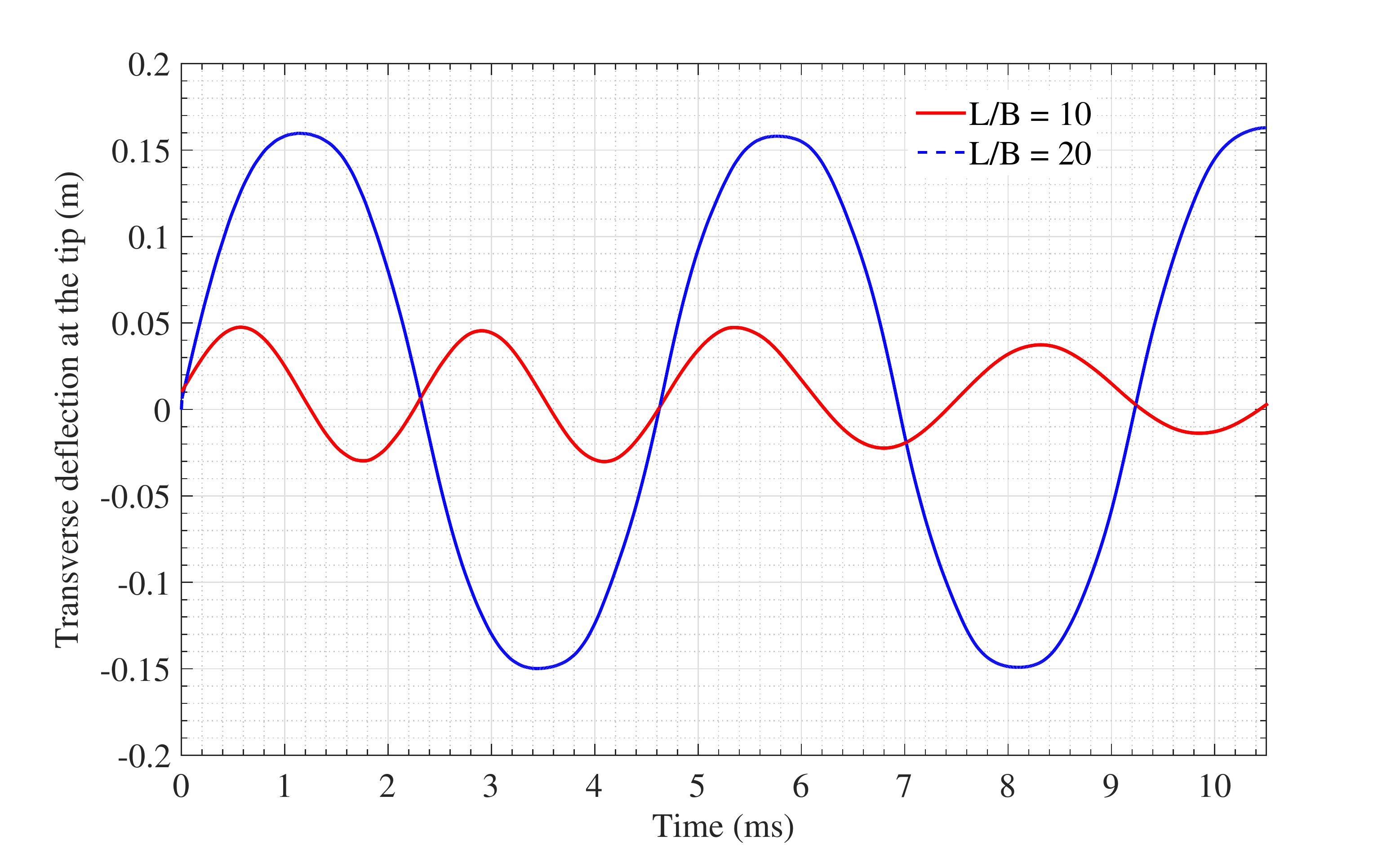}}
\caption{Transverse displacement time histories at the tip of the cantilevr plate}\label{P3_hist}
\end{figure}

In order to investigate the performance of the adaptive kernel in a situation where structure undergoes excessive deformation, next a similar plate problem but with different L/B ratio (L = 200mm and B = 10mm) and initial condition $V_f$ = 0.05 m/s is considered. When simulation is performed with the standard cubic kernel, computation suffers an unphysical numerical fracture (Figure \ref{long_beam_st}). Adaptive kernel, on the other hand, continues to perform in a stable manner as illustrated in Figure  \ref{beam_long_adb}. The computed time period (T) is 4.62 ms which is very close to the theoretical value of 4.56 ms. Time histories of the transverse displacement at the tip (for both L/B = 10 and L/B = 20) obtained via the adaptive kernel are shown in Figure \ref{P3_hist}. 

\subsection{Collision of two rubber rings} 
\begin{figure}[h!]
	\centering
		\begin{subfigure}[t]{0.5\textwidth}
			\centering
				\includegraphics[trim={0 0 0 0},width=0.5\textwidth]{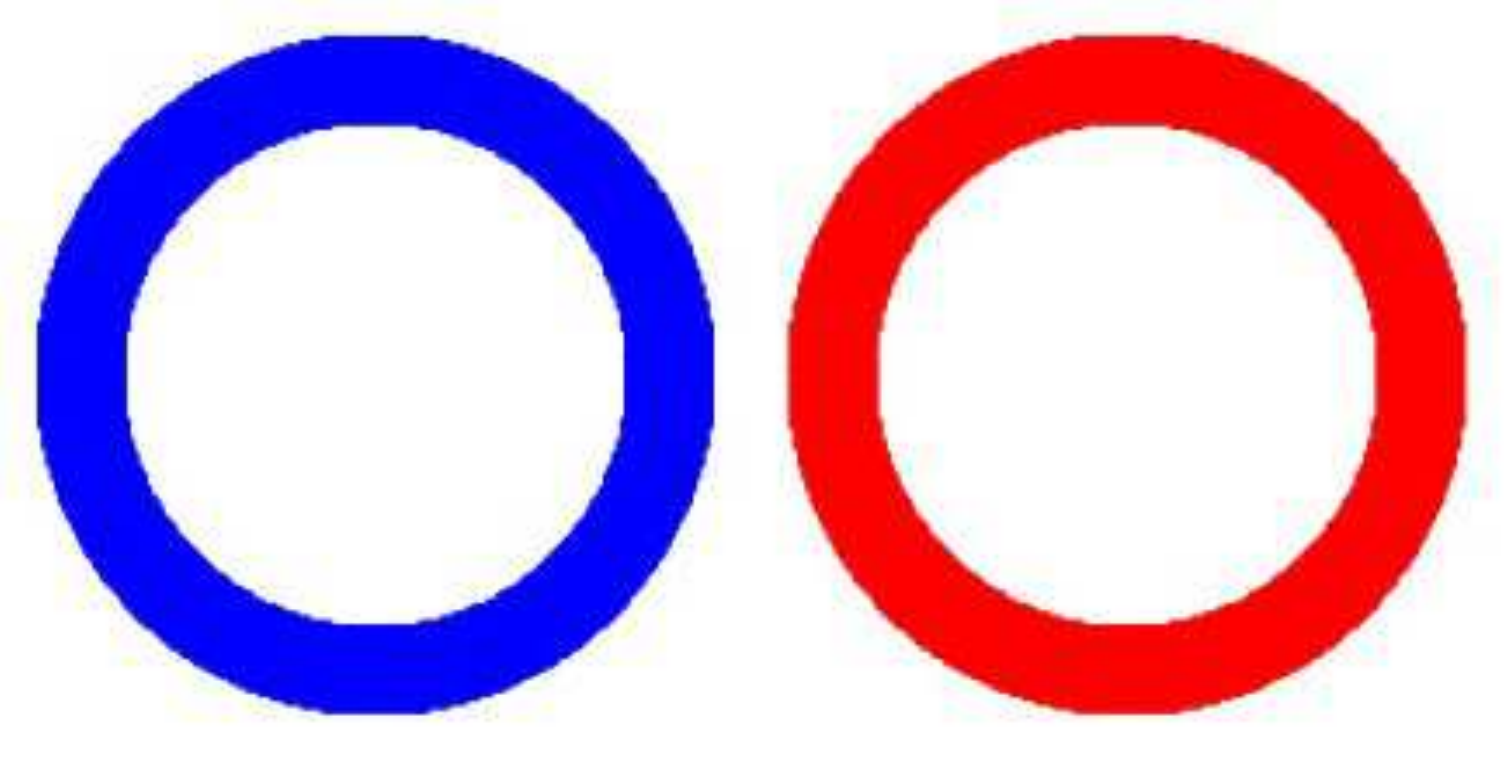}\label{P4_st1}
				\caption{$t = 0 s$}
		\end{subfigure}%
		\begin{subfigure}[t]{0.5\textwidth}
			\centering
				\includegraphics[trim={0 0 0 0},width=0.4\textwidth]{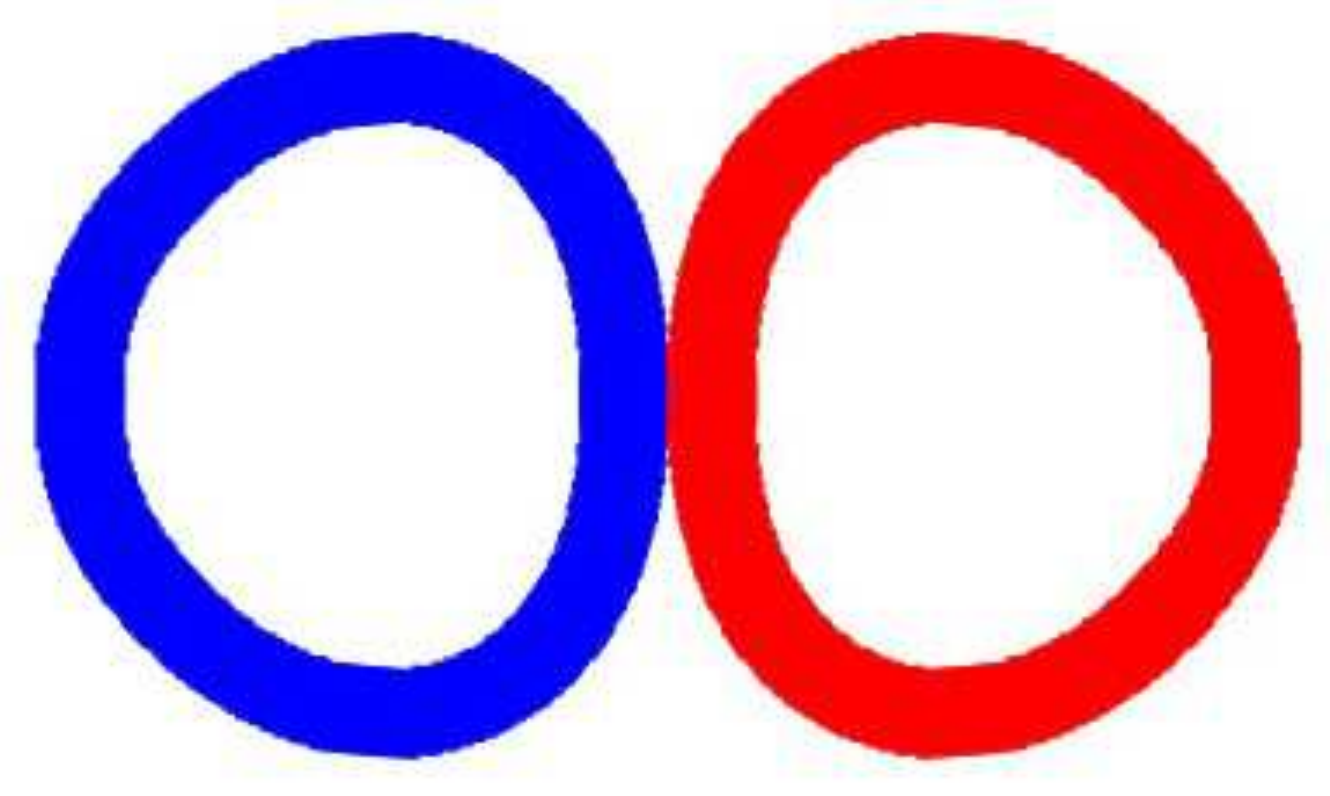}\label{P4_st21}
				\caption{$t = 200 \mu s$}
		\end{subfigure}\\
		\begin{subfigure}[t]{0.5\textwidth}
			\centering
				\includegraphics[trim={0 0 0 0},width=0.4\textwidth]{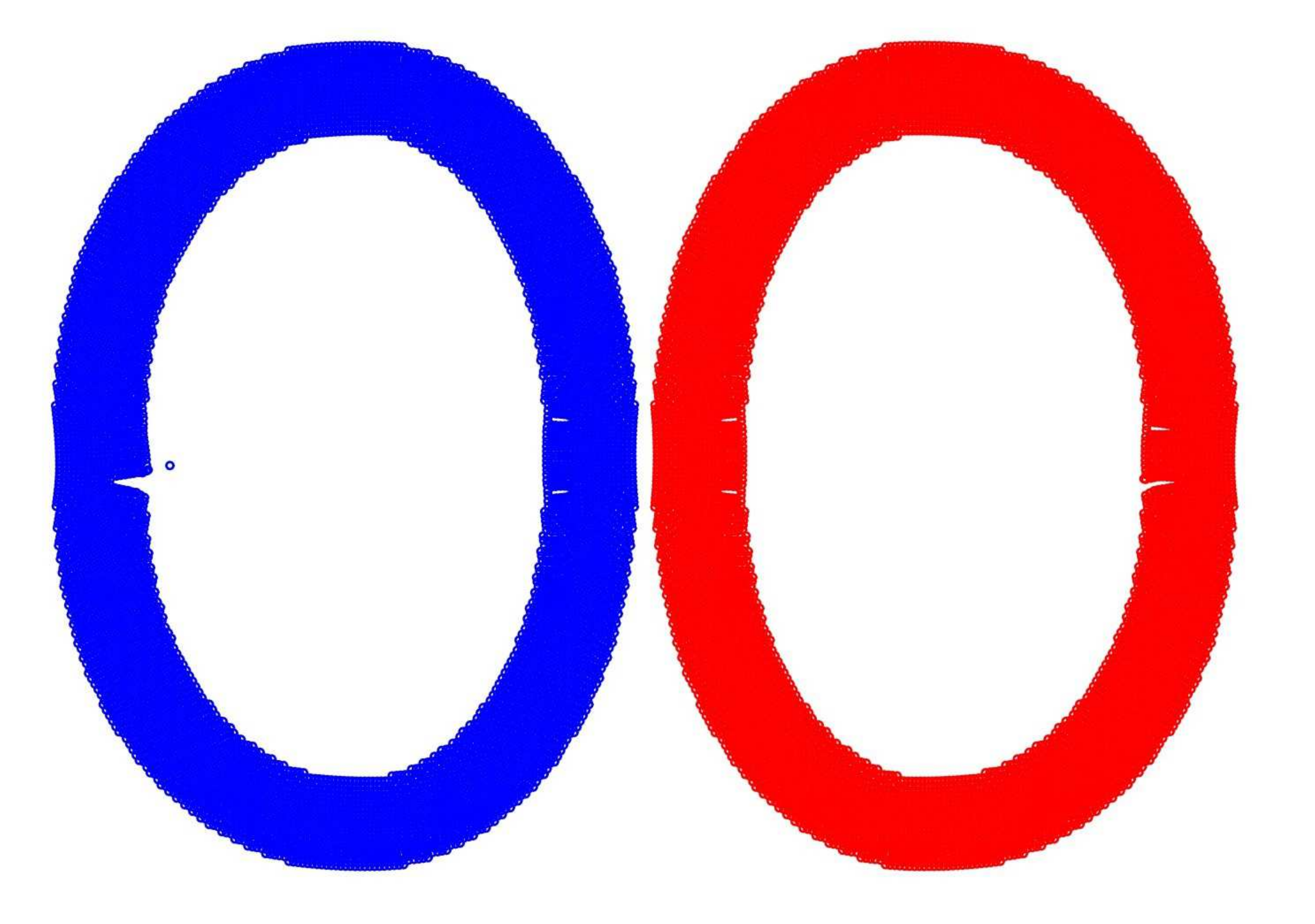}\label{P4_st41}
				\caption{$t = 400 \mu s$}
		\end{subfigure}%
		\begin{subfigure}[t]{0.5\textwidth}
			\centering
				\includegraphics[trim={0 0 0 0},width=0.4\textwidth]{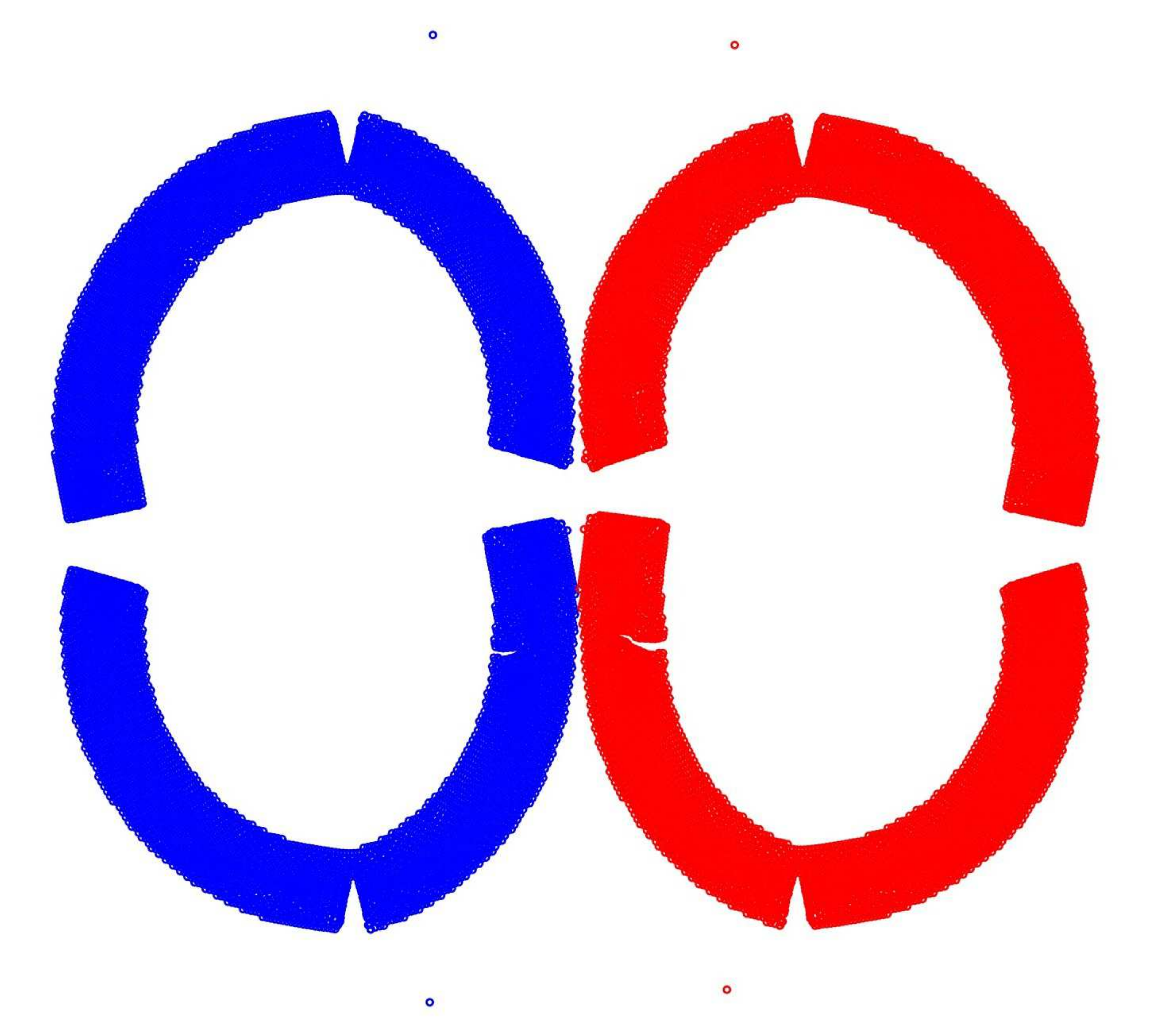}\label{P4_st61}
				\caption{$t = 600 \mu s$}
		\end{subfigure}
	\caption{Configurations of the rubber rings at different time instants obtained via standard SPH}\label{P4_st}
\end{figure}

\begin{figure}[h!]
	\centering
		\begin{subfigure}[t]{0.5\textwidth}
			\centering
				\includegraphics[trim={0 0 0 0},width=0.4\textwidth]{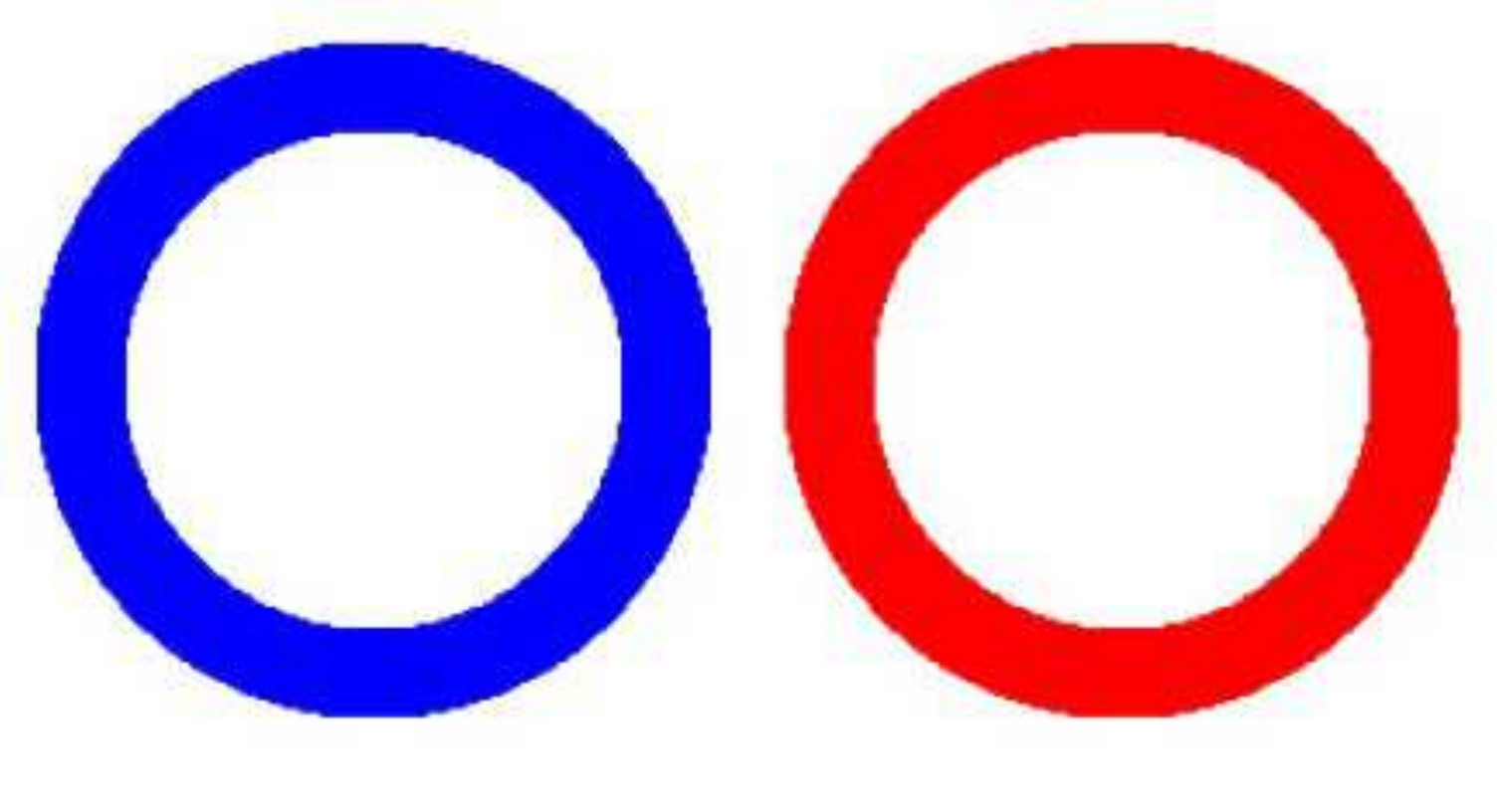}\label{P4_adp1}
				\caption{$t = 0 s$}
		\end{subfigure}%
		\begin{subfigure}[t]{0.5\textwidth}
			\centering
				\includegraphics[trim={0 0 0 0},width=0.4\textwidth]{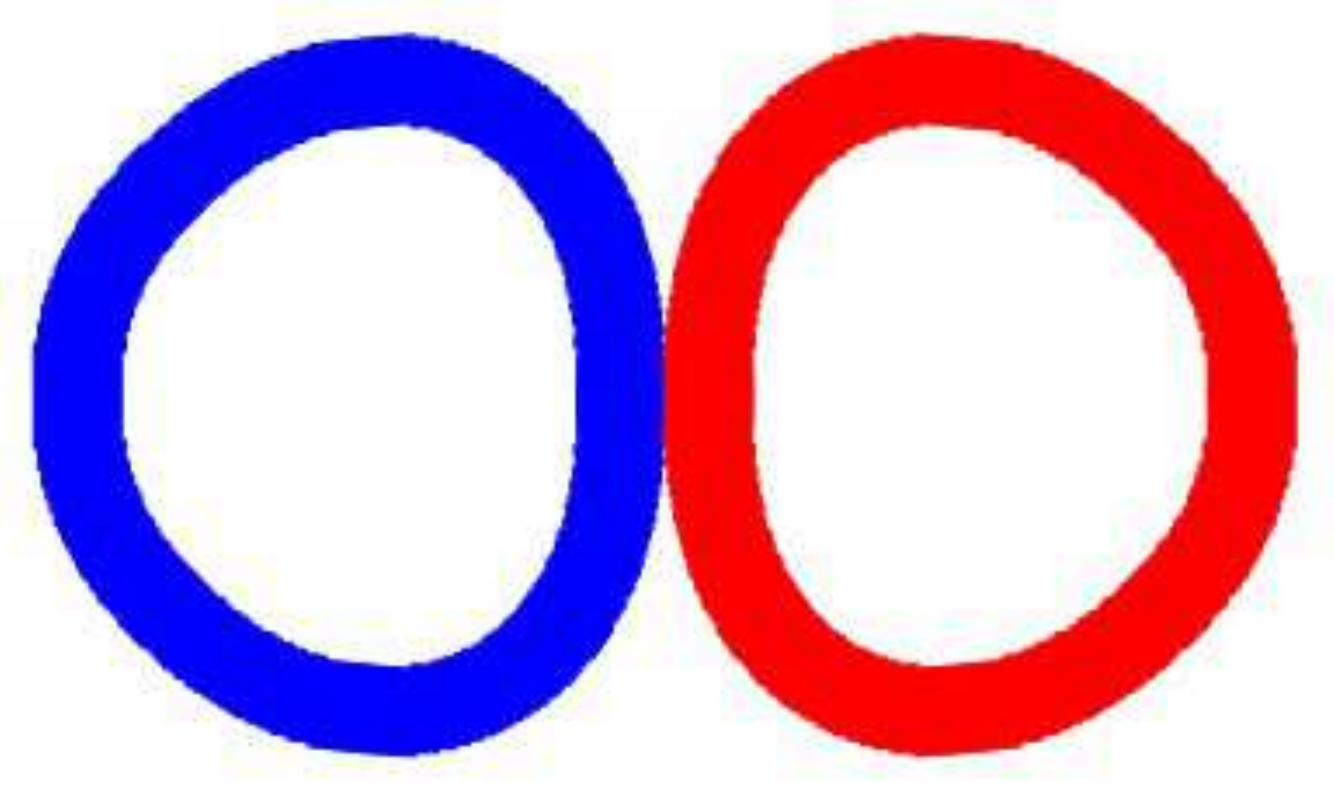}\label{P4_adp21}
				\caption{$t = 200 \mu s$}
		\end{subfigure}\\
		\begin{subfigure}[t]{0.5\textwidth}
			\centering
				\includegraphics[trim={0 0 0 0},width=0.4\textwidth]{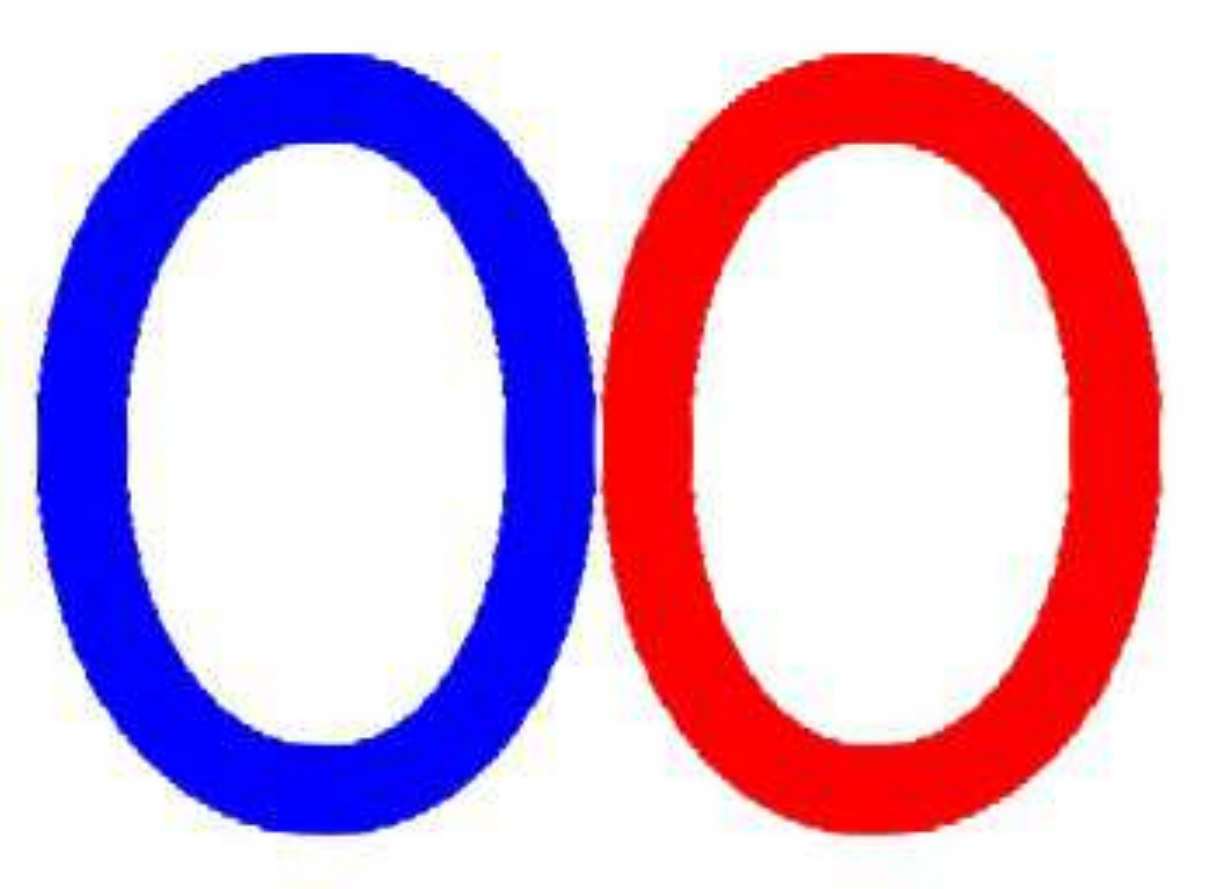}\label{P4_adp41}
				\caption{$t = 400 \mu s$}
		\end{subfigure}%
		\begin{subfigure}[t]{0.5\textwidth}
			\centering
				\includegraphics[trim={0 0 0 0},width=0.4\textwidth]{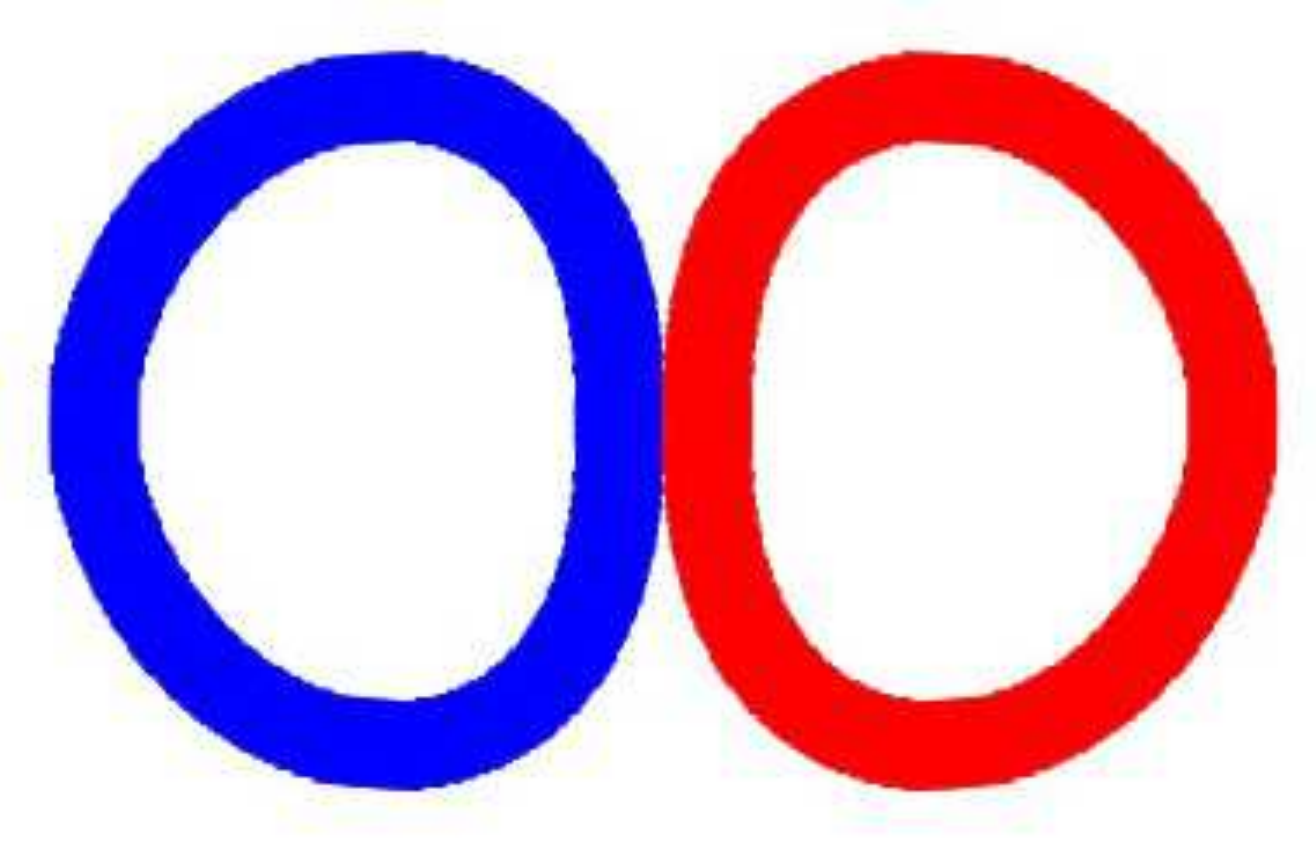}\label{P4_adp61}
				\caption{$t = 600 \mu s$}
		\end{subfigure}
		\begin{subfigure}[t]{0.5\textwidth}
			\centering
				\includegraphics[trim={0 0 0 0},width=0.4\textwidth]{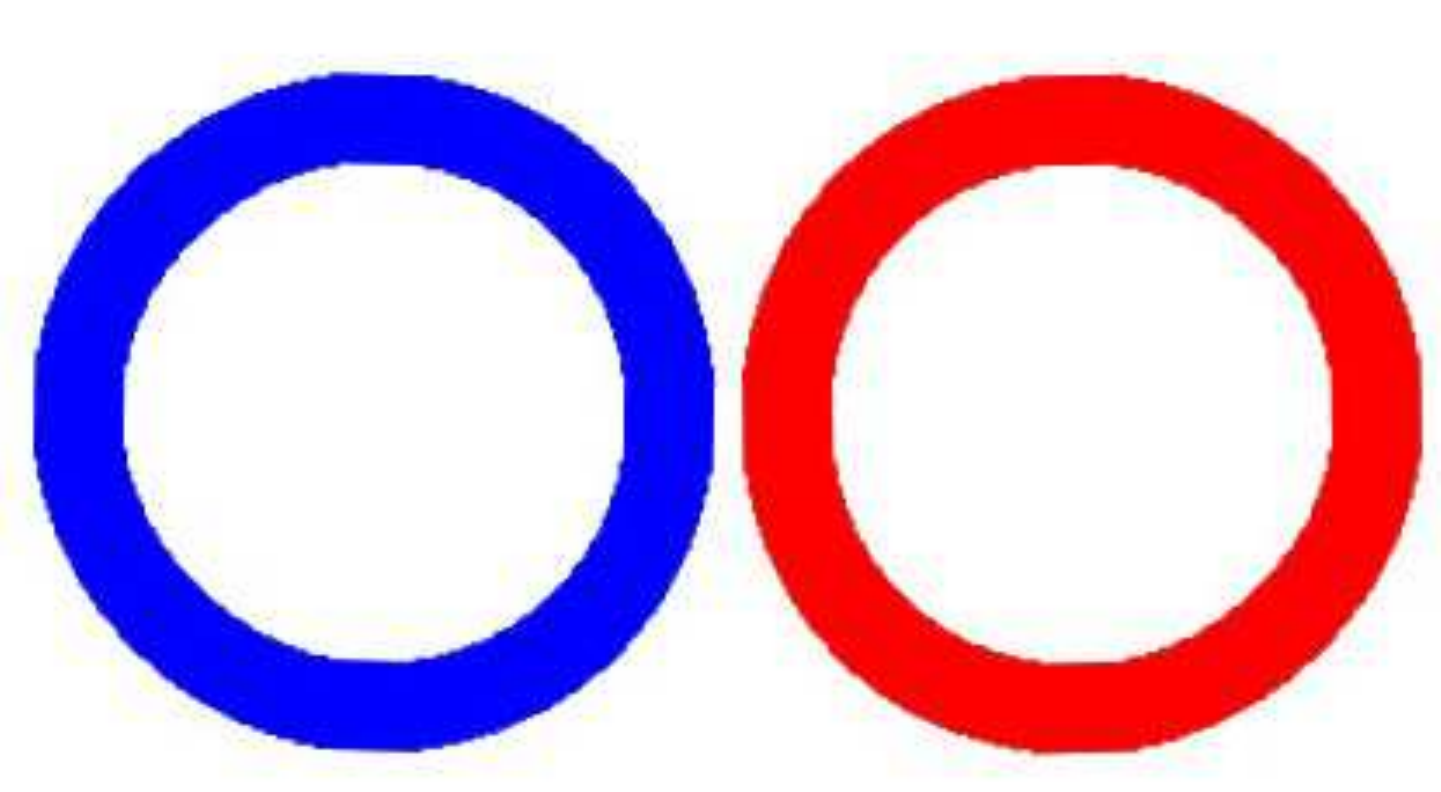}\label{P4_adp81}
				\caption{$t = 800 \mu s$}
		\end{subfigure}%
		\begin{subfigure}[t]{0.5\textwidth}
			\centering
				\includegraphics[trim={0 0 0 0},width=0.5\textwidth]{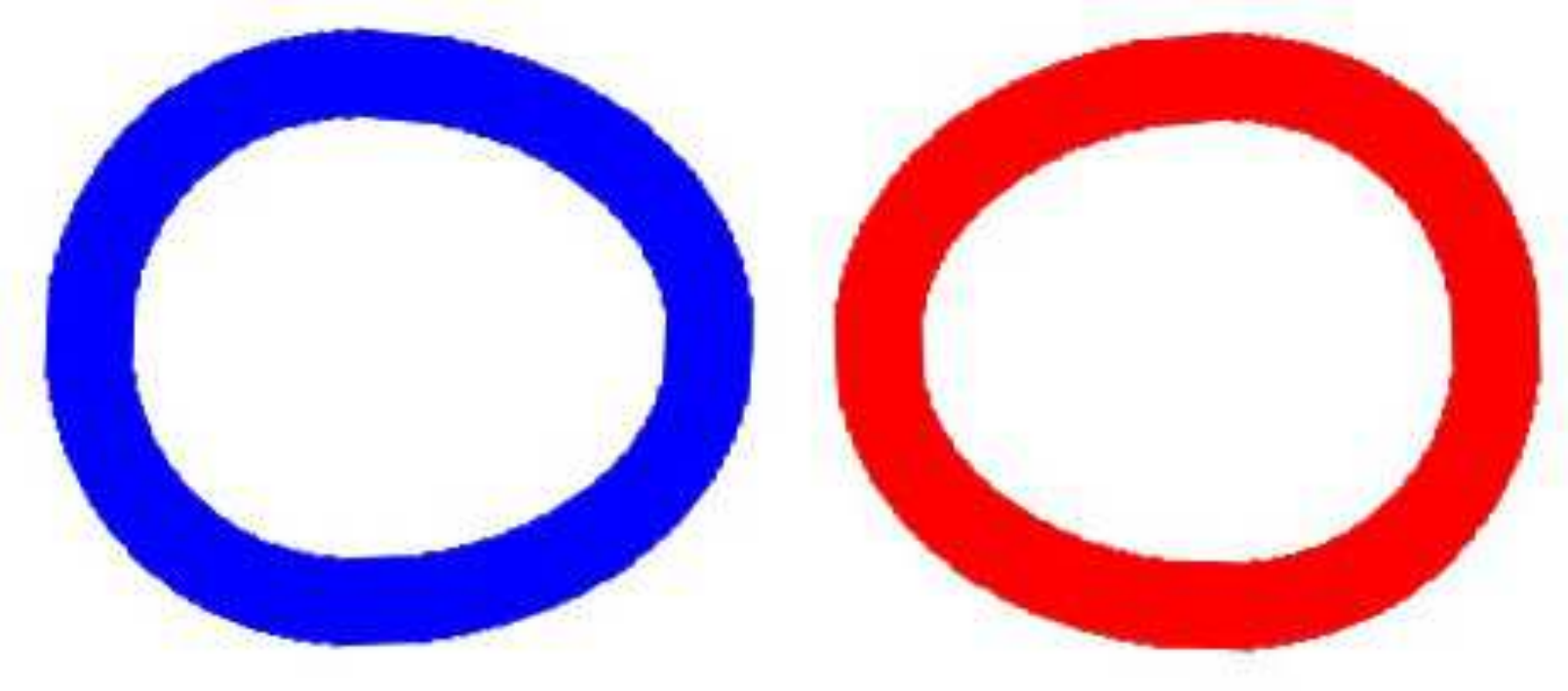}\label{P4_adp101}
				\caption{$t = 1000 \mu s$}
		\end{subfigure}%
	\caption{Configurations of the rubber rings at different time instants obtained via SPH with adaptive kernel}\label{P4_adp}
\end{figure}

As the last example, elastic collision of two rubber rings with inner diameter 30mm and outer diameter 40mm at a velocity 50m/s is considered. Material properties and other computational data are given in Table \ref{rubber_value}. The deformed configurations of the rings at different time instants are shown in Figure \ref{P4_st} and \ref{P4_adp}. Instability is clearly visible in SPH with standard Cubic B-spline kernel (Figure \ref{P4_st}). However on implementing the adaptive kernel (Figure \ref{P4_adp}) the instability is removed from the system.  

\begin{table}[hbtp!]
\centering
\caption{Parameters for the simulation of rubber collision}\label{rubber_value}
\begin{tabular}{cccccccc}
\hline
                           & \multicolumn{3}{c}{Material Proerties}      & \multicolumn{2}{c}{Discretization} & \multicolumn{2}{c}{Simulation}                                              \\
\multirow{2}{*} & $\rho$     & $E$   & \multirow{2}{*}{$\nu$} & $\Delta p$          & $\Delta t$          & \multirow{2}{*}{$\gamma_1$} & \multirow{2}{*}{$\gamma_2$}         \\
                           & ($kg/m^3$) & (GPa) &                        & (mm)                & (sec)         &                            &                             \\ \hline
                      & 1010       & 0.73  & 0.4                    & 0.50                & $2.5\times 10^{-8}$         & 1.0                        & 1.0                                      \\ \hline
\end{tabular}
\end{table}

\section{Closure}
Like any other computational tool, SPH also has its share of limitations. The \textit{tensile instability} is perhaps the major stumbling block among them. The cause of the instability lies in the core of the formulation, which uses a bell-shaped compactly supported kernel function for discretizing the strong form of the governing Equations. Swegle's \citep{swegle1994analysis} analysis suggests that stability can be enforced by either modifying the stress tensor or choosing a suitable kernel such that their interaction does not induce any spurious negative stiffness. The second approach is adopted in this study.

Herein an adaptive algorithm is proposed where the shape of the kernel function is continuously modified depending on the state of stress such that the condition associated with \textit{tensile instability} does not arise. Towards this, the B-spline basis function constructed over a symmetric knot vector is taken as the kernel function, and its shape is adapted by changing the location of the intermediate knot. Theoretical framework and implementation of the algorithm are discussed in detail. One dimensional dispersion analysis is also performed to have a better insight into the effect of the intermediate knot on the stability. The most distinguishing feature of the proposed algorithm is that it does not require tuning of any user-defined parameters, unlike common practices available in the literature. Moreover, the additional computational effort needed for the algorithm is also insignificant.        

The efficacy of the algorithm is demonstrated through some benchmark elastic dynamics problems. The proposed approach shows great potential in removing tensile instability at an affordable computational effort. The applicability may further be extended to different class of problems which may constitute the topic for future work.

\section*{Acknowledgement}
The authors would like to acknowledge the Naval Research Board, DRDO, India for their support of this work.

\section*{References}


\bibliographystyle{elsarticle-num-names}

\bibliography{asph-archive}

\end{document}